\documentclass[prd,a4,superscriptaddress,nofootinbib]{revtex4}
\usepackage{bm,amsmath,amssymb,epsfig,natbib}
\usepackage{hyperref}
\usepackage{ifthen}

\begin{document}
\newcommand{\nside}{N_{\text{side}}}
\newcommand{\arcmin}{\text{arcmin}}
\newcommand{\begm}{\begin{pmatrix}}
\newcommand{\enm}{\end{pmatrix}}
\newcommand{\threej}[6]{{\begm #1 & #2 & #3 \\ #4 & #5 & #6 \enm}}
\newcommand{\Coff}{{\hat{C}_l^\off}}
\newcommand{\lmax}{l_\text{max}}
\newcommand{\lmin}{l_{\text{min}}}
\newcommand{\fsky}{f_{\text{sky}}}
\newcommand{\off}{{\text{off}}}
\newcommand{\chieff}{\chi^2_{\text{eff}}}
\renewcommand{\ell}{l}
\renewcommand{\bar}[1]{#1}
\newcommand{\nHI}{{n_{HI}}}
\newcommand{\clh}{\mathcal{H}}
\newcommand{\ud}{{\rm d}}

\def\eprinttmp@#1arXiv:#2 [#3]#4@{
\ifthenelse{\equal{#3}{x}}{\href{http://arxiv.org/abs/#1}{#1}}{\href{http://arxiv.org/abs/#2}{arXiv:#2} [#3]}}

\renewcommand{\eprint}[1]{\eprinttmp@#1arXiv: [x]@}
\newcommand{\adsurl}[1]{\href{#1}{ADS}}
\renewcommand{\bibinfo}[2]{\ifthenelse{\equal{#1}{isbn}}{%
\href{http://cosmologist.info/ISBN/#2}{#2}}{#2}}

\newcommand\ba{\begin{eqnarray}}
\newcommand\ea{\end{eqnarray}}
\newcommand\be{\begin{equation}}
\newcommand\ee{\end{equation}}
\newcommand\lagrange{{\cal L}}
\newcommand\cll{{\cal L}}
\newcommand\clx{{\cal X}}
\newcommand\clz{{\cal Z}}
\newcommand\clv{{\cal V}}
\newcommand\clo{{\cal O}}
\newcommand\cla{{\cal A}}
\newcommand{\uD}{{\mathrm{D}}}
\newcommand{\calE}{{\cal E}}
\newcommand{\calB}{{\cal B}}
\newcommand{\curl}{\,\mbox{curl}\,}
\newcommand\del{\nabla}
\newcommand\Tr{{\rm Tr}}
\newcommand\half{{\frac{1}{2}}}
\renewcommand\H{{\cal H}}
\newcommand\K{{\rm K}}
\newcommand\mK{{\rm mK}}
\newcommand{\clk}{{\cal K}}
\newcommand{\bq}{\bar{q}}
\newcommand{\bv}{\bar{v}}
\renewcommand\P{{\cal P}}
\newcommand{\numfrac}[2]{{\textstyle \frac{#1}{#2}}}
\newcommand{\la}{\langle}
\newcommand{\ra}{\rangle}
\newcommand{\rar}{\rightarrow}
\newcommand{\Rar}{\Rightarrow}
\newcommand\gsim{ \lower .75ex \hbox{$\sim$} \llap{\raise .27ex \hbox{$>$}} }
\newcommand\lsim{ \lower .75ex \hbox{$\sim$} \llap{\raise .27ex \hbox{$<$}} }
\newcommand\bigdot[1] {\stackrel{\mbox{{\huge .}}}{#1}}
\newcommand\bigddot[1] {\stackrel{\mbox{{\huge ..}}}{#1}}
\newcommand{\Mpc}{\text{Mpc}}
\newcommand{\Al}{{A_l}}
\newcommand{\Bl}{{B_l}}
\newcommand{\eAl}{e^\Al}
\newcommand{\ix}{{(i)}}
\newcommand{\ixp}{{(i+1)}}
\renewcommand{\k}{\beta}
\newcommand{\HD}{\mathrm{D}}
\newcommand{\mCh}{\hat{\bm{C}}}
\newcommand{\mCf}{{{\bm{C}}_{f}}}
\newcommand{\mCXf}{{{\bm{C}}_{Xf}}}
\newcommand{\mMXf}{{{\bm{M}}_{f}}}
\newcommand{\cov}{\text{cov}}
\newcommand{\Cfl}{{C_f}_l}
\renewcommand{\vec}{\text{vec}}
\newcommand{\muK}{\mu \rm{K}}

\newcommand{\vecp}{\text{vecp}}
\newcommand{\vecL}{\text{vecl}}

\newcommand{\mCfl}{{\mC_{f}}_l}
\newcommand{\mCgl}{{\mC_{g}}_l}

\newcommand{\Ch}{\hat{C}}
\newcommand{\Bt}{\tilde{B}}
\newcommand{\Et}{\tilde{E}}
\newcommand{\bld}[1]{\mathrm{#1}}
\newcommand{\mLambda}{\bm{\Lambda}}
\newcommand{\mA}{\bm{A}}
\newcommand{\mB}{\bm{B}}
\newcommand{\mBp}{\mB_n}

\newcommand{\mC}{\bm{C}}
\newcommand{\mD}{\bm{D}}
\newcommand{\mE}{\bm{E}}
\newcommand{\mF}{\bm{F}}
\newcommand{\mg}{\bm{g}}

\newcommand{\mQ}{\bm{Q}}
\newcommand{\mU}{\bm{U}}
\newcommand{\mX}{\bm{X}}
\newcommand{\mV}{\bm{V}}
\newcommand{\mP}{\bm{P}}
\newcommand{\mR}{\bm{R}}
\newcommand{\mW}{\bm{W}}
\newcommand{\mI}{\bm{I}}
\newcommand{\mH}{\bm{H}}
\newcommand{\mM}{\bm{M}}
\newcommand{\mN}{\bm{N}}
\newcommand{\mMh}{\hat{\mM}}

\newcommand{\mS}{\bm{S}}
\newcommand{\mzero}{\bm{0}}
\newcommand{\mL}{\bm{L}}
\newcommand{\btheta}{\bm{\theta}}
\newcommand{\bphi}{\bm{\psi}}
\newcommand{\va}{\mathbf{a}}
\newcommand{\vX}{\mathbf{X}}
\newcommand{\vchi}{\bm{\chi}}

\newcommand{\vXh}{\hat{\vX}}

\newcommand{\vS}{\mathbf{S}}
\newcommand{\vm}{\mathbf{m}}

\newcommand{\vN}{\mathbf{N}}
\newcommand{\vXhat}{\hat{\mathbf{X}}}
\newcommand{\vb}{\mathbf{b}}
\newcommand{\vA}{\mathbf{A}}
\newcommand{\vAt}{\tilde{\mathbf{A}}}
\newcommand{\ve}{\mathbf{e}}
\newcommand{\vE}{\mathbf{E}}
\newcommand{\vB}{\mathbf{B}}
\newcommand{\vl}{\mathbf{l}}
\newcommand{\vp}{\mathbf{p}}
\newcommand{\vXf}{\mathbf{X}_f}
\newcommand{\vEt}{\tilde{\mathbf{E}}}
\newcommand{\vBt}{\tilde{\mathbf{B}}}
\newcommand{\vEw}{\mathbf{E}_W}
\newcommand{\vBw}{\mathbf{B}_W}
\newcommand{\vx}{\mathbf{x}}
\newcommand{\vXt}{\tilde{\vX}}
\newcommand{\vXb}{\bar{\vX}}
\newcommand{\vTb}{\bar{\vT}}
\newcommand{\vTt}{\tilde{\vT}}
\newcommand{\vY}{\mathbf{Y}}
\newcommand{\vBwr}{{\vBw^{(R)}}}
\newcommand{\RW}{{W^{(R)}}}
\newcommand{\mUt}{\tilde{\mU}}
\newcommand{\mVt}{\tilde{\mV}}
\newcommand{\mDt}{\tilde{\mD}}

\newcommand{\healpix}{HEALPix}

\title{Likelihood Analysis of CMB Temperature and Polarization Power Spectra}

\author{Samira Hamimeche}
\email{samira@ast.cam.ac.uk}
 \affiliation{Institute of Astronomy, Madingley Road, Cambridge, CB3 0HA, UK.}

\author{Antony Lewis}
\homepage{http://cosmologist.info}
\affiliation{Institute of Astronomy, Madingley Road, Cambridge, CB3 0HA, UK.}

\date{\today}

\begin{abstract}
Microwave background temperature and polarization observations are a powerful way to constrain cosmological parameters if the likelihood function can be calculated accurately.
The temperature and polarization fields are correlated, partial sky coverage correlates power spectrum estimators at different $l$, and the likelihood function for a theory spectrum given a set of observed estimators is non-Gaussian. An accurate analysis must model all these properties.  Most existing likelihood approximations are good enough for a temperature-only analysis, however they cannot reliably handle a temperature-polarization correlations. We give a new general approximation applicable for correlated Gaussian fields observed on part of the sky. The approximation models the non-Gaussian form exactly in the ideal full-sky limit and is fast to evaluate using a pre-computed covariance matrix and set of power spectrum estimators.
We show with simulations that it is good enough to obtain correct results at $l\agt 30$ where an exact calculation becomes impossible. We also show that some Gaussian approximations give reliable parameter constraints even though they do not capture the shape of the likelihood function at each $l$ accurately. Finally we test the approximations on simulations with realistically anisotropic noise and asymmetric foreground mask.
\end{abstract}

\maketitle

\pagenumbering{arabic}

\section{Introduction}
The Cosmic Microwave Background (CMB) anisotropies are a powerful cosmological probe as they depend simply on the primordial inhomogeneities, content and  geometry of the Universe.  If the perturbations are Gaussian, the full-sky power spectra of the CMB anisotropies and their polarization contain all of the cosmological information.  Information in the polarization power spectra can help to break degeneracies that are present if only temperature information is used, and also helps to reduce cosmic variance uncertainty. Parameter constraints can therefore be significantly improved by using polarization information even if the data is significantly noisier than the temperature.


An accurate joint likelihood analysis of the CMB temperature and polarization data is crucial to estimate cosmological parameters reliably. In principle this is straightforward at linear order if the primordial perturbations are Gaussian as the distribution can be calculated exactly. However calculating the likelihood exactly from partial sky data with anisotropic noise is computationally prohibitive except at low $l$ because large matrices need to be inverted.  Most analyses therefore rely on approximations to the likelihood function at high $l$, using only the information in a set of estimators for the power spectra and a covariance estimated (or calibrated) from simulations. An alternative approach not considered further here would be to use the Gibbs sampling approach of Ref.~\cite{Wandelt:2003uk,Larson:2006ds}, though this has serious convergence problems of its own~\cite{Chu:2004zp}.

If the likelihood of the theory power spectrum $C_l$ as a function of the measured estimators $\Ch_l$ were Gaussian the likelihood could be calculated straightforwardly from the measured $\Ch_l$. However the distribution is non-Gaussian because for a given temperature power spectrum $C_l$, the $\Ch_l$, a sum of squares of Gaussian harmonic coefficients, have a (reduced) $\chi$-squared distribution. At large $l$ the distribution does tend to Gaussian by the central limit theorem; for example the mean and maximum likelihood values of $\Ch_l$ converge as $1/l$. However the precision with which we can hope to measure the cosmological parameters also improves at $1/l_{\text{max}}$, so the relative bias due to the non-Gaussianity is potentially independent of $l$. On all scales the distribution must be modeled carefully to get unbiased cosmological parameter constraints.

The importance of the non-Gaussianity of the temperature likelihood function at low $l$ is well known, and there are several well established likelihood approximations to model it~\cite{Bond:1998qg,Verde:2003ey,Smith:2005ue,Percival:2006ss}. Current polarization data only contributes interesting information at low $l$ where an exact likelihood can be used~\cite{Gorski94,Slosar:2004fr,Page:2006hz}, however in the future the small-scale polarization signal will be less noise dominated and contain useful information. On small scales the likelihood function cannot be computed exactly in reasonable time, and the likelihood function is significantly more complicated than for the temperature because the temperature and polarization fields are correlated. The only existing attempt to model the polarized likelihood function at high $l$, Ref.~\cite{Percival:2006ss}, relies on variable transformations that are not guaranteed to be well defined, and is untested in practice. We give a new general well-defined likelihood approximation that can be used with partial-sky Gaussian polarized CMB data, or any other set of correlated Gaussian fields observed on part of the sky. It is exact in the full-sky limit, and can easily be calibrated from simulations. We also discuss under what circumstances a Gaussian likelihood approximation is reliable.

The layout of the paper is as follows.  In section~\ref{fullsky_like} we present a brief overview of the exact full-sky likelihood function for isotropic noise, and discuss the accuracy required in general for unbiased parameter estimation. We start section~\ref{like_approx} with a review of various temperature likelihood approximations available in the literature, discuss the accuracy of the various Gaussian approximations, and move on to derive a new general likelihood approximation (Eq.~\eqref{new_approx}) that is exact in the full-sky limit.  In section~\ref{tests} we test the approximations by comparing with the exact likelihood function for azimuthal sky cuts and consistency with the binned likelihood. Finally in section~\ref{anisotropic_parameters} we check the approximations with realistically anisotropic noise and demonstrate consistent parameter estimation from simple Planck-like simulations. Some mathematical and analysis details are described in the appendices: appendix~\ref{matrices} gives identities relating expressions with symmetric matrices to expressions with a vector of components; appendix~\ref{goodness} calculates the non-Gaussian correction to the full-sky effective chi-squared; appendix~\ref{multimap} gives results for the likelihood function when using cross-power spectrum estimators from different maps; appendix~\ref{cutsky} reviews the basic Pseudo-$C_l$ estimator and exact likelihood formalism and appendix~\ref{Planck} slightly generalizes previous hybrid Pseudo-$C_l$ estimators for anisotropic noise and gives details of our Planck-like test simulations.

We assume Gaussianity and statistical isotropy of the fields, and focus on the idealized case of pure CMB observations without the complications of foregrounds, point sources, non-linear effects, anisotropic beams, and other observational artefacts. Generalizing our work to more realistic situations will be crucial for application to real data. If the fluctuations turn out to be significantly non-Gaussian or anisotropic a more complicated analysis may also be required.


\section{Exact Full-Sky Likelihood Function}
\label{fullsky_like}
Observations on the full-sky can be decomposed into spherical harmonics $Y_{lm}$, for example the temperature at position $\Omega$ can be written
\be
T(\Omega)=\sum_{l m} a^T_{l m} Y_{l m}(\Omega).
\ee
The polarization field can be expanded analogously in terms of $E$ and $B$ harmonics with opposite parity, see e.g. Ref.~\cite{Kamionkowski:1996ks}. If the CMB field is Gaussian, as expected in linear theory, the corresponding harmonic components $a^T_{lm}$, $a^E_{lm}$ and $a^B_{lm}$ are Gaussian variables with zero mean. The CMB power spectrum $C_l^{XY}$ determines the variance, which is independent of $m$ if we assume statistical isotropy, so that
\be
\la|a^T_{lm}|^2\ra = C_l^{TT} \qquad \la|a^E_{lm}|^2\ra = C_l^{EE} \qquad \la|a^B_{lm}|^2\ra = C_l^{BB}.
\ee
The temperature and $E$-polarization fields are expected to be correlated, so there is an additional correlation power spectrum
$\la|a^T_{lm} a^E_{lm}{}^*|\ra = C_l^{TE}$, but for a parity-invariant ensemble the B-polarization is expected to be uncorrelated to the other fields and the other cross-correlation power spectra are zero.

Since we only observe one sky, we cannot measure the power spectra directly, but instead form the rotationally-invariant estimators, $\Ch_{l}^{XY}$, for full-sky CMB maps given by
\be
\label{Chatdef}
\Ch_{l}^{XY} \equiv \frac{1}{2l+1} \sum_{m=-l}^{l} a_{l m}^{X} a_{lm}^Y{}^*.
\ee
The expectation values of these estimators are the true power spectra, $\la \Ch_l^{XY}\ra = C^{XY}_l$.

To keep things general we consider $n$ (correlated) Gaussian fields, and define an $n$-dimensional vector $\va_{lm}$ of the harmonic coefficients at each $l$ and $m$. In the case of the CMB $\va_{lm} = (a^T_{lm}, a^E_{lm}, a^B_{lm})^T$. The covariance matrix at each $l$ is defined as
\be
\mC_l \equiv \la \va_{lm} \va_{lm}^\dag \ra,
\ee
and the equivalent estimator is
\be
\mCh_l \equiv \frac{1}{2l+1} \sum_{m} \va_{lm}\va_{lm}^\dag.
\ee
Since the $\va_{lm}$ are assumed to be Gaussian and statistically isotropic, they have independent distributions (for $|m|\ge 0$)
and the probability of a set of $\va_{lm}$ at a given $l$ is given by
\be
-2\ln \left( P(\{\va_{lm}\}|\mC_l)\right) = \sum_{m=-l}^l \left[ \va_{lm}^\dag \mC_l^{-1} \va_{lm} + \ln |2\pi\mC_l| \right]= (2l+1)\left(\Tr[\mCh_l \mC_l^{-1} ] + \ln |\mC_l|\right) + \text{const}.
\label{exactlike}
\ee
The fact that this likelihood for $\mC_l$ depends only on the $\Ch_l^{XY}$ (components of the matrix $\mCh_l$) shows that on the full-sky the CMB data can losslessly be compressed to a set of power spectrum estimators that contain all the relevant information about the posterior distribution. In other words $\mCh_l$ is a sufficient statistic for the likelihood.
Integrating out all the $\{\va_{lm}\}$ with the same $\mCh_l$ (or normalizing with respect to $\mCh_l$) gives a Wishart distribution\footnote{Technically $(2l+1)\mCh_l \sim W_n(2l+1,\mC_l)$} for $\mCh_l$ (for a thorough review see Ref.~\cite{Gupta99}):
\be
P(\mCh_l|\mC_l)  \propto \frac{|\mCh_l|^\frac{2l-n}{2}}{|\mC_l|^{\frac{2l+1}{2}}}e^{-(2l+1)\text{Tr} (\mCh_l \mC_l^{-1})/2}.
\ee
The likelihood function for $\mC_l$ given the observed $\mCh_l$ is $\cll(\mC_l|\mCh_l) \propto P(\mCh_l|\mC_l)$, an inverted Wishart distribution. It is straightforward to show that the likelihood has a maximum when $\mC_l = \mCh_l$, so $\mCh_l$ is the maximum likelihood estimator.
When $n=1$, for example when only the temperature is considered, the Wishart distribution reduces to
\be
\label{exact_TT}
-2\ln P(\Ch_l|C_l) = (2l+1)\left[\Ch_l/C_l + \ln (C_l) - \frac{2l-1}{2l+1} \ln (\Ch_l)\right] + \text{const}.
\ee
Considered as a function of $\Ch_l$ this is a (reduced) $\chi$-squared distribution with $2l+1$ degrees of freedom; it has mean $\la \Ch_l\ra = C_l$, but maximum at $\Ch_l= C_l(2l-1)/(2l+1)$. This skewness is also apparent in the likelihood distribution $\cll(C_l|\Ch_l)\propto P(\Ch_l|C_l)$, which peaks at $C_l =\Ch_l$ but has mean value $\Ch_l (2l+1)/(2l-3)$. The mean value of $C_l$ calculated from the estimators should be above the $\Ch_l$, which is why using a quadratic approximation symmetric in $C_l$ (with mean at $C_l=\Ch_l$) potentially biases results by $\clo(1/l)$ at each $l$.

For $n$ correlated Gaussian fields, there are in general $n(n+1)/2$ distinct power spectra $[\mC_l]_{ij}=\la a^{(i)*}_{lm} a^{(j)}_{lm}\ra$, and on the full-sky their estimators have covariance given by
\be
\label{gen_cov}
\cov([\mCh_l]_{ij}, [\mCh_l]_{pq} ) = \frac{1}{2l+1}\left( [\mC_l]_{ip}[\mC_l]_{jq}  + [\mC_l]_{iq} [\mC_l]_{jp} \right).
\ee
It is sometimes convenient to work with vectors rather than matrices, so that $\vX_l \equiv \vecp(\mC_l)$ is a vector of the $n(n+1)/2$ distinct elements of $\mC_l$, and similarly for the estimators. The corresponding covariance matrix is $\mM_{l} \equiv \la (\vXh_l-\vX_l)(\vXh_l-\vX_l)^T\ra$. For symmetric matrices $\vA$ and $\vB$ a useful and somewhat unobvious identity is (see Appendix~\ref{matrices}):
\be
\label{matrix_identity}
\vecp(\vA)^T \mM_{l}^{-1} \vecp(\vB) = \frac{2l+1}{2}\Tr\left[ \vA \mC_l^{-1} \vB \mC_l^{-1}\right],
\ee
which can be used to relate results involving $\mC_l$ to results involving $\vX_l$. In particular by writing $\Tr[\mCh_l\mC_l^{-1}] = \Tr[\mC_l\mC_l^{-1}\mCh_l\mC_l^{-1}]$ we can write the Wishart distribution in terms of the covariance $\mM_l=\mM_l(\vX_l)$ as
\begin{eqnarray}
-2\log P(\vXh_l|\vX_l) &=& 2\vXh_l^T\mM_l^{-1}\vX_l + \frac{2l+1}{n+1}\log|\mM_l|-\frac{2l-1}{n+1}\log|\mMh_l|+\text{const}\\
&=& 2(\vXh_l-\vX_l)^T\mM_l^{-1}\vX_l + \frac{2l+1}{n+1}\log|\mM_l|-\frac{2l-1}{n+1}\log|\mMh_l|+\text{const},
\end{eqnarray}
where we used $\log|\mM_l| = (n+1)\log|\mC_l|+\text{const}$ and $\mMh_l=\mM_l(\vXh_l)$.

We now briefly review the standard Bayesian argument to link the function $P(d|\alpha)$ for the data $d$ given parameters $\alpha$, to the posterior $P(\alpha|d)$, the distribution of the parameters given the data.  Bayes theorem states that the posterior probability of $\alpha$ given the data is:
\be
P(\alpha|d)=\frac{P(d|\alpha) P(\alpha)}{P(d)}\propto \cll(\alpha|d)P(\alpha),
\ee
where the prior $P(\alpha)$ gives information we already know about the models.  In the case of linear CMB power spectra, the $\mC_l$ can be computed essentially exactly from a set of parameters using standard Boltzmann codes.  The probability distribution function of a set of parameters given observed data $\{\mCh_l\} \equiv d$ is therefore given on the noise-free full-sky by:
\be
P(\alpha|\{\mCh_{l}\})\propto \cll(\{\mC_{l} (\alpha)\}|\{\mCh_l\})P(\alpha) =\prod_{l}\cll(\mC_{l}(\alpha)|\mCh_{l}) P(\alpha).
\ee
Since the prior depends on the models under consideration, in this paper we analyse the methods for estimating the likelihood $\cll(\{\mC_{l}\}|\{\mCh_{l}\})$, which is the required input to cosmological parameter estimation codes such as CosmoMC\footnote{\url{http://cosmologist.info/cosmomc/}}.
When analysing the likelihood function it is often convenient to normalize so that $\ln \cll=0$ when $\mC_l=\mCh_l$, i.e. to use
\be
-2\ln\cll(\{\mC_{l}\}|\{\mCh_l\})=\sum_l  (2l+1)\left\{ \Tr[\mCh_l \mC_l^{-1} ] - \ln |\mCh_l\mC_l^{-1}| -n\right\}.
\ee
The expected value for this log likelihood is about $n(n+1)/2$ per $l$, corresponding to the $n(n+1)/2$ distinct components of $\mC_l$. For a more detailed analysis and discussion of `chi-squared' goodness-of-fit see the Appendix~\ref{goodness}.

If there are multiple maps, for example from different frequencies and detectors, cross-map $\mCh_l$ estimators can be used to avoid noise bias. If a set of cross-estimators is used the exact full-sky likelihood function is somewhat different from the above, as discussed in Appendix~\ref{multimap}. However in the limit of many maps the distribution becomes Wishart. In the limit in which there are enough maps that the information loss from using only cross-estimators is small, the approximations developed in this paper should therefore also be applicable.

When the underlying fields are non-Gaussian, the analysis in this paper does not apply directly. However in many cases it is likely to be a good approximation to use the same likelihood approximations but with the covariance replaced with its full non-Gaussian version including 4-point terms. Non-Gaussianity associated with mode-coupling (e.g. from non-linear evolution) can also change the effective number of modes at a given scale. For example the $B$-mode CMB polarization power spectrum is generated by lensing of an $E$ field by a relatively small number of lensing convergence modes. This leads to strong correlations between $l$, and a drastically reduced number of modes compared to $\lmax^2$ expected for Gaussian fields. Ref.~\cite{Smith:2005ue} have shown that using a likelihood approximation designed for analysing Gaussian fields, but allowing for the full covariance from the non-Gaussianity, can give acceptable results. They also demonstrate the importance of modeling the non-Gaussianity of the likelihood function accurately when analysing fields that depend on a small number of underlying modes.


\subsection{Required accuracy}
\label{tolerance}

To assess how accurately we need to be able to model the likelihood we need to know how biases on the posterior $C_l$ translate into constraints on parameters. The simplest case is instructive: consider estimating an amplitude parameter $A$, where $\mC_l = A\mCfl$ for some fiducial fixed spectrum $\mCfl$. For zero noise and a range of $l$ with $\lmin\le l \le \lmax$, we have
\be
-2\ln\cll(A|\{\mCh_l\})= \sum_l  (2l+1)\left\{ \frac{1}{A}\Tr[\mCh_l \mCfl^{-1} ] -  \ln |\mCh_l\mCfl^{-1}| +n \log A -n\right\},
\ee
and the maximum likelihood value is
\be
\hat{A} = \frac{\sum_l (2l+1) \Tr[\mCh_l \mCfl^{-1} ]}{n\sum_l(2l+1)}.
\label{Ahat}
\ee
If $\mCfl$ is the underlying true model then $\la \hat{A}\ra = 1$ and the Fisher variance is
\be
\sigma^2_A\equiv -\left.\left\la \frac{\ud^2}{\ud A^2} \ln\cll(A|\{\mCh_l\}) \right\ra^{-1}\right|_{A=1} =  \frac{2}{n\sum_l (2l+1)} = \frac{2}{n((\lmax+1)^2 - \lmin^2)}\sim\frac{2}{n\lmax^2}
\ee
for a range of $l$ satisfying $\lmax \gg \lmin$. We therefore need any biases to give $\Delta\hat{A} \ll \sqrt{(2/n)}/\lmax$ in order for the bias on $\hat{A}$ to be small compared to its error bar. If we have an $l$-dependent bias $\delta\mC_l$, the bias on $\hat{A}$ from Eq.~\eqref{Ahat} is small compared to its error if
\be
|\la \delta\hat{A}\ra| = \frac{|\sum_l (2l+1)\Tr(\mC_l^{-1}\delta\mC_l)|}{n\sum_l (2l+1)} \ll \frac{\sqrt{2/n}}{\lmax}.
\ee
The tolerated bias scales as $1/(2l+1)$, so this criterion will be satisfied for $\lmax\gg\lmin$ for any systematic error with
\be
\frac{1}{n}|\Tr(\mC_l^{-1}\delta\mC_l)| \ll \sqrt{\frac{2}{n}} \frac{1}{2l+1}\sim \sqrt{\frac{1}{2n}} \frac{1}{l}.
\ee
For a multiplicative bias $\delta \mC_l = B_l\mC_l$ this criterion is $|B_l| \ll (2n)^{-1/2}/l$.
Alternatively if $B_l$ is a constant the requirement is $|B_l| \ll \sqrt{(2/n)}/\lmax$.
We shall loosely refer to $1/(l\sqrt{n})$ as the `systematic error', and require biases to be much smaller than this, which is appropriate for nearly full-sky observations. For a realistic experiment with effective sky coverage $\fsky$ the bias can be $\sim \fsky^{-1/2}$ times larger.

In the presence of noise the situation is more complicated. For one field with $C_l\rightarrow C_l+N_l$, using the Gaussian approximation we require
\be
\left|\sum_l  B_l \frac{(2l+1)C_l^2}{(C_l+N_l)^2}\right| \ll \sqrt{2\sum_l  \frac{(2l+1)C_l^2}{(C_l+N_l)^2}}.
\ee
The bias should be smaller than the systematic error $\sim 1/l$ where $N_l\ll C_l$, but there is greater tolerance where the noise is important.

\section{Likelihood approximations}
\label{like_approx}

\subsection{Single-field likelihood approximations}
\label{Tapprox}
To approximate the likelihood on the cut-sky, the usual approach when analysing the CMB temperature is to develop a form for the log likelihood that is quadratic in some function of the $C_l$, and hence can easily be generalized to the cut-sky using an estimate of the $C_l$ covariance matrix. Here we summarize some common approximations in their full-sky form.

At large $l$, Eq.~\eqref{exact_TT} is approximated by a symmetric Gaussian distribution where the variance is determined by the estimators themselves~\cite{Bond:1998qg}:
\be
\label{TTSGauss}
-2\ln \cll_{S}(C_{l}|\Ch_{l}) = \frac{2l + 1}{2}\left[ \frac{\Ch_{l}-C_{l}}{\Ch_{l}}\right]^{2}.
\ee
This approximation is well known to produce a poor fitting to the true likelihood function at low $l$ \cite{Bond:1998qg}; being symmetric it biases posterior $C_l$ low compared to the true likelihood function.
Approximating the exact likelihood of Eq.~\eqref{exact_TT} with a second order expansion in $\Ch_l/C_l -1$ gives the same form but with $\Ch_l$ replaced by $C_l$ in the denominator:
\be
-2\ln \cll_Q(C_{l}|\Ch_{l}) = \frac{2l + 1}{2}\left[ \frac{\Ch_{l}-C_{l}}{C_{l}}\right]^{2}.
\ee
This distribution is closer to the true likelihood, being skewed in the right direction, however it is still a poor approximation in general, this time biasing the posterior $C_l$ high. It is often somewhat misleadingly referred to as the `Gaussian approximation', even though it does not have the determinant term required for $P(\Ch_l|C_l)$ to be a normalized Gaussian distribution\footnote{For this reason we denote it $\cll_Q$ - for a quadratic approximation - rather than $\cll_G$ used by some other authors.}.
Another possibility is
\be
\label{TTfidgauss}
-2\ln \cll_{f}(C_{l}|\Ch_{l}) = \frac{2l + 1}{2}\left[ \frac{\Ch_{l}-C_{l}}{\Cfl}\right]^{2},
\ee
where $\Cfl$ is some fixed fiducial model assumed to be smooth and close to the model $C_l$ under consideration.
This is more interesting as although the shape of the likelihood is wrong at any given $l$, as we shall see when summed over a range of $l$ it can give results consistent with the exact likelihood function. It is equivalent to a Gaussian approximation since the determinant term is a constant when using a fixed fiducial model.
Adding a $C_l$-dependent determinant term to the quadratic approximation can also produce valid results; we refer to this as Gaussian$_{D}$, given by
\be
-2\ln \cll_D(C_{l}|\Ch_{l}) = \frac{2l + 1}{2}\left[ \frac{\Ch_{l}-C_{l}}{C_{l}}\right]^{2} + \ln|C_{l}|.
\ee
See Section~\ref{sec:Gauss} for more details of this approximation.

Beyond these quadratic/Gaussian approximations, other approximations that have been used include the log-normal distribution where the log-likelihood is quadratic in the log of the power~\cite{Bond:1998qg}
\be
-2 \ln \cll_{\text{LN}}(C_{l}|\Ch_{l}) = \frac{2l + 1}{2} \left[\ln\left(\frac{\Ch_{l}}{C_{l}}\right)\right]^2.
\ee
This distribution is also somewhat biased~\cite{Verde:2003ey,Smith:2005ue}: it only matches the exact full-sky result to second order in $\Ch_l/C_l-1$.

A weighted combination of the quadratic and the log-normal distributions can be a more accurate approximation to the exact likelihood, being correct to third order in $\Ch_l/C_l-1$.  This approximation was adopted in the analysis of the one, three and five-year WMAP data at high $l$~\cite{Verde:2003ey}:
\be
\ln \cll_{\text{WMAP}}(C_{l}|\Ch_{l}) = \frac{1}{3}\ln \cll_Q(C_{l}|\Ch_{l}) + \frac{2}{3} \ln \cll_{LN}(C_{l}|\Ch_{l}).
\ee

Ref.~\cite{Smith:2005ue} suggest even better approximations of the form
\be
-2\ln \cll(C_{l}|\Ch_{l}) \approx (2l+1)\frac{9}{2}\left(\frac{2l+\alpha}{2l+1}\right)^{1/3}\left[ \left(\frac{\Ch_l}{C_l}\right)^{1/3} - \left( \frac{2l+\alpha}{2l+1}\right)^{1/3}\right]^2 + (1-\alpha) \ln C_l,
\label{generalthird}
\ee
where $\alpha$ is one (referred to as `$-1/3$' approximation) or minus one (referred to as `$1/3$'-approximation). The value $\alpha=1/3$ corresponds to taking the distribution of $\Ch^{1/3}_l$ to be Gaussian. These approximations are correct to third order in $\Ch_l/C_l-1$, and also very nearly correct to fourth order.



\subsection{Gaussian approximation for correlated fields}
\label{sec:Gauss}
For a model $\mCfl$ with corresponding full-sky $\vXh_l$ covariance $\mM_{f l}$, a Gaussian approximation to the likelihood function is given by
\begin{eqnarray}
-2\ln\cll_f(\mC_l|\mCh_l) &=& (\vX_l-\vXh_l)^T \mM_{fl}^{-1}(\vX_l-\vXh_l) + \log|\mM_{f l}|\\ &=& \frac{2l+1}{2}\Tr\left[ (\mC_l-\mCh_l) \mCfl^{-1} (\mC_l-\mCh_l) \mCfl^{-1} \right] +(n+1)\log|\mCfl|.
\label{GaussC}
\end{eqnarray}
In the second line we used Eq.~\eqref{matrix_identity}.
If $\mCfl$ is fixed (independent of $\mC_l$) the determinant factors can be dropped, giving the generalization of the approximation for one field given in Eq.~\eqref{TTfidgauss}. It is worth studying this approximation more carefully as it turns out to be very good for smooth models even if the shape of the likelihood function at each $l$ is not accurate. To see this, consider how the total likelihood varies with  a parameter $\theta$,
\be
-2 \frac{\partial\ln\cll_f(\theta|\mCh_l)}{\partial\theta} = \sum_l (2l+1)\Tr\left[ \frac{\partial\mC_l}{\partial\theta} \mCfl^{-1}(\mC_l-\mCh_l)\mCfl^{-1}\right],
\ee
and compare with the equivalent result for the exact likelihood function
\be
-2 \frac{\partial\ln\cll(\theta|\mCh_l)}{\partial\theta} = \sum_l (2l+1)\Tr\left[ \frac{\partial\mC_l}{\partial\theta} \mC_l^{-1}(\mC_l-\mCh_l)\mC_l^{-1}\right].
\ee
This will be zero for the maximum likelihood value $\hat{\theta}$, and if $\mCf_l \propto \mC_l(\hat{\theta})$ then $\hat{\theta}$ will also maximize the approximate likelihood function $\cll_f$. In other words the approximation returns the exact best-fit value as long as the fiducial model is proportional to the best-fit model. If the true model and the fiducial model are both smooth functions of $l$, this will often be approximately true locally, even if it is not strictly true everywhere. An error in the normalization of $\mCfl$ would effect the error bar on $\hat{\theta}$. However since we can easily choose a fiducial model with fractional difference $< \clo(1/\sqrt{l})$, this would only be a small fractional error on the error. The numerical values of the log likelihoods typically differ by $\clo(\ln(\lmax))$ (assuming the fiducial model is accurate to $\clo(1/l)$; c.f. discussion in Appendix~\ref{goodness}), but $\cll_f$ is otherwise generally a good approximation for smooth models.

Note that the above comments only apply to the Gaussian approximation using a fixed fiducial model. If instead we make the covariance $\mM_l$ a function of $\mC_l$ the best-fit model would differ from the exact result due to additional terms in the derivative from the change in the covariance with parameters. However the Gaussian approximation is still quite accurate, and unbiased in an average sense. To see this first consider the simple case of estimating an amplitude parameter $A$, where the exact result for the best-fit value was given in Eq.~\eqref{Ahat}, or in terms of $\vX_l$ by
\be
\hat{A} = 1+\frac{2\sum_l \Delta \vX_l \mM_l^{-1} \vX_l}{n\sum_l (2l+1)},
\ee
where $\Delta \vX_l \equiv \vXh_l-\vX_l$. Using the Gaussian approximation with $\mM_{f l}=\mM_l(\vX_l)$ and expanding we instead get the best-fit value
\be
\hat{A}' = \hat{A} + \frac{ \sum_l \left[\Delta \vX_l\mM_l^{-1}\Delta\vX_l - n(n+1)/2\right]}{n\sum_l(2l+1)/2} - \left[\frac{\sum_l \vX_l\mM_l^{-1}\Delta\vX_l}{n\sum_l (2l+1)/2}\right]^2
+ \clo(\Delta_l^{-1/2}l^{-3/2}),
\ee
where $\Delta_l$ is the size of the range of $l$ under consideration (assuming $l\gg 1$).
The second term has expectation value zero in the true model, and typical variation of order $\clo(\Delta_l^{-1/2}l^{-1})$. The third term is of order $\clo(\Delta_l^{-1}l^{-1})$. So in almost all realizations with $\Delta_l \gg 1, l\gg 1$ we have $\hat{A}'=\hat{A} + \clo(\Delta_l^{-1/2}l^{-1})$. The Gaussian approximation is therefore almost certainly good to within the required error of $\clo(1/l)$ as long as $\Delta_l \gg 1$. However unless $\Delta_l$ is large it won't be much better than required: local features are likely to be more problematic than the overall amplitude (determined from $\Delta_l=\lmax$).
More generally we can consider the expectation of the log likelihood
\be
-2\left\la \ln\cll_f(\{\vX_l\}|\{\vXh_l\}) \right\ra_t =\sum_l \left\{(\vX_l-\vX^{(t)}_{l})^T \mM_{f l}^{-1}(\vX_l-\vX^{(t)}_l) + \Tr\left[\mM_{f l}^{-1}\mM_l^{(t)}\right]+\log|\mM_{f l}|\right\},
\ee
compared to the exact result
\be
-2\left\la \ln\cll(\{\vX_l\}|\{\vXh_l\}) \right\ra_t =\sum_l (2l+1)\left\{\Tr[\mC_l^{(t)}\mC_l^{-1}] + \log|\mC_l|\right\}.
\ee
The exact mean log likelihood has a maximum at the true model, when $\vX_l = \vX_l^{(t)}$. This is however also true of the Gaussian approximation, both when $\mM_{f l}$ is for a fixed fiducial model, and also when we allow it to vary with parameters $\mM_{f l}=\mM_l(\vX_l)$. To the extent that $\mC_l$ are constant in $l$, so that summing over $l$ effectively averages the log likelihood, we therefore expect the Gaussian approximations to be nearly unbiased.

In the case when $\mM_{f l}=\mM_l(\vX_l)$ the reliability of the Gaussian approximation depends critically on the inclusion of the determinant term. For example dropping the determinant, the mean approximate log likelihood for $A$ where $\vX_l=A\vX_l^{(t)}$ is
\be
-2\la \ln\cll_Q(A|\{\vXh_l\}) \ra_t = \sum_l \left\{ \frac{(2l+1)n}{2}\frac{(1-A)^2}{A^2} + \frac{n(n+1)}{2A^2}\right\}.
\ee
For large $\lmax$ the maximum is at $\hat{A}\sim 1 + (n+1)/\lmax$ rather than $1$, so we expect $A$ to be  biased high by the order of the expected error, confirming that $\cll_Q$ is not a good approximation to the likelihood.
If a fixed fiducial model is used then the determinant does not affect the likelihood, and we have
\begin{eqnarray}
-2\la \ln\cll_f(A|\{\vXh_l\}) \ra_t &=& \sum_l  \left\{(1-A)^2 \vX_l^{(t)}{}^T \mM_{f l}^{-1}\vX_l^{(t)} + \Tr\left[  \mM_{f l}^{-1} \mM_l^{(t)} \right]\right)\\
&\propto& (1-A)^2 + \text{const},
\end{eqnarray}
which has a minimum in agreement with the exact likelihood function ($\hat{A}=1$) regardless of the choice of fiducial model (though the variance of $A$ would be wrong by the order of the fractional error in the fiducial model).

The case where $\mM_{f l} = \mM_l(\vXh_l)$ is harder to analyse, but it is not a good approximation because the covariance is then correlated with the $\mCh_l$ (so the contribution of high-fluctuating $\mCh_l$ is down-weighted by larger covariance there).

\subsection{Noise, binning and the Gaussian approximation}

\label{sec:binning}
In the presence of isotropic uncorrelated noise $n_{lm}$ with known power spectrum $N_l$, the observed field $a_{lm}+n_{lm}$ is just another Gaussian field with power spectrum $C_l + N_l$. The likelihood functions are then exactly the same as without noise, where $C_l$ and $\Ch_l$ are replaced with their values including noise.

Consider a toy problem where we wish to constrain the amplitude of the power spectrum $A$ over some range of scales over which the power spectrum is flat. If there are $n_m$ Gaussian modes, and we estimate the power spectrum in $n_b$ equal bins, each bin will have $\nu \equiv n_m/n_b$ modes. If each mode has independent Gaussian noise with known variance $N$,
each $\Ch_b$ estimator then has a $\chi^2$ distribution with $\nu$ degrees of freedom and mean $A+N$. The posterior mean of $A$ will differ systematically from its maximum likelihood $\Ch_b-N$ by $\sim (A+N)/\nu$, which we can take as an estimate of the bias obtained in each bin by using a Gaussian approximation. Using all the bins we can constrain $A$ to within an error of $\sim (A+N)/\sqrt{n_m}$. The criterion for the bias to be much smaller than the error bar is then $n_b \ll \sqrt{n_m}$. Perhaps surprisingly this is independent of the noise: when this inequality is violated a Gaussian approximation would be biased for a given bin, even if the signal is noise dominated. Of course if the bin width is increased so that the signal to noise in each bin remains constant, then the Gaussian approximation for the binned estimates does improve as the noise increases.

In the case of observations of the CMB over a fraction $\fsky$ of the sky, with useful signal at $\lmin \alt l \alt \lmax$, the number of modes is $n_m \sim \fsky(\lmax^2-\lmin^2)$, so for the Gaussian approximation to be good for each bin we need the number of bins $n_b \ll \fsky^{1/2}\lmax$ (assuming $\lmax^2\gg\lmin^2$). This is violated by the natural full-sky binning into $\lmax$ bins, one at each $l$ (which has optimal $l$-resolution), regardless of how large $\lmax$ is.  For partial sky observations with bin-width $\Delta_l^{(b)}$ in $l$, you would need $\Delta_l^{(b)} \gg \fsky^{-1/2}$ for the Gaussian approximation to be reliable. However often we do not actually need each bin to be individually unbiased, so this criterion can in practice be relaxed.

Binning different $l$s together makes the distribution more Gaussian, so binning full-sky $C_l$ into bands of width $\Delta_l^{(b)}\gg 1$ would allow any of the quadratic likelihood approximation to be used with very small bias at each bin. For basic vanilla models it is straightforward to assess the impact of binning on parameter constraints: we generated a toy full-sky simulation at Planck sensitivity~\cite{unknown:2006uk}, generated samples of the posterior parameter values from the exact likelihood function using CosmoMC~\cite{Lewis:2002ah}, and then importance sampled using the exact likelihood function on binned values of the $C_l$ (keeping the $l< 30$ spectrum un-binned where in realistic cases the likelihood could also be calculated exactly). Using the quadratic approximation $\cll_Q$ in this case (with $\Delta_l^{(b)}=1$) biases parameters like the spectral index by around 1-sigma compared to the exact result; however using $\cll_f$ with a sensible fiducial model produces unbiased constraints (see previous subsection).
Binning with a width $\Delta_l^{(b)}=50$ degrades parameter error bars by only $\alt 10\%$ for basic models; this would be sufficient to make the bias a tiny fraction ($\sim 1/\Delta_l^{(b)}$) of the error bar on each bin. Bins of $\Delta_l\sim 10$ would likely be wide enough to render the error from a quadratic likelihood approximation small relative to other systematic errors. The cost of doing this is that some $l$-resolution of the acoustic peak structure is lost, and any non-standard models with features that vary over a few $l$ could not be analysed reliably (for example see Ref.~\cite{Martin:2003sg}).

As we shall show, modelling the non-Gaussian distribution accurately is straightforward, and in any case a Gaussian approximation is often adequate, so for full-sky observations there is no need to degrade the data by binning. Note that binning may however be useful for other reasons, for example to increase the accuracy with which the covariance can be estimated from a fixed number of simulations, or to improve the optimality of the cut-sky $C_l$ estimator. Since almost all theoretical power spectra are very smooth in $l$, binning is likely to lose little information as long as the bins are narrow compared to the width of any features.


\subsection{Partial Sky Likelihood function}


When observations are obtained over part of the sky, or part of the sky is obscured by foregrounds or there is anisotropic noise, the maximum-likelihood estimators $\Ch_l$ can no longer be measured directly. The CMB is still expected to be Gaussian however, so in principle there is an exact pixel-based likelihood function of the form
\be
\cll(\{C_l \}| \vp) \propto \frac{e^{-\vp^T \mC_p^{-1} \vp /2}}{|\mC_p|^{1/2}},
\ee
where $\vp$ is a vector of pixel values and $\mC_p$ is the pixel-pixel covariance (a function of $\{C_l\}$). Equivalently the CMB fields can be expanded in a set of modes that are orthogonal and complete over the observed sky, and the likelihood in terms of these mode coefficients will also be Gaussian~\cite{Gorski94,Mortlock00,Lewis:2001hp}. Neither likelihood function can be expressed solely in terms of a set of maximum-likelihood power spectrum estimators, so an optimal analysis does not allow radical compression. The problem with using the exact likelihood function is that the number of pixels goes like $\lmax^2$, so the Cholesky decomposition required to calculate $\mC_p^{-1}\vp$ will scale like $\lmax^6$, which is prohibitive for $\lmax$ larger than a few hundred and slow for $l\agt 30$. Gibbs sampling methods avoid doing large matrix inversions, but still have exponential convergence problems if an exact analysis is attempted for general $C_l$. A sensible strategy is therefore to use an exact likelihood only at low $l$ where it is numerically feasible, and to use an approximate analysis at higher $l$~\cite{Efstathiou:2003dj,Slosar:2004fr,Hinshaw:2006ia}. The most obvious way to do this is to compress the high-$l$ data into a set of cut-sky power spectrum estimators, and then find an approximate likelihood function that is a function only of these estimators. There is some evidence that doing this is close to optimal, and it has the advantage of being fast. This means that numerous practical complications can be accounted for simply by adding additional terms to the covariance matrix estimated from simulations.

There are various possible estimators for the cut-sky power spectrum that can be used, varying from maximum likelihood to a variety of quadratic estimators. At high $l$ quadratic estimators can be close to the maximum likelihood and we focus here on the widely used Pseudo-$C_l$ methods~\cite{Tegmark:1996qt,Wandelt:2000av,Hivon:2001jp,Hansen:2002zq,Efstathiou:2003dj,Brown:2004jn,Efstathiou:2006eb,Smith:2006vq} that are in many cases equivalent to methods based on correlation functions~\cite{Szapudi:2000xj,Chon:2003gx}. In principle the statistical distribution of these estimators could be calculated exactly~\cite{Wandelt:2000av}, but only at prohibitive numerical cost in general. We therefore look for a fast likelihood approximation that is a function only of the set of cut-sky estimators $\{\mCh_l\}$, an estimate of their covariance (e.g. from simulations or calculated), and knowledge of the noise contribution $\{ \mN_l\}$. One of the aims of this work is to quantify whether such a likelihood approximation is good enough to obtain reliable and nearly-optimal parameter constraints. As our guide for modeling the non-Gaussian shape of the likelihood function we will use the known form in the full-sky limit; we aim for our approximation to be exact when the  $\{\mCh_l\}$ are calculated on the full-sky with isotropic noise.

\subsection{New likelihood approximation for correlated fields}
\label{sec:newlike}

We now derive a new likelihood approximation that can be used with $\Ch_l$ estimators calculated from correlated Gaussian fields. It is exact on the full-sky, and should give reasonable results even for non-standard models that are not necessarily very smooth functions of $l$. The approximation involves a fiducial model so that the covariance can easily be pre-computed. However errors in the fiducial model are automatically corrected, in that the result remains exact on the full-sky however wrong the fiducial model is. We assume that the matrix of estimators $\mCh_l$ is positive definite, which may break down for some estimators at low $l$.

Given the observed estimators $\mCh_l$ for the covariance of $n$ Gaussian fields, the full-sky likelihood function can be written
\begin{eqnarray}
-2\log \cll(\mC_l|\mCh_l) &=& (2l+1) \left\{\text{Tr}\left[ \mCh_l \mC^{-1}_l
  \right] - \log |\mC_l^{-1} \mCh_l | -n \right\} \\
 &=& (2l+1) \left\{\text{Tr}\left[ \mC^{-1/2}_l \mCh_l \mC^{-1/2}_l
  \right] - \log |\mC_l^{-1/2} \mCh_l \mC_l^{-1/2} | -n \right\}
 \label{exsymm}
 \\  &=& (2l+1) \sum _i \left[D_{l,ii} - \log (D_{l,ii})-1\right].
\end{eqnarray}
 The symmetric form is defined using the Hermitian square root and
$\mC^{-1/2}_l\mCh_l\mC^{-1/2}_l = \mU_l\mD_l\mU_l^T$ for orthogonal $\mU_l$ and diagonal $\mD_l$. In the presence
of instrumental noise the $\mC_l$ and $\mCh_l$ should include the noise variance.

To generalize to the cut-sky we want to make this look quadratic, so we write
\begin{eqnarray}
-2\log \cll(\mC_l|\mCh_l) &=& \frac{2l+1}{2}\sum_i [g(D_{l,ii})]^2=\frac{2l+1}{2}  \text{Tr}\left[ \mg(\mD_l)^2  \right]
\label{eigs_exact}
\end{eqnarray}
where
$$
g(x) \equiv \text{sign}(x-1)\sqrt{ 2(x - \ln(x) - 1}),
$$
and $[\mg(\mD_l)]_{ij}=g(D_{l,ii})\delta_{ij}$. Although the $\text{sign}$ of the function is irrelevant for consistency with the exact full-sky result, this choice ensures consistency with the Gaussian approximation and that $g(x)$ is a smooth function at $x=1$.
We now want to relate this quadratic form to a version that is quadratic in the matrix elements.
To do this we use Eq.~\eqref{matrix_identity} in the form
\be
\frac{2l+1}{2} \text{Tr}\left[(\mCfl^{-1/2}\mCgl\mCfl^{-1/2})^2 \right] = {\vX_g}_l^T \mMXf_l^{-1} {\vX_g}_l,
\label{Gauss_approx}
\ee
where ${\vX_g}_l\equiv \vecp({\mC_g}_l)$ (dimension $n(n+1)/2$) is the vector of distinct elements of ${\mC_g}_l$,  and $\mMXf_l$ is the covariance of $\vXh$ evaluated for $\mC_l = \mCf_l$.
We therefore write the exact result of Eq.~\eqref{eigs_exact} as
\begin{eqnarray}
-2\log \cll(\mC_l|\mCh_l)
= \frac{2l+1}{2} \text{Tr}\left[(\mCfl^{-1/2}\mCgl\mCfl^{-1/2})^2 \right]
={\vX_g}_l^T \mMXf_l^{-1} {\vX_g}_l,
\end{eqnarray}
where $\mCgl\equiv \mCfl^{1/2}\mU_l \mg(\mD_l)\mU_l^T\mCfl^{1/2}$ for some fiducial model $\mCfl$.
This can then be generalized to our final cut-sky approximation where the estimators at different $l$ may be correlated:
\be
-2\log \cll(\{\mC_l\}|\{\mCh_l\}) \approx \vX_g^T \mMXf^{-1} \vX_g =
\sum_{ll'} [\vX_g]_l^T [\mMXf^{-1}]_{ll'} [\vX_g]_{l'}.
\label{new_approx}
\ee
Here $\mMXf$ is the fiducial model covariance block matrix with $n(n+1)/2 \times n(n+1)/2$ blocks labeled by $l$ and $l'$, and $\vX_g$ is a $(\lmax-\lmin+1)n(n+1)/2$-row block vector:
\begin{eqnarray}
\,[\mMXf]_{ll'}&=&\la (\vXh_l-{\vX}_l)(\vXh_{l'}-{\vX}_{l'})^T\ra_f
\\
 \,[\vX_g]_l &=& \vecp\left( \mC_{f l}^{1/2} \mg[\mC_l^{-1/2}\mCh_l \mC_l^{-1/2}] \mC_{f l}^{1/2}\right),
\end{eqnarray}
where the matrix function $\mg$ applied to a symmetric positive definite matrix is defined by application of $g$ to its eigenvalues.
On the full-sky with isotropic noise $[\mMXf]_{ll'} = \delta_{ll'}\mMXf_l$ and the approximation is exact.
It is fast to evaluate because $\mMXf^{-1}$ is independent of $\mC_l$ and hence can be pre-computed. Remaining diagonalizations on the small matrices at each $l$ are fast. In principle the fiducial model $\mCfl$ could also be chosen to be equal to $\mCh_l$ or $\mC_l$, but for most purposes using a fixed smooth theoretical fiducial spectrum that is a good fit to the data is likely to be most convenient. For a general correlation structure $\mMXf$ has $[(\lmax-\lmin+1)n(n+1)/2]^2$ elements (but is symmetric). Remember that here $\mC_l$ and $\mCh_l$ include the noise contribution, so for a pure-theory (zero-noise) $\mC_l^{\text{th}}$ the approximation requires an (effective) noise $\mN_l$ at each $l$, a covariance matrix, and the set of estimators $\{\mCh_l\}$.

If $\mC_l$ is block diagonal, as in the case of CMB polarization with $B$ modes, the exact full-sky likelihood is separable in the blocks. On the cut-sky the estimators for the blocks may however be correlated; in particular a sky cut will correlate $E$ and $B$-mode polarization estimators. The approximation can be applied with full $[(\lmax-\lmin+1)n(n+1)/2]$ vectors, or the approximation can be applied to a truncated vector including only terms in each block. For example we could use $\vX_l = [C_l^{TT},C_l^{TE},C_l^{EE},C_l^{BB}]^T$, with covariance allowing for correlations between $E$ and $B$ power spectra, but ignoring any potential information in components like $\hat{C}_l^{TB}$ (the full-sky likelihood is independent of $\hat{C}_l^{TB}$, but this may not be the case when there are couplings between $T$, $E$ and $B$). If the smaller vector is used the transformation to $\vX_g$ can be calculated for each block separately.

For a single Gaussian field the approximation is simply
\be
-2\log \cll(\{C_l\}|\{\Ch_l\}) \approx \sum_{ll'}[g(\Ch_l/C_l)\Cfl] [{M_{f}}^{-1}]_{ll'} [{C_f}_{l'} g(\Ch_{l'}/C_{l'})].
\label{singlefieldnewlike}
\ee
\subsubsection{Generalization}

On the full-sky, and in some generalizations, the distribution of the estimators $\Ch_l$ scales approximately with $C_l$, so that $P(\Ch_l|C_l)\ud\Ch_l = S_l(\Ch_l/C_l) (\ud \Ch_l)/C_l$ for some function $S(x)$. The full-sky likelihood function considered above is of this form. In general $S(x)$ can differ from the full-sky form, and could be estimated approximately from simulations using a given fiducial $C_l$. The likelihood function is then given by $\cll(C_l|\Ch_l) \propto S_l(\Ch_l/C_l)/C_l$. We can then use the same likelihood approximations as above, where for each $l$
\begin{equation}
g(x) = \text{sign}(x-x_m)\sqrt{2\sigma^2 \log\left[\frac{x_m S_l(x_m)}{x S_l(x)}\right]},
\end{equation}
$x_m$ is the value of $x$ that maximizes $x S_l(x)$,
and $\sigma^2 = \text{var}(x)$ (on the full-sky $x_m=1$, $\sigma^2 = 2/(2l+1)$). With multiple fields a similar argument applies as long as the likelihood function can be written in terms of $\mC_l^{-1/2} \mCh_l\mC_l^{-1/2}$. The function $S_l(x)$ can then be estimated from the distribution of the diagonal elements of $\mC_l^{-1/2} \mCh_l\mC_l^{-1/2}$ at fixed $\mC_l$.

The exact distribution of single-field pseudo-$C_l$s is discussed in Ref.~\cite{Wandelt:2000av} for azimuthally symmetric sky cuts. Even in this simple case with no noise the marginalized distribution at each $l$ is of a different functional form from the full-sky result, similarly for the corresponding $\Ch_l$-estimators. Using Pseudo-$C_l$ estimators with our approximation using $g(x) = \text{sign}(x-1)\sqrt{ 2(x - \ln(x) - 1})$ amounts to approximating the marginalized distribution of the $\Ch_l$ as $\chi^2$ with $\nu_l$ degrees of freedom, where $\nu_l = 2C_l^2/\text{var}(\Ch_l)$. At high $l$ and for small cuts with uniform weighting outside the cut $\nu_l \sim (2l+1)\fsky^2$~\cite{Hinshaw:2003ex}; for binned estimators that are nearly uncorrelated, $\nu_l \sim (2l+1)\Delta_l\fsky$~\cite{Hivon:2001jp,Challinor:2004pr}.

\subsubsection{Gaussian approximation}

The Gaussian approximations of Section~\ref{sec:Gauss} generalize straightforwardly to a  $(\lmax-\lmin+1)n(n+1)/2$-vector of cut-sky estimators $\vXh$ with a covariance matrix $\mM_f$,
\be
-2\log \cll_f(\{\vX_l\}|\{\vXh_l\}) = (\vXh-\vX)^T\mM^{-1}_f(\vXh-\vX) + \log|\mM_f|.
\ee
Note that even with no correlations between $l$ this cannot be written as a matrix variate normal distribution in the form of Eq.~\eqref{GaussC} because a general $\mM$ has many more degrees of freedom than the exact full-sky matrix where $\mM_l$ (a symmetric $n(n+1)/2\times n(n+1)/2$ matrix) can be expressed in terms of the smaller matrix $\mC_l$ (an $n\times n$ symmetric matrix). From the discussion in Section~\ref{sec:Gauss} we expect the Gaussian approximations to be accurate for $\lmax\gg 1$ in almost all cases where parameter variations produce changes that are smooth in $l$.


\section{Testing the likelihood approximations}
\label{tests}


For accurate parameter estimation we need to be able to constrain the theory $C_l$ accurately as a function of $l$ given the estimators $\Ch_l$. On the full-sky the likelihood approximations can easily be compared to the exact likelihood function.
We fit an amplitude parameter $A$ where  $\cll(A|\{\Ch_{l}\}) = \cll(\{C_{l}=AC_{l}^{in}\}|\{\Ch_{l}\})$, over some range of $l$ using some fiducial model $C_l^{in}$. The $\Ch_l$ are simulated using $C_l^{in}$, so that on average the best-fit value of $A$ is $A=1$. Since in almost all models the theory power spectra $C_l$ are smooth functions of $l$, and we wish to check that off-diagonal correlations are being accounted for correctly, we chose to fit over a range $\Delta l=10$ in $l$.  This was done for $l=(\lmin\rightarrow \lmax=\lmin+\Delta l-1)$, i.e., bins of size $\Delta l$ with $\lmin$ and $\lmax$ being the lower and upper values of $l$ in each bin, respectively, as a function of $\lmin$.

Using a standard search routine\footnote{Fortran 90 numerical recipes: Golden Section Search.}, we searched for the best fit value of $A$, $\hat{A}$.  In other words, for the exact likelihood and each approximation, we numerically extracted the amplitude that would maximize the likelihood.  We then estimate the variance of this estimated maximum likelihood value of $A$ compared to the true maximum likelihood in that realization, $\langle(\hat{A}_{i} - \hat{A}_{\text{Exact}})^{2}\rangle_{\text{simulations}}$. This gives a measure of any error introduced by the approximation. Note that since we are using a range of $\Delta l=10$ in $l$, the best-fit value of $A$ depends on the likelihood approximation at each $l$ value, and in particular probes the full range of deviations of $\Ch_l$ from $C_l$ expected from cosmic variance.


To quantify whether an approximation is good enough, we consider how well we need to know the amplitude of the $C_l$ as a function of $l$ to get unbiased results on an amplitude parameter. We consider the noise-free case.
 The cosmic variance error on a single $l$, is $\sqrt{\frac{2C_{l}^{2}}{(2l +1)}}$.  Since we are averaging over a range $\Delta l = 10$, the cosmic variance error we can obtain on $A$ from a single bin will be reduced by a factor of $\Delta l$, hence a fractional error of $\sim \sqrt{\frac{2}{(2\lmin + 1)\Delta l}}$ from one band. However as discussed in Section~\ref{tolerance} for unbiased results from the full spectrum we
need a fractional average systematic error on the $C_l$ much smaller than $\Delta C_l/C_l \ll n^{-1/2}/l$.
We therefore require likelihood approximations that give values that are unbiased to better than the systematic error.

\subsection{Full-sky tests}

On the full-sky with isotropic noise estimators at different $l$ are uncorrelated: the likelihood function is $\cll(A|\{\Ch_{l}\}) = \prod_{l=\lmin}^{\lmax} \cll(C_{l}=A C_{l}^{in}|\Ch_{l})$, where $\cll$ can take the form of the exact likelihood or any of the approximations described in section~\ref{Tapprox}.

\begin{figure}[htbp]
\centering
\psfig{file=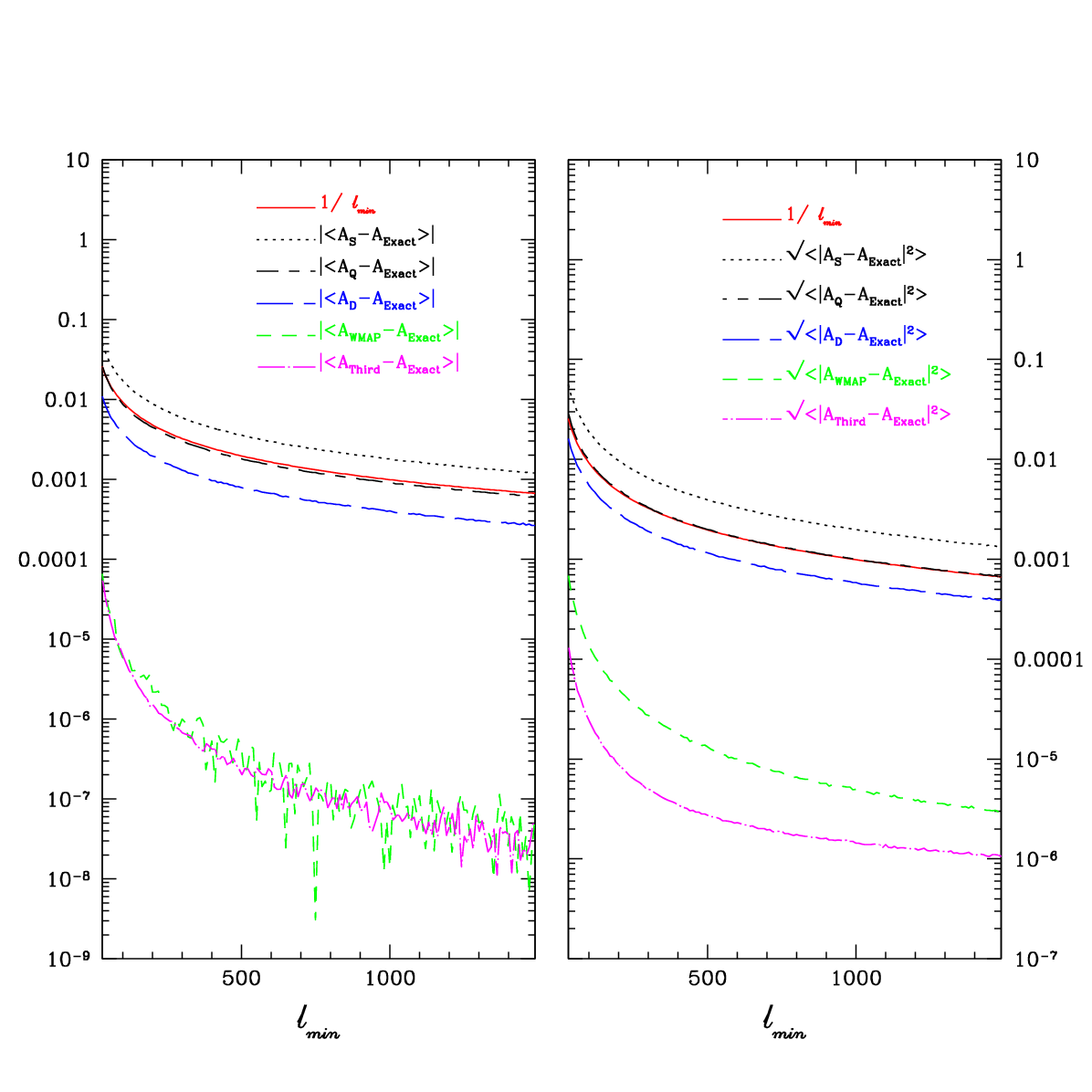, width=14cm}\caption{The plot compares various likelihood approximations on the full-sky for the case of a single field (temperature only) and no noise.  The left-hand panel shows the difference between best-fit posterior amplitude of a $\Delta_l=10$ bin with the likelihood approximations and the exact likelihood over 10000 simulations where $A_{\text{Exact}}$ is the best fit amplitude of the exact likelihood and $A_{S}$, $A_{Q}$, $A_{D}$ and $A_{\text{WMAP}}$ are the best fit amplitude of the symmetric Gaussian, quadratic,  Gaussian$_{D}$ and WMAP approximations respectively.  The right-hand panel shows the root-mean-square difference.  These two quantities are compared to the systematic error tolerance.  Only the symmetric Gaussian and the quadratic approximations are clearly not good enough.  The fiducial Gaussian and new likelihood approximations are not show as they are exactly unbiased in this simple test case with correct fiducial model.}
\label{fig:Tfullsky}\end{figure}

In Fig.~\ref{fig:Tfullsky} we show the results for the temperature likelihood approximations on the full-sky.  We calculate on average over simulations the difference between the posterior amplitudes, $|\la\hat{A}_{i}-\hat{A}_{\text{Exact}}\ra|$ (to probe bias) and the variance  $\la|\hat{A}_{i}-\hat{A}_{\text{Exact}}|^{2}\ra$ (to probe posterior differences in each realization).  We require both quantities to be smaller than $1/l$, where $\hat{A}_{i}$ is the best-fit value from one of the likelihood approximation given in section~\ref{Tapprox}.  As expected, the symmetric Gaussian distribution $\cll_S$ shows a very poor fitting as its variance is larger than the systematic error.  The quadratic approximation $\cll_Q$ gives results almost identical to the systematic error and hence is not a good enough approximation.  The Gaussian$_{D}$ results are probably good enough, but the WMAP-approximation and approximation developed in Ref.\cite{Smith:2005ue} are much better.  The fiducial Gaussian approximation is exactly unbiased in this simple test and is not shown. Any of these last four approximations should be adequate for temperature parameter estimation, at least assuming cut-sky accuracy with realistic noise follows the full-sky behaviour.  The new likelihood approximation by construction is also exactly correct in this full-sky case.

\subsection{Cut-sky tests}

We now move on to test the approximations on the cut-sky. In particular we want to check that any bias on parameter constraints is much smaller than the posterior error, and that the likelihood function has the right shape.
To do this we calculate simple Pseudo-$C_l$ estimators for azimuthal cuts with isotropic noise where the exact likelihood function can also be computed in reasonable time. Although idealized, realistic cuts are often approximately azimuthal due to the disc-shape of the galaxy, and consistency in this simple case is clearly necessary (if not strictly sufficient) to justify the use of a given likelihood approximation. An azimuthal cut introduces most of the qualitative differences in a cut-sky analysis,  namely correlations between different $l$ and not-exactly Wishart distributions of the $\mCh_l$.  The detailed derivations of the Pseudo-$C_l$ estimators, the covariance matrix and the exact likelihood for correlated fields are reviewed in Appendix~\ref{cutsky}. We test the more general case of anisotropic noise and asymmetric cuts later in Section~\ref{anisotropic_parameters}.

\subsubsection{Single-field Results}

\begin{figure}[htbp] \centering \psfig{file=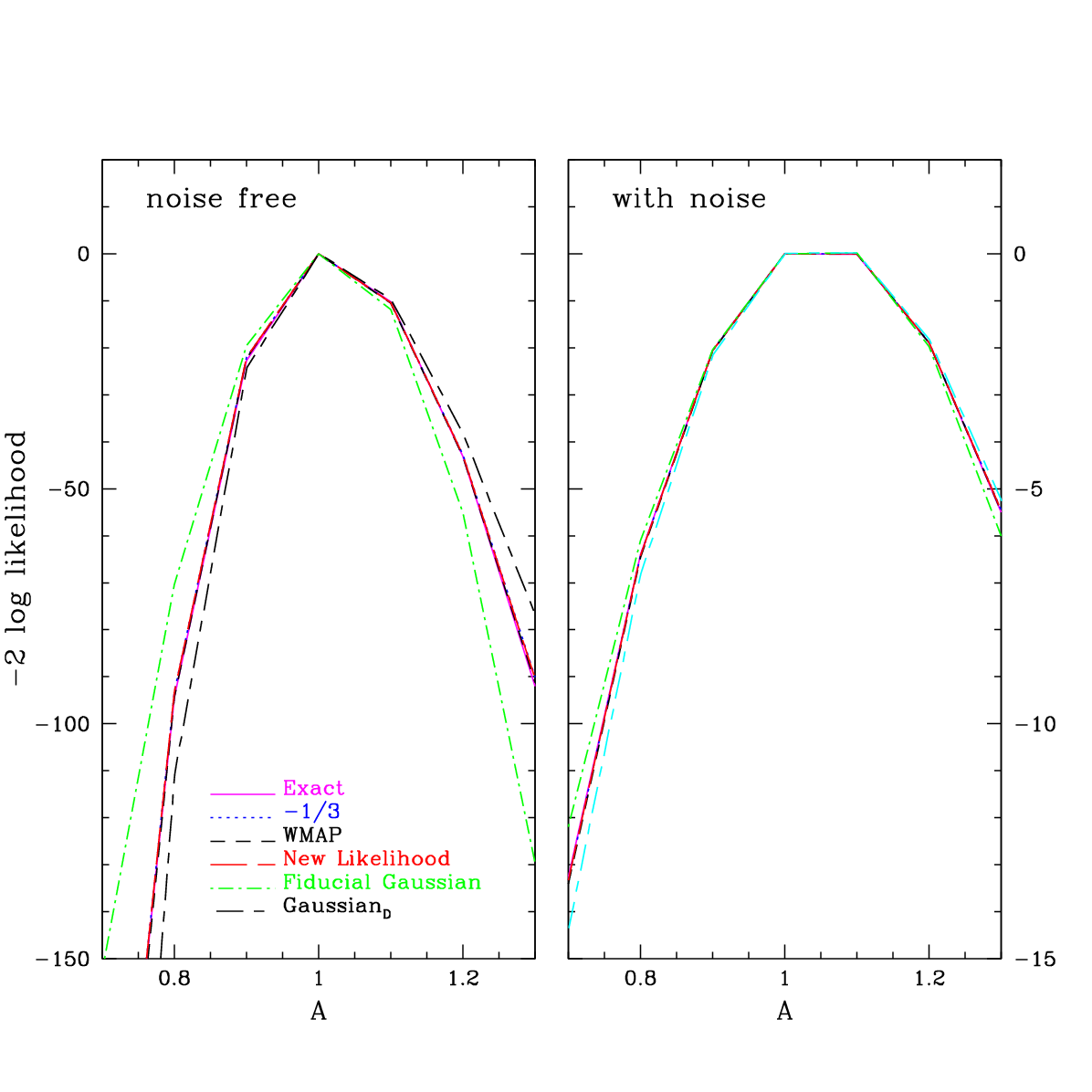, width=12cm}\caption{Single field likelihood approximation results for the likelihood as a function of bin amplitude, $A$.  The plot compares the likelihood approximations to the exact likelihood for an azimuthal galactic cut with $\fsky=0.862$, $\lmax=600$ and bin located at 200 $\leq$ $l$ $\leq 209$ for one realization.}
\label{fig:TCutsky} \end{figure}

The approximations used in the analysis of the temperature power spectra in the cut-sky are given below, where $C_l$ are taken to include noise and $[\mM^{-1}]_{ll'}$ is the inverse of the covariance matrix [$\mM=\mM(\vX)$, $\hat{\mM}=\mM(\vXh)]$ when only a single field is considered:
\begin{align}  \begin{split}
-2 \ln \cll_{WMAP} &= \frac{1}{3} \sum_{l l'} \biggl( \Ch_{l}-C_{l} \biggr) [\mM^{-1}]_{l l'} \biggl( \Ch_{l'}-C_{l'} \biggr)
 + \frac{2}{3} \sum_{l l'} \ln \biggl( \frac{\Ch_{l}}{C_{l}} \biggr) C_{l} [\mM^{-1}]_{l l'} C_{l'} \ln \biggl( \frac{\Ch_{l'}}{C_{l'}} \biggr).
\end{split} \end{align}
\be
-2 \ln \cll_{-1/3} = 9 \sum_{l l'} (\Ch_{l}^{-1/3} - C_{l}^{-1/3}) C_{l} \Ch_{l}^{1/3} [\hat{\mM}^{-1}]_{l l'} \Ch_{l'}^{1/3} C_{l'} (\Ch_{l'}^{-1/3} - C_{l'}^{-1/3}),
\ee
\be
-2 \ln \cll_{D} = \sum_{l l'} \biggl( \Ch_{l}-C_{l} \biggr) [\mM^{-1}]_{l l'} \biggl( \Ch_{l'}-C_{l'} \biggr) + \log |\mM| ,
\ee
\begin{align}  \begin{split}
-2 \ln \cll_{f} &= \sum_{l l'} \biggl( \Ch_{l}-C_{l} \biggr) [\mM_{f}^{-1}]_{l l'} \biggl( \Ch_{l'}-C_{l'} \biggr),
\end{split} \end{align}
where $[\mM_{f}]_{l l'}$ is the covariance of some fiducial model, similar to the one used in New Likelihood (see Eq.~\eqref{singlefieldnewlike}).

Fig.~\ref{fig:TCutsky} shows the exact likelihood and the approximations presented in this subsection as a function of the posterior amplitude for a bin in one simulation.  We consider both cases of noise-free and noisy power spectra.  The approximations compare well to the exact result in both cases, though the results for the Gaussian approximations are not the right shape far away from the peak.  Simulations were performed for azimuthal cuts with $\fsky=0.862$\footnote{that is a Galactic cut of $20^{o}$.}.  We have also fixed $\Ch_{l}$ at $l \le 30$ to $\Ch_{l}=C_{l}$ to prevent occasional negative values in the simulations.

\subsubsection{Correlated-field Results}

\begin{figure}[htbp] \centering \psfig{file=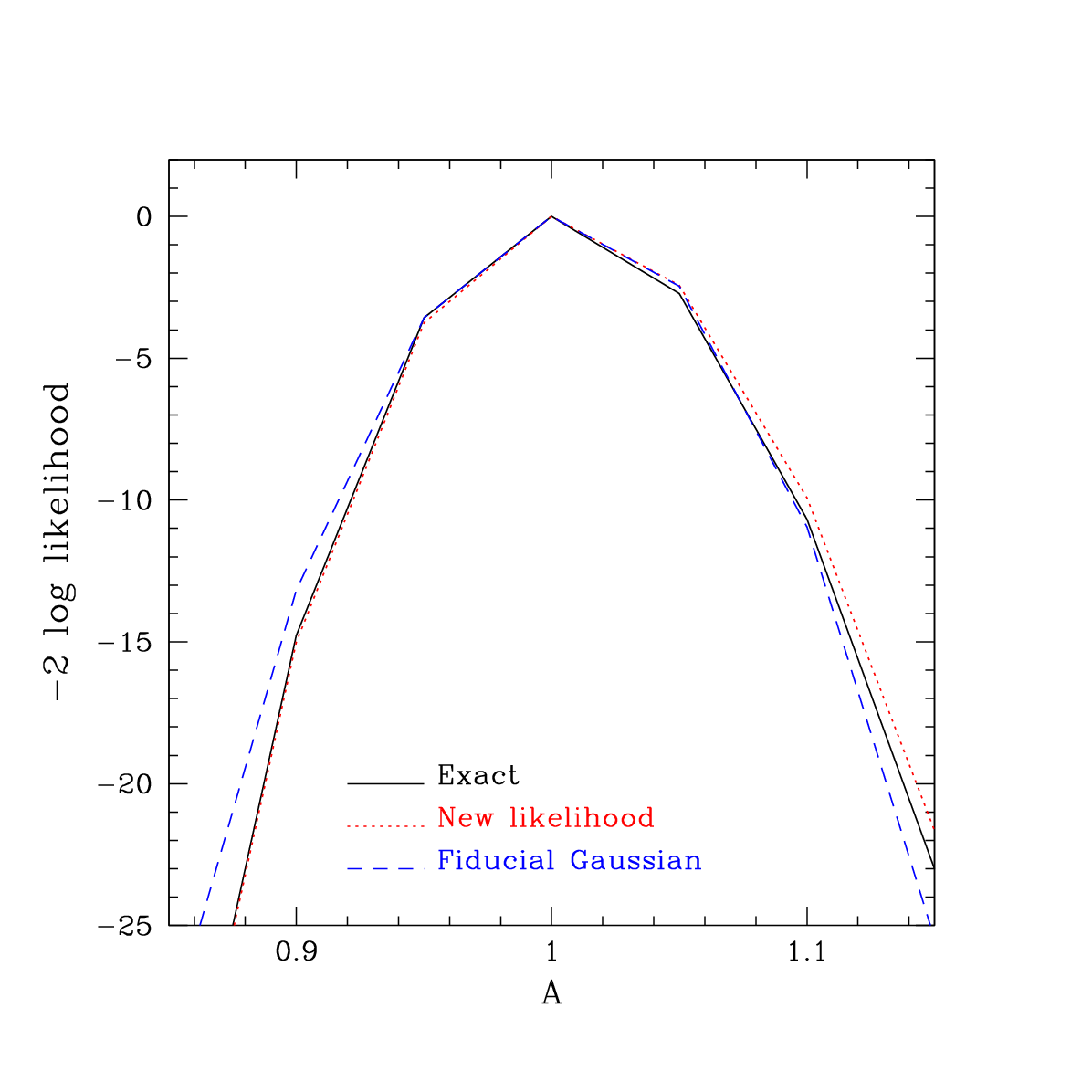, width=12cm}\caption{  The likelihood as a function of bin amplitude, A, for the temperature and polarization fields in one realization.  The black (solid) line is the exact likelihood, the red (dotted) line is the new likelihood and the blue (dashed) line is the fiducial Gaussian distribution.  Unlike the fiducial Gaussian distribution which only agrees well around the peak, the new likelihood captures the shape of the exact one well.  We used an azimuthal cut with $\fsky=0.862$, $\lmax=500$ and bin at 150 $\leq$ $l$ $\leq 159$.  Noise is isotropic and uncorrelated and the $E$ and $B$ modes noise is twice the $T$ noise.}
\label{fig:polcutlike} \end{figure}

\begin{figure}[htbp] \centering \psfig{file=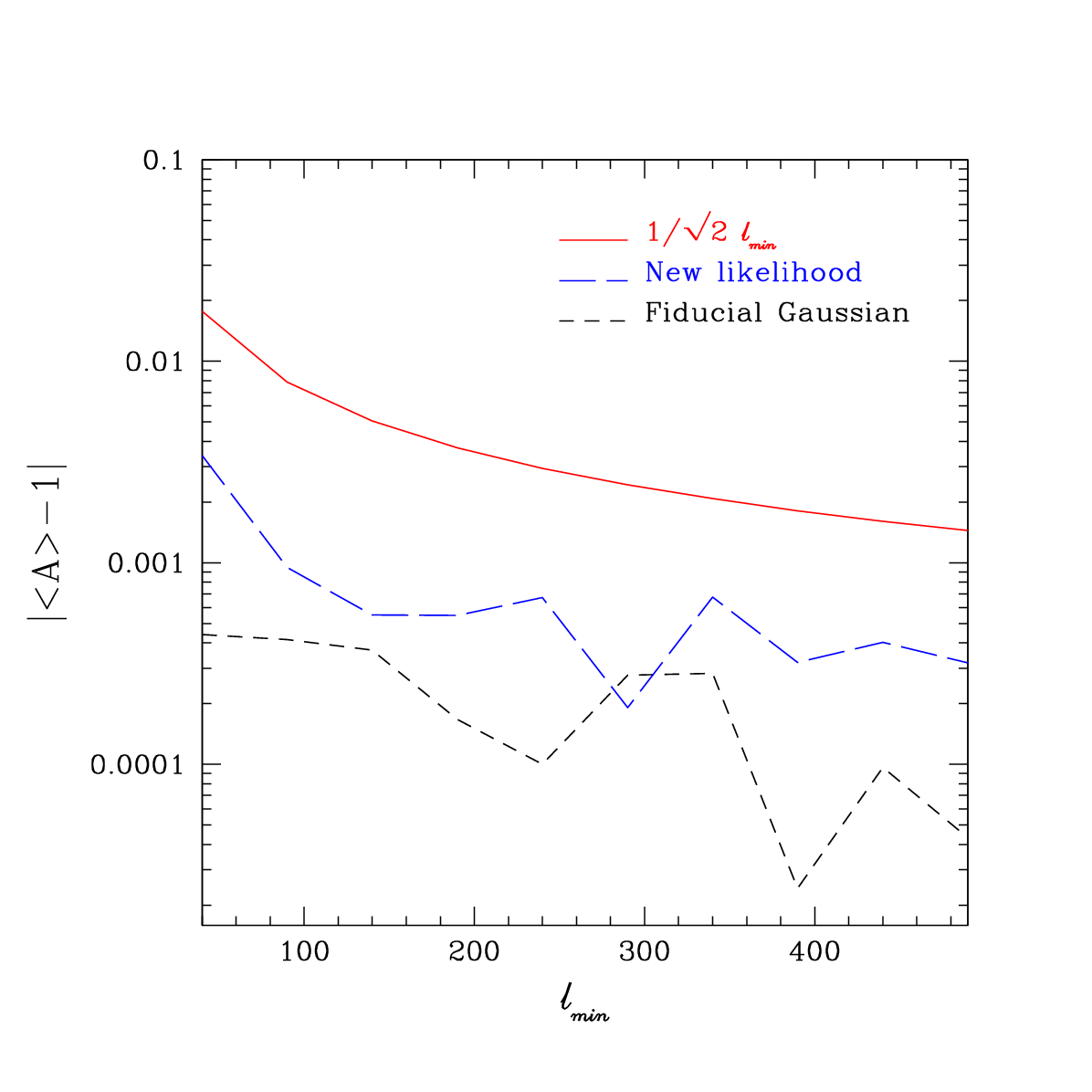, width=12cm}\caption{The difference between the average of the posterior amplitude and the true input model compared to the systematic error (red (solid) line).  The green (long-dashed) and the black (dashed) lines represent the differences for the new likelihood and fiducial Gaussian, respectively.  The curves clearly do not show any significant bias in the posterior amplitudes.  The averages were taken over 5000 simulations (realizations) for $\lmax=800$.  Simulations were performed for spin-0 $T$ and $E$-mode only and for azimuthal cuts with $\fsky=0.862$ and a bin-size set to 10.}
\label{fig:AmplAverage} \end{figure}

To obtain unbiased results on an amplitude parameter from $n$ noise-free correlated fields we need the systematic fractional bias on the amplitude to be $\ll 1/l\sqrt{n}$. With more than one field there is of course a lot more freedom than simply a change in amplitude. Nonetheless it is a useful first test as many important parameters, such as those governing the primordial power spectrum, affect the $\mC_l$ essentially through an $l$-dependent scaling. If there is an apparent systematic error $\delta\mC_l$ in the $\mC_l$ spectrum, the criterion for an unbiased amplitude is $\Tr[\mC^{-1}_l\delta\mC_l]/n \ll 1/l\sqrt{n}$. Since in practice polarization observations are likely to be noise dominated compared to the temperature for the near future, any approximation that satisfies this criterion will be more than adequate. We should however also test for accuracy of the likelihood to other changes in the spectrum, for example the degree of cross-correlation, as an amplitude scaling is a very special (if relevant) case.

We first test the approximate likelihood function compared to the exact result (see Eq.~\eqref{newlike} for exact likelihood function used); the result is shown in the Fig.~\ref{fig:polcutlike}.  The new likelihood approximation compares quite well with the exact likelihood, though it is slightly broader due to the loss of information from compressing the data into a set of pseudo-$C_l$ power spectrum estimators $\mCh_l$. The fiducial Gaussian approximation shows significant deviations from the shape of the exact likelihood far from the peak.

For a quick analysis, the tests in the rest of this section were performed for spin-0 $T$ and $E$-mode only, i.e. the $E$-polarization was simulated as a scalar field similar to temperature so that $E$-$B$ mixing may be ignored (but $T$-$E$ correlations correctly accounted for).  For all simulations, we also fix $\Ch_{l}$ at $l \le 30$ to $\Ch_{l}=C_{l}$ to avoid negative estimators and use a bin-size of $\Delta_{l}=10$.

The first consistency check is that on average over simulations $|\la\hat{A}\ra -1| \ll 1/l\sqrt{n}$: this is sufficient to check that there is no significant bias in the posterior amplitude. We ran simulations for an azimuthal cut with $\fsky=0.826$ with the results shown in  Fig.~\ref{fig:AmplAverage}. The new likelihood and the fiducial Gaussian approximations appear to be unbiased.

\begin{figure}[htbp] \centering \psfig{file=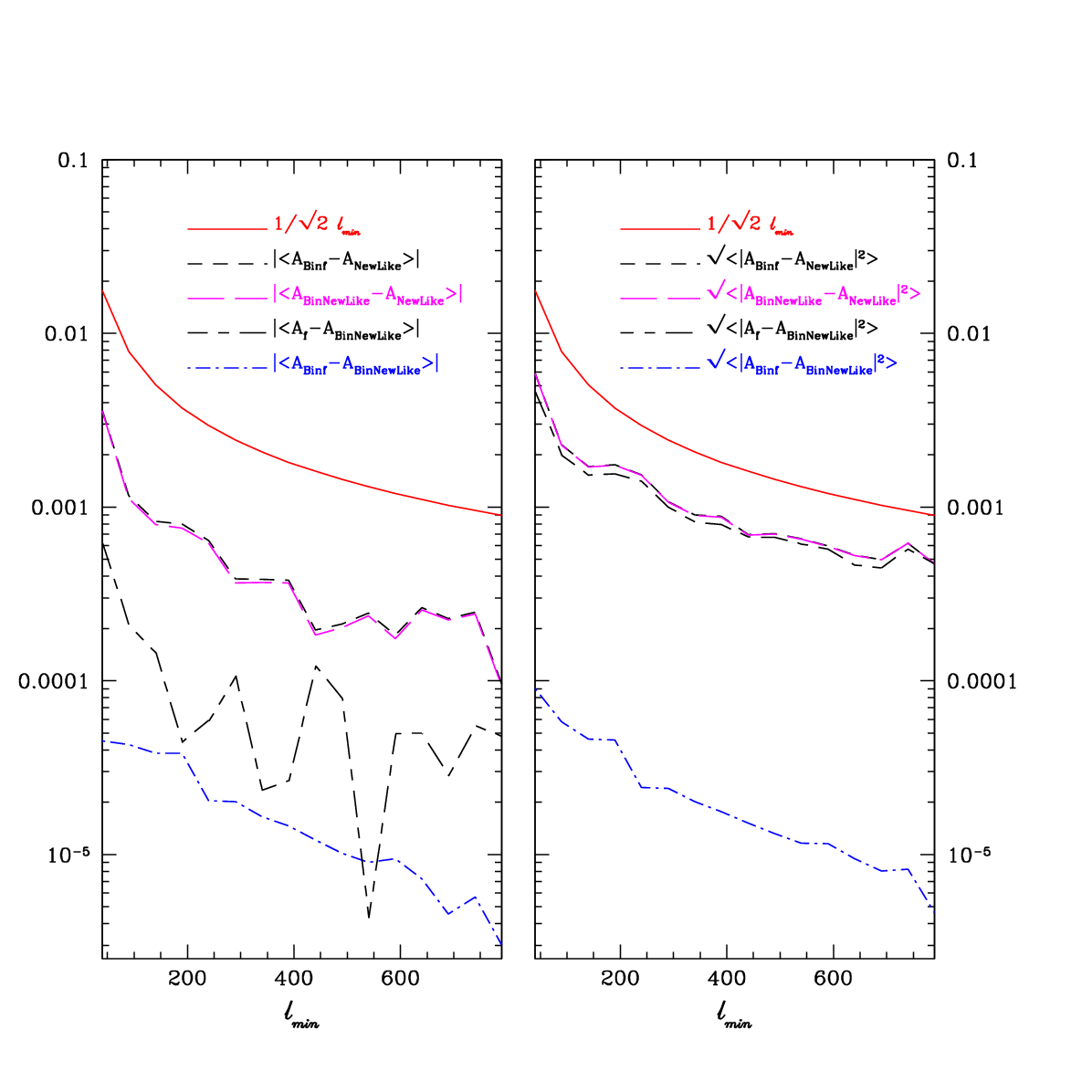, width=14cm}\caption{Comparison  between various binned and un-binned likelihood approximations.  The left plot shows the average of the difference between the posterior amplitudes of these likelihoods and the right plot shows the variance, both compared to the systematic error (red (solid) line).  The black (dashed), the green (long-dashed), the cyan (dashed long-dashed) and the blue (dotted dashed) lines represent the comparison between binned fiducial Gaussian and new likelihood, binned new likelihood and new likelihood, fiducial Gaussian and binned new likelihood and binned fiducial Gaussian and binned new likelihood, respectively.  Averages were taken over 200 simulations (realizations) for $\lmax=800$.  Simulations were performed as previously mentioned. Results are all consistent to the required accuracy.}
\label{fig:ExpectVarFid} \end{figure}

\begin{figure}[htbp] \centering \psfig{file=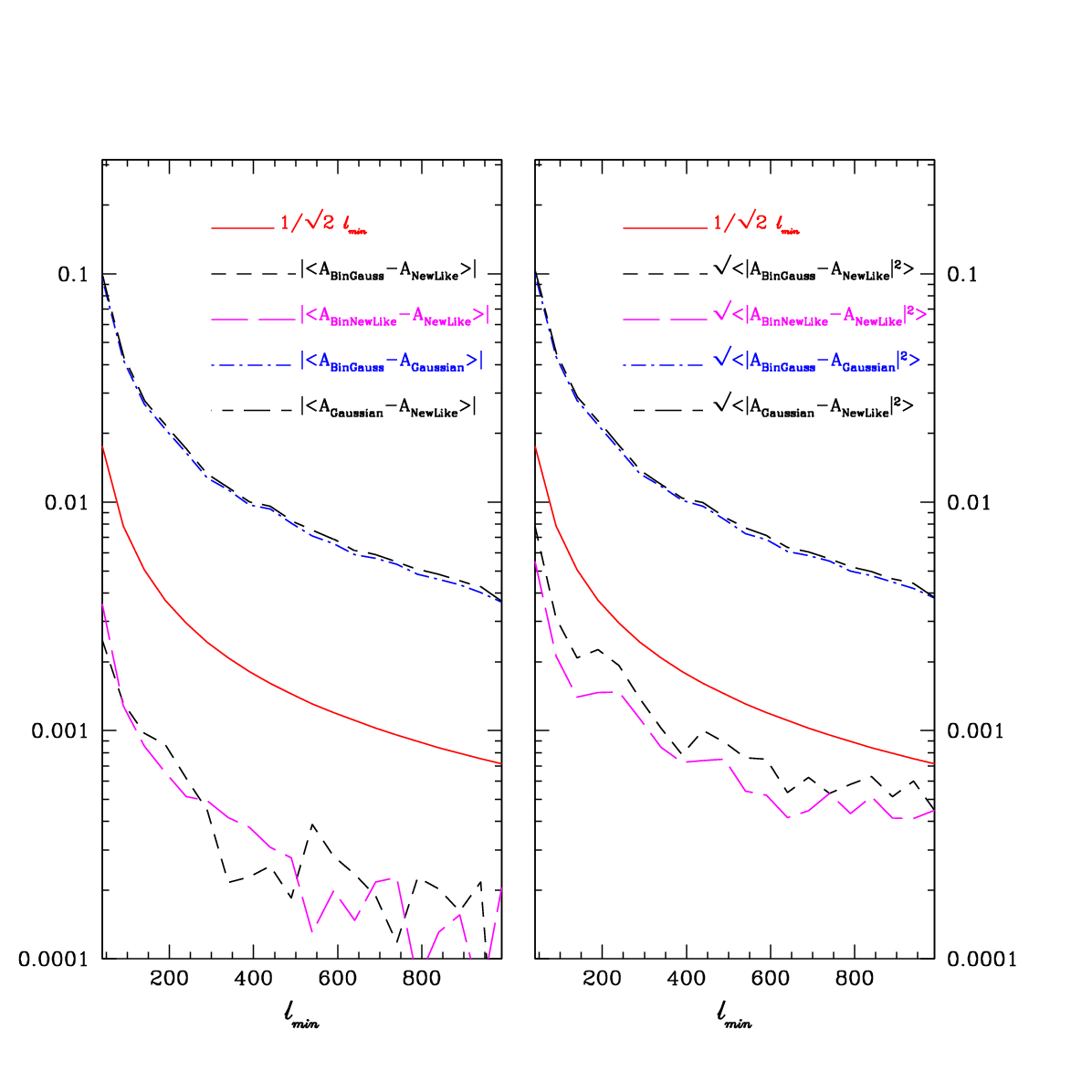, width=14cm}\caption{Similar comparison as in Fig.6 but using the symmetric Gaussian approximation $\cll_S$.  Unlike the binned and un-binned fiducial Gaussian, the binned and un-binned symmetric Gaussian approximations $\cll_S$ show significant bias.  Averages are over 100 simulations (realizations) for $\lmax=1000$.}
\label{fig:ExpectVarSGauss} \end{figure}

We can also check the consistency of the likelihood function by comparing the binned and un-binned likelihood: as discussed in Section~\ref{sec:binning} the likelihood function for bins with $\Delta_l^{(b)}\gg 1$ should be accurately Gaussian. For a smooth power spectrum binning can be performed with very little loss of information, and so the likelihood $P(\{\mC_b\}|\{\mCh_b\})$ can be calculated essentially exactly in the Gaussian approximation. We can check that this is consistent with the likelihood approximation evaluated using each $l$; if it is, then we are using the information in the $\mCh_l$ essentially optimally, at least when the spectrum is very smooth (even if compressing the sky into a set of $\mCh_l$-estimators is not optimal).  Similar to the full-sky single-field analysis, we calculate on average over simulations the difference between the posterior amplitudes, $|\la\hat{A}_{a}-\hat{A}_{b}\ra|$ and the variance which is the square of the difference, $\la|\hat{A}_{a}-\hat{A}_{b}|^{2}\ra$.  We again require both quantities to satisfy the criterion set earlier, i.e. $|\la\hat{A}_{a}-\hat{A}_{b}\ra|$, $\la|\hat{A}_{a}-\hat{A}_{b}|^{2}\ra^{1/2} \ll 1/l\sqrt{n}$.  Fig.~\ref{fig:ExpectVarFid} compares fiducial Gaussian, binned fiducial Gaussian, new likelihood, binned new likelihood  and Gaussian$_{D}$.  The plot clearly demonstrates that these approximations would produce the same results and are good enough to be used in analysing CMB data.  Fig.~\ref{fig:ExpectVarSGauss} shows the comparison between the $\cll_S$ approximation (Gaussian with variance given by $\mCh_l$), binned $\cll_S$ Gaussian, new likelihood and binned new likelihood.  This shows that $\cll_S$ is strongly biased when used with un-binned estimators, but when the data is binned it can produce consistent results as expected.

\begin{figure}[htbp] \centering \psfig{file=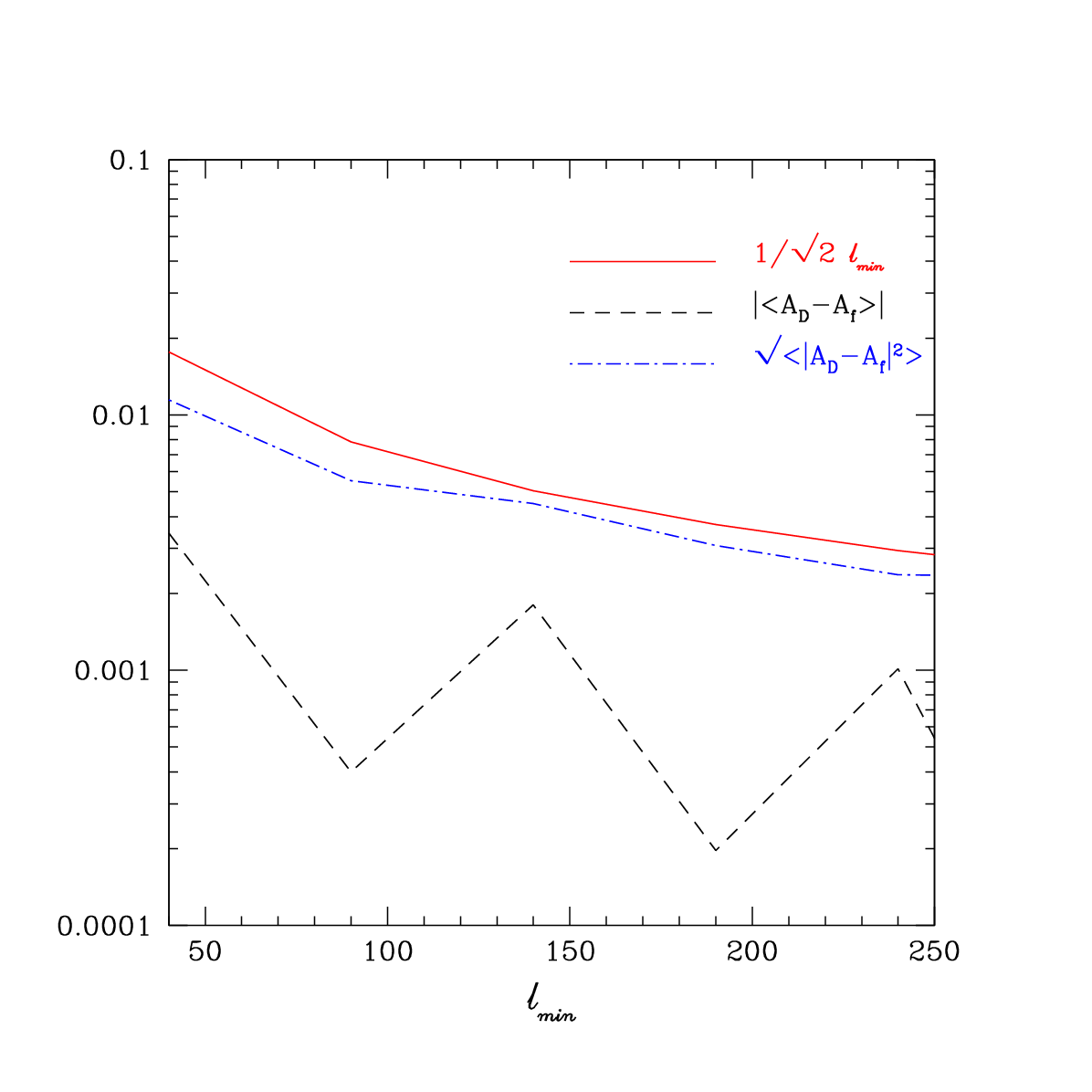, width=12cm}\caption{A test over 20 simulations to compare the Gaussian$_{D}$ and Gaussian$_{f}$ distributions for $\lmax=300$ and bin width $\Delta_l=10$.}
\label{fig:CompareGaussd} \end{figure}

The Gaussian approximation with varying covariance, Gaussian$_{D}$, is significantly slower to compute than the other approximations. It is compared to the fiducial-model Gaussian in Fig.~\ref{fig:CompareGaussd} for a small number of simulations. Since the fiducial-Gaussian result is unbiased this shows that Gaussian$_{D}$ is also unbiased to good enough (though not excellent) accuracy in this case.



\section{Parameter estimation tests with anisotropic noise}
\label{anisotropic_parameters}
\begin{figure}
\epsfig{file=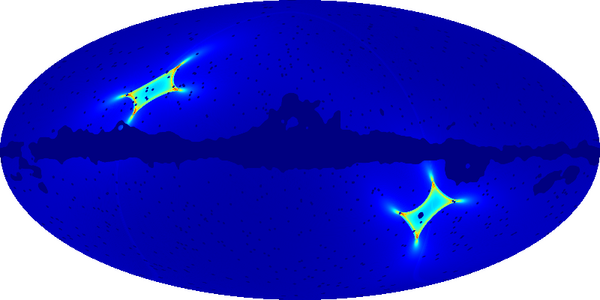, width=12cm}\caption{Smoothed regularized inverse-noise weight map with WMAP kp2 cut as used by our test Plank-like simulation analysis. Noise is lowest in the cuspy regions around the ecliptic poles. The cut gives zero weight to regions around the galactic plane and numerous point sources. Noise and cut are smoothed with a $7\arcmin$-fwhm Gaussian.}
\label{fig:planck_noise} \end{figure}
So far we have been using azimuthally symmetric cuts and assuming that the noise is isotropic. Isotropic noise is particularly simple case because the variance of the $\hat{C}_l$ estimators scales as $\propto (C_l+N_l)^2$ to a good approximation. When the noise is anisotropic, as in realistic observations, this is no longer the case in general, and it is important to test the likelihood approximations in this more realistic situation. For example using a fiducial model covariance in our approximation of Eq.~\eqref{new_approx} was motivated in the case where everything is a function only of $(C_l+N_l)$. In general it may be necessary to instead evaluate the covariance for each theoretical model to correctly account for the more complicated scaling of the covariance with the signal. This could be done for example by re-scaling a sum of covariance matrices calculated for noise-only, signal-only and signal plus noise realizations in some fiducial model. Although perfectly tractable, we shall see that in the case of Planck the simpler fiducial model approximation appears to be adequate.

We test the likelihood approximations by performing parameter estimation using single sky maps simulated corresponding to an idealization of the combined Planck 143Ghz channels with $7\arcmin$ symmetric Gaussian beam~\cite{unknown:2006uk}. The Planck satellite scanning strategy samples points near the ecliptic poles more densely than near the equator, and so there is a large ($\sim 100$ factor) range of noise values across the sky~\cite{Ashdown:2007ta}. In addition we use the `kp2'
map\footnote{\url{http://lambda.gsfc.nasa.gov/}}~\cite{Bennett:2003bz} as a semi-realistic sky cut to simulate masking out the galaxy and point sources. Details of our simulation, hybrid Pseudo-$C_l$ analysis and covariance model (following Ref.~\cite{Efstathiou:2003dj}) are given in Appendix~\ref{Planck}. In the high signal to noise regime the hybrid estimator uses an approximate inverse-noise weighted map with sky cut. As shown in Fig.~\ref{fig:planck_noise} this is highly anisotropic. This inverse-noise weighted map is combined with a uniform-weighted map to give $\mC_l$ estimators that are fairly close to optimal on all scales with $l\agt 30$. For our simple test we assume a noise level average equivalent to the number for the 143Ghz channel quoted in the Planck science case~\cite{unknown:2006uk}.
We take the polarization and temperature pixel noise to be uncorrelated and proportional, with the polarization noise a factor of four larger than the temperature.

We use the range $30\le l \le 2000$ for test parameter estimation from simulations; the low $l$ likelihood is problematic because the Pseudo-$C_l$ estimators are not guaranteed to be positive definite, and the covariance structure becomes complicated due to E/B mixing effects on the cut-sky. It may be possible to obtain reliable results from the $\tilde{C}_l$ directly (without inverting to the unbiased estimators), using maximum-likelihood or other more optimal estimators, however at low $l$ the likelihood function can also be calculated essentially exactly in reasonable computational time, so here we focus on the higher $l$ region where an exact analysis is intractable. Investigation of the low $l$ likelihood function for Planck-like noise, how to combine with higher-$l$ approximations, and dealing with real-world complications such as foregrounds is beyond the scope of this paper.\footnote{If only temperature is used then the new likelihood approximation works reliably with Pseudo-$C_l$ estimators down to $l=2$ in almost all realizations.}
\begin{figure}
\epsfig{file=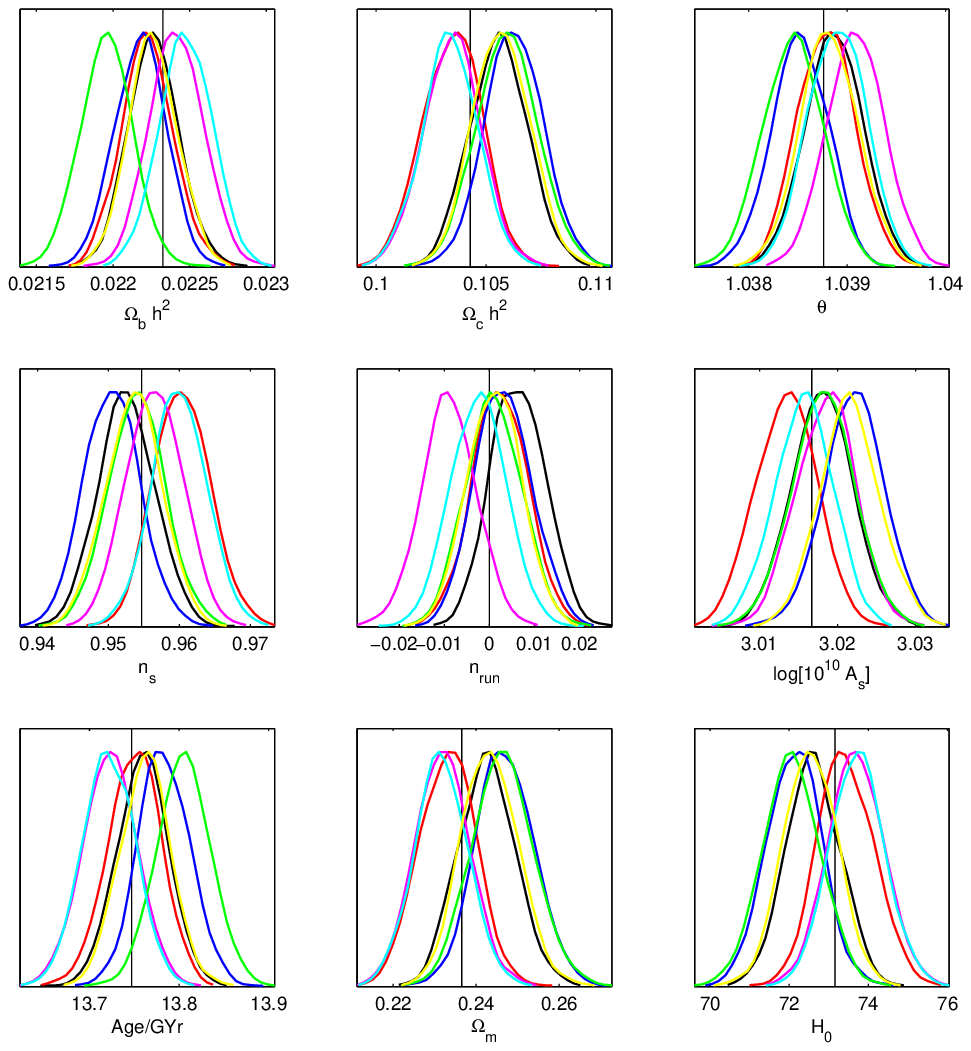, width=12cm}\caption{Parameter constraints from six idealized Planck-like single map simulations with anisotropic noise as described in the text. The 1-dimensional marginalized posteriors are from using the new likelihood approximation with hybrid Pseudo-$C_l$ temperature, E-polarization and cross-correlation estimators at $l>30$. The optical depth was fixed, and the simulation input parameters are shown with vertical lines. Very similar results are obtained if the noise-dominated $B$-polarization estimators are included with no tensor modes.}
\label{fig:planck_params} \end{figure}

From the simulated $\hat{\mC}_l$-estimators we calculate the likelihood function of a given theoretical model using a likelihood approximation. This is used in the CosmoMC\footnote{\url{http://cosmologist.info/cosmomc/}; new CMB likelihood module at \url{http://cosmologist.info/cosmomc/CMBLike.html}} parameter estimation code to sample from the posterior parameter distribution~\cite{Lewis:2002ah}. For our tests we consider a vanilla adiabatic flat $\Lambda$-CDM model, with baryon density $\Omega_b h^2$, dark matter density $\Omega_c h^2$, amplitude, spectral index and running of the primordial power spectrum ($A_s$, $n_s$ and $n_{\text{run}}$), and the parameter $\theta$, 100 times an approximation of the ratio of the sound horizon to the angular diameter distance at recombination. The age, Hubble parameter ($H_0 \text{km} s^{-1} \text{Mpc}^{-1}$) and matter density relative to critical $\Omega_m$ are derived parameters. Since we are only considering the likelihood at $l\ge 30$ we fix the optical depth to reionization; our simulated parameter constraints are therefore tighter than expected from a full realistic analysis.

Figure~\ref{fig:planck_params} shows the consistent marginalized parameter constraints obtained when using the new likelihood approximation to analyse a set of sky simulations. Very similar constraints are obtained whether noise-dominated $B$ power spectrum estimators are included or not, at least when there are no tensor modes. The new likelihood approximation seems to work well with realistically anisotropic noise.

\begin{figure}
\epsfig{file=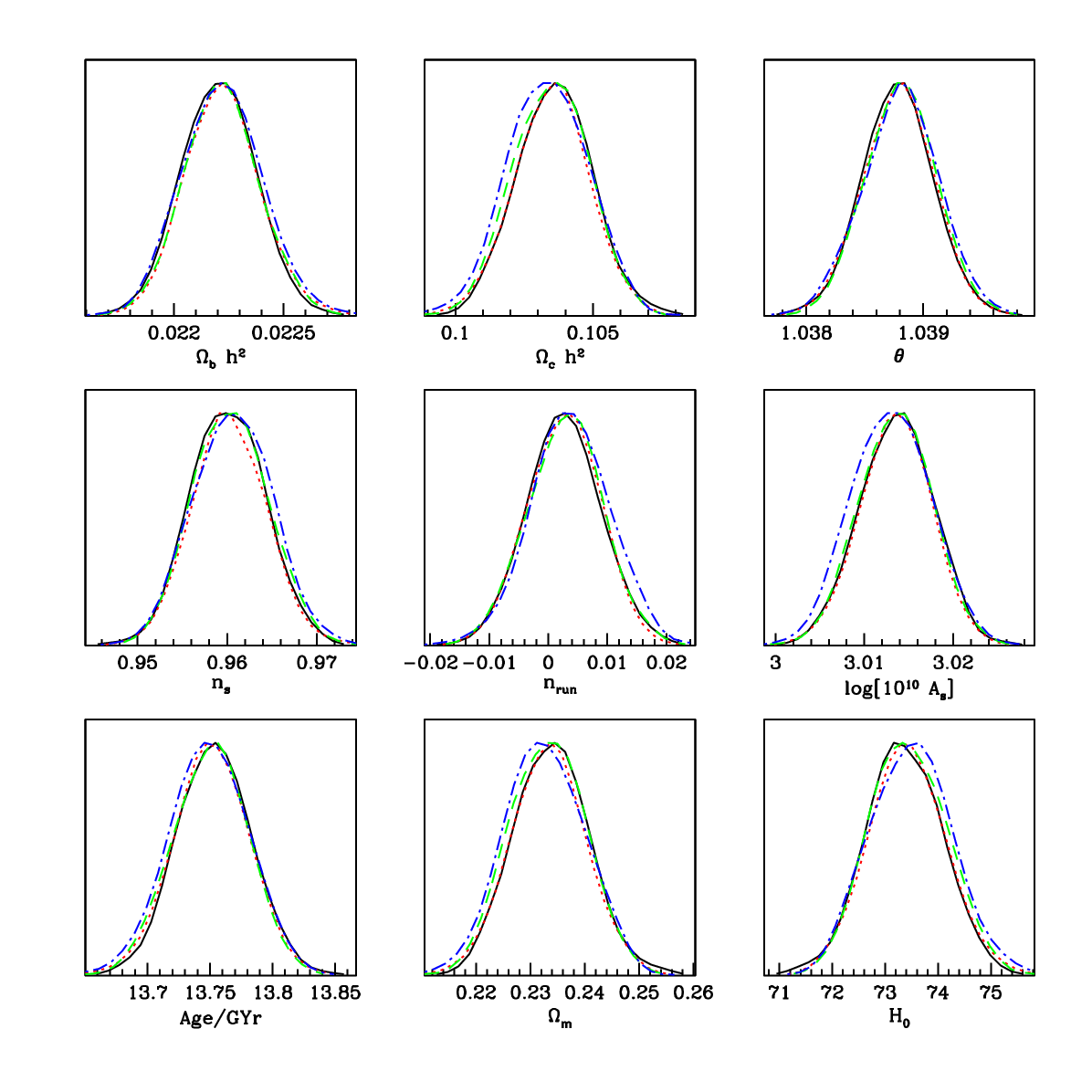, width=12cm}\caption{Parameter constraints from a single idealized Planck-like simulations with anisotropic noise.  The 1-dimensional marginalized posteriors are from using the new likelihood and the fiducial Gaussian approximations, and compare the results obtained when assuming an exactly correct fiducial model or using a wrong $n_s=1$ model.  The red (dotted) line is the new likelihood with the right model, the black (solid) line is new likelihood with the wrong model, which agree very well.  The green (short-dashed) line is the fiducial Gaussian with the right model and the blue (dotted-short dashed) line is the fiducial Gaussian with the wrong model.  The new likelihood results are consistent but the fiducial Gaussian results are slightly affected by the choice of the model.}
\label{fig:new_fid_compare} \end{figure}

Since in reality we will not know a priori exactly what fiducial model to choose, it is important that results be robust to choosing a slightly wrong model.
Figure~\ref{fig:new_fid_compare} compares the results from one simulation using the new likelihood approximation compared to using the fiducial-model Gaussian approximation; the fiducial models have $n_s=1$ (wrong) and $n_s\sim 0.955$ (true), a difference of many sigma at Planck sensitivity. All the results are broadly consistent, but the fiducial-model Gaussian approximation shows some dependence on the choice of fiducial model. The new likelihood approximation results are more independent of the choice of fiducial model, and so appear to be more robust as expected\footnote{Some of the difference here is due to changes in the hybrid pseudo-$C_l$ estimators when the fiducial model is changed.}. The values of the goodness-of-fit parameter $\chieff$ (see Appendix~\ref{goodness}) are also much more stable for the new approximation compared to the fiducial Gaussian; the new likelihood approximation best-fits differ by $\Delta\chieff\sim 4$, but the fiducial-model Gaussian approximations differ by $\Delta\chieff\sim 400$. With a fiducial model chosen to be sensibly closer to the maximum likelihood model both numbers should be significantly smaller.

Although detailed analysis of secondary signals is beyond the scope of this paper, in the Appendix~\ref{lensing} we show that with Planck noise levels our likelihood approximations also work when applied to lensed CMB fields and the covariance is estimated simply by using the lensed power spectra.

\section{Conclusions}
In this paper we have attempted to find solutions to the problems facing the likelihood analysis of the CMB temperature and polarization estimators on small scales.  With realistic data we need to be able to calculate the likelihood accurately from partial sky observations. Previous attempts have established some excellent approximations to model the non-Gaussianity of the temperature likelihood function.  However, no good general approximation has been derived to model the polarized likelihood.  At large $l$ computing the likelihood function exactly is computationally prohibitive and the correlation between the temperature and polarization fields makes it more complicated than for the temperature field only.  We gave a new general approximation that can account for this correlation and is exact on the full-sky.  This new approximation is fast to evaluate as it involves a pre-computed covariance independent of $\mC_l$, and appears to be more than adequate to obtain robust parameter constraints from clean small-scale CMB temperature and polarization data.

In summary, our conclusions regarding the modelling of the likelihood function of power spectrum estimators are:


\begin{itemize}
\item
In the case of binned power spectra, the number of modes per bin ($n_m/n_b$) must be much larger than the number of bins ($n_b$) for
non-Gaussian corrections to the likelihood function to be unimportant in all cases;
i.e. $n_b \ll \sqrt{n_m}$ is required to ensure that parameter bias is much smaller than the error bar.

\item A Gaussian approximation with fixed fiducial-model covariance gives unbiased results for smooth power spectra at high $l$, but error bars have some dependence on the choice of the fiducial model.  Goodness-of-fit estimators $\chieff$ can be misleading even for small differences between the fiducial and true model.

\item A Gaussian approximation with covariance that varies with parameters can give reliable results at high $l$ for smooth spectra, but only if the determinant-term is consistently included; the quadratic approximation without determinant, $\cll_Q$, is biased in general.

\item The new likelihood approximation presented in Section~\ref{sec:newlike} appears to work well for power spectrum estimators with correlated fields and can give nearly optimal results when applied to good power spectrum estimators. It is fast to evaluate as it relies on a pre-computed fiducial-covariance matrix, but is insensitive to small errors in the fiducial model. We recommend it for future work.

\item Most likelihood approximations with binned estimators ($n_b \ll \sqrt{n_m}$) can produce consistent results by the central limit theorem; for smooth power spectra consistency of parameter constraints with those from binned power spectra is a good check.
\end{itemize}

Since the new likelihood approximation is based on estimators and a covariance matrix, it is likely to generalize well to more realistic data where additional uncertainties, non-Gaussianities and correlations can be accounted for via changes to the estimator covariance. It is also likely to produce good results down to low $l$ if positive-definite estimators are used, though this has not been the focus of this paper. Complications such as correlated noise may be well encapsulated in the covariance of a set of maximum-likelihood (or similar) estimators, giving a fast alternative to much slower brute-force likelihood calculations. If the approximation is nearly correct, importance sampling techniques could be used to correct the results with a much smaller number of high-accuracy calculations.

We have not touched at all on the complications of foreground modelling, point sources, non-linear and non-Gaussian anisotropies (e.g. due to SZ), beam uncertainties, or a plethora of other real-world complications. Extending our work to account for these will be crucial for the correct interpretation of future data.

\section{Acknowledgements}
SH gratefully acknowledges the support of the Algerian Ministry of Higher Education and Scientific Research (MESRS).  AL acknowledges a PPARC/STFC Advanced fellowship and thanks Anthony Challinor, Steven Gratton, George Efstathiou, Dipak Munshi, Mark Ashdown and Carlo Contaldi for discussion. Some of the results in this paper have been derived using the \healpix~\cite{Gorski:2004by} package.

\appendix

\section{Useful results for matrix vectorization}
\label{matrices}
In this appendix we review some results from matrix theory relating equations involving matrices to those involving vectors of their components, and establish Eq.~\eqref{matrix_identity} in the main text. For further details and references see e.g. Ref.~\cite{Gupta99}.

The elements of a general matrix $\mA$ can be assigned column-wise into a vector $\vec(\mA)$. For matrices $\mA$ and $\mB$
\be
\label{Trequiv}
\Tr\left[ \mA^T \mB\right] = \vec(A)^T \vec(B).
\ee
The Kronecker product of an $m\times n$ matrix $\mA$ with and $p\times q$ matrix $\mB$ is defined to be the $mp \times nq$ matrix
\be
\mA\otimes\mB = \begm A_{11}\mB & A_{12}\mB &\dots & A_{1n}\mB \\ A_{21}\mB & A_{22}\mB & \dots & A_{2n}\mB
\\ \vdots & \vdots & & \vdots
\\ A_{m1}\mB & A_{m2}\mB & \dots & A_{mn}\mB
\enm.
\ee
Using this we can write
\be
\vec( \mA \mB \mC) = (\mC^T\otimes \mA) \vec(\mB),
\ee
and using Eq.~\eqref{Trequiv} this implies
\be
\label{Tr4}
\Tr\left[ \mA^T \mD \mE \mF\right] = \vec(A)^T (\mF^T\otimes \mD)\vec(\mE).
\ee
For a symmetric $n\times n$ matrix there are only $n(n+1)/2$ distinct elements, and we define $\vecp(\mA)$ to be the corresponding vector of distinct components of $\mA$
\be
\vecp(\mA) = \begm A_{11}, & A_{21}, & \dots & A_{n1}, & A_{22}, & A_{32}, & \dots \enm^T.
\ee
The matrix $n^2\times n(n+1)/2$ matrix $\mBp$ is defined so that for a general square matrix $\mA$
\be
\vecp(\mA) = \mBp^T \vec(\mA)= \mBp^T \vec(\mA+\mA^T)/2.
\ee
For example, a $2\times 2$ matrix $\mA$ has $\mBp^T\vec(\mA) = \begm A_{11}, & (A_{12}+A_{21})/2, & A_{22}\enm^T$. The pseudo-inverse $\mBp^+ \equiv (\mBp^T \mBp)^{-1}\mBp^T$ can be used to construct $\vec(\mA)$ from $\vecp(\mA)$ when $\mA$ is symmetric:
\be
\vec(\mA) = (\mBp^+)^T \vecp(\mA).
\ee
Applying Eq.~\eqref{Tr4} to symmetric matrices $\mA$ and $\mD$ we then have
\be
\Tr\left[ \mA \mC \mD \mE\right] = \vecp(\mA)^T \mBp^+ (\mE\otimes \mC) (\mBp^+)^T \vecp(\mD).
\ee
Using the results that $(\mA\otimes\mB)^{-1}=\mA^{-1}\otimes \mB^{-1}$ (for non-singular matrices) and
$\mBp \mBp^+ (\mC\otimes\mC) = (\mC\otimes\mC) \mBp \mBp^+$ it follows that
$\mBp^T (\mC\otimes \mC) \mBp \mBp^+ (\mC^{-1}\otimes \mC^{-1}) (\mBp^+)^T =\mI$ and hence
\be
\label{TrInv}
\Tr\left[ \mA \mC^{-1} \mD \mC^{-1}\right] = \vecp(\mA)^T \left[\mBp^T (\mC\otimes \mC) \mBp\right]^{-1} \vecp(\mD).
\ee
The $\mCh_l$ covariance matrix of Eq.~\eqref{gen_cov} is defined by
\be
\mM_l \equiv \la \vecp(\mCh_l-\mC_l) \vecp(\mCh_l-\mC_l)^T\ra  = \mBp^T \la \vec(\mCh_l-\mC_l) \vec(\mCh_l-\mC_l)^T\ra  \mBp,
\ee
where since $\mCh_l \equiv \sum_m \va_{lm} \va_{lm}^\dag/(2l+1)$ we have
\be
\vec(\mCh_l) = \frac{1}{2l+1}\sum_m \va_{lm}\otimes \,\va_{lm}^*.
\ee
Using  $(\mA\otimes\mB)^T = \mA^T\otimes \mB^T$  gives
\begin{eqnarray}
\vec(\mCh_l)\vec(\mCh_l)^T &=& \frac{1}{(2l+1)^2} \sum_{m m'}(\va_{lm}\otimes \,\va_{lm}^*)(\va_{lm'}^\dag\otimes\, \va^T_{l m'}).
\end{eqnarray}
The general result (for appropriately sized matrices) that $(\mA\otimes\mB)(\mC\otimes\mD) = (\mA\mC)\otimes(\mB\mD)$ lets us use the expectation value  $\la \va_{lm}\va_{lm'}^\dag\ra = \delta_{mm'} \mC_l$. However to do the different contractions we need to use the symmetry $\vec(\mCh_l) = \mBp\mBp^+\vec(\mCh_l)$ with the fact that for vectors $\va$, $\vb$ the ordering can be changed after symmetrization using $\mBp\mBp^+(\va\otimes\vb) =\mBp\mBp^+(\vb\otimes\va) $. This then gives
\begin{eqnarray}
\la\vec(\mCh_l)\vec(\mCh_l)^T\ra &=& \frac{1}{(2l+1)^2} \sum_{m m'}\mBp\mBp^+\la(\va_{lm}\otimes \,\va_{lm}^*)(\va_{lm'}^\dag\otimes\, \va^T_{l m'})\ra\nonumber\\
&=&  \frac{2}{(2l+1)^2}\mBp\mBp^+\sum_{m m'} \la\va_{lm}\va_{lm'}^\dag\ra\otimes\la\va_{lm}\va^\dag_{lm'}\ra^* + \vec(\mC_l)\vec(\mC_l)^T.
\end{eqnarray}
Hence the covariance is given by\footnote{This equation is missing the $\mBp \mBp^+$ factor in the published version, we thank Anthony Challinor for pointing this out.}
\be
\la \vec(\mCh_l-\mC_l) \vec(\mCh_l-\mC_l)^T\ra = \frac{2}{2l+1}\mBp\mBp^+(\mC_l\otimes \mC_l),
\ee
so that $\mM_l = 2 \mBp^T (\mC_l\otimes \mC_l) \mBp/(2l+1)$. Then from Eq.~\eqref{TrInv} we have
\be
\Tr\left[ \mA \mC_l^{-1} \mD \mC_l^{-1}\right] = \frac{2}{2l+1}\vecp(\mA)^T \mM_l^{-1} \vecp(\mD),
\ee
establishing Eq.~\eqref{matrix_identity}. As a special case
\be
\vecp(\mC_l)^T\mM_l^{-1}\vecp(\mC_l) = \frac{(2l+1)n}{2}.
\ee
If $\mC$ has eigenvectors $\{\ve_i^c\}$ with eigenvalues $\{\lambda_i^c\}$ then
\be
(\mC\otimes\mD) (\ve_i^c\otimes \ve_j^d) = (\mC\ve_i^c)\otimes(\mD\ve_j^d) = \lambda_i^c\lambda_j^d  (\ve_i^c\otimes \ve_j^d),
\ee
so the determinant is $|\mC\otimes\mD| = \prod_{ij}\lambda_i^c\lambda^d_j = |\mC|^{n}|\mD|^{n}$.
Also using $\mBp\mBp^+(\mC\otimes\mC) = (\mC\otimes\mC) \mBp\mBp^+$, we have
\be
\mBp^T(\mC\otimes\mC)\mBp \mBp^+ (\ve_i\otimes \ve_j)=
[\mBp^T(\mC\otimes\mC)\mBp(\mBp^T\mBp)^{-1}] \mBp^T (\ve_i\otimes \ve_j) = \lambda_i\lambda_j\mBp^T (\ve_i\otimes \ve_j).
\ee
So there are $n(n+1)/2$ distinct eigenvectors $\mBp^T (\ve_i\otimes \ve_j)$ of $[\mBp^T(\mC\otimes\mC)\mBp(\mBp^T\mBp)^{-1}]$, and hence
\be
\left|\mBp^T(\mC\otimes\mC)\mBp(\mBp^T\mBp)^{-1}\right| = \prod_i\prod_{j\ge i} \lambda_i\lambda_j = |\mC|^{n+1}.
\ee
The matrix $\mBp^T\mBp$ is diagonal with $n$ unit entries and $n(n+1)/2-n=n(n-1)/2$ that are a half, so $|\mBp^T\mBp|=2^{-n(n-1)/2}$ and hence
\be
|\mBp^T(\mC\otimes\mC)\mBp| =  2^{-n(n-1)/2}|\mC_l|^{n+1}.
\ee
The covariance matrix therefore has determinant
\be
|\mM_l| =  \frac{2^n}{(2l+1)^{n(n+1)/2}} |\mC_l|^{n+1}.
\ee

\section{Full-sky goodness of fit}
\label{goodness}
Often  people like to quote a chi-squared value as a crude measure of how well the data fit a given model. In the context of the full-sky CMB, where the $\va_{lm}$ are Gaussian, we could define
\begin{equation}
\chi^2 \equiv \sum_l (2l+1) \text{Tr}\left[ \mCh_l \mC^{-1}_l \right]
\end{equation}
so that $P(\{ \va_{lm} \}|\mC) \propto e^{-\chi^2/2}$.
This is minimized ($\chi^2=0$) when the $\va_{lm}$ take their maximum likelihood values (zero). The mean is $\la \chi^2\ra = \sum_l (2l+1)n$ and variance $\sum_l 2(2l+1)n$.

Alternatively, we could define an `effective' chi-squared, measuring the goodness of fit of the $\{\mC_l\}$ to $\{\mCh_l\}$~\cite{Verde:2003ey}:
\begin{eqnarray}
\chieff \equiv -2 \ln(P(\{\mC_l\}|\{\mCh_l\}) = \sum_l (2l+1) \left\{\text{Tr}\left[ \mCh_l \mC^{-1}_l
  \right] - \log |\mCh_l \mC_l^{-1}  | -n \right\}
\end{eqnarray}
(to within a $\mC_l$-independent constant).
 This is normalized so that if $\mC_l = \mCh_l$ then $\chieff=0$. To assess the goodness of fit we could compare $\chieff$ to the value expected if $\mC_l$ were the true model.  The expectation value under the Wishart distribution can be calculated by performing a Cholesky decomposition  into a lower triangular matrix $L$, where $\mC^{-1/2}_l \mCh_l \mC^{-1/2}_l = L L^T$, and using the independence of $L_{ij}$ (the off-diagonal elements being Gaussian distributed, the diagonal elements chi-squared)~\cite{Gupta99}. The result is
\begin{equation}
\la \chieff \ra = \sum_l  (2l+1)\left\{ n \ln(l+1/2) -\sum_{i=1}^n \psi(l+1-i/2)  \right\},
\end{equation}
where $\psi(x) \equiv \ud (\ln\Gamma(x))/\ud x$. For $l\gg n$ we have
\begin{equation}
 (2l+1)\left\{ n \ln(l+1/2) -\sum_{i=1}^n \psi(l+1-i/2)\right\}  = \frac{n(n+1)}{2} + \frac{1}{12}\frac{n(2n^2+3n-1)}{2l+1} + \clo(1/l^2),
\end{equation}
so for a large range of $l$ with $n\ll \lmin \le l \le \lmax$ we have
\begin{eqnarray}
\la\chieff\ra \approx   (\lmax-\lmin+1) \frac{n(n+1)}{2} + \frac{1}{24} n(2n^2+3n-1) \ln (\lmax/\lmin).
\end{eqnarray}
The first term is just what we would expect for a Gaussian distribution in  $\vXhat_l$, the $n(n+1)/2$ distinct components $\mCh_l$. The second term is the logarithmic leading-order correction. For $\lmin=30$, $\lmax=2000$ it is $\sim 0.7$ (for $n=1$), $\sim 4.6$ (for $n=2$) and $\sim 13.7$ (for $n=3$).
The variance can be calculated similarly, giving
\begin{eqnarray}
\text{var}(\chieff) &=& \sum_l (2l+1)\left\{ (2l+1) \sum_{i=1}^n \psi'(l+1-i/2) - 2 n \right\}\\
&=& \sum_l \left\{ n(n+1) + \frac{1}{3}\frac{n(2n^2+3n-1)}{2l+1} + \clo(1/l^2)\right\}\\
&\approx& 2\la\chieff\ra + \frac{1}{12} n(2n^2+3n-1)\ln (\lmax/\lmin),
\end{eqnarray}
where the prime denotes the derivative.

Note that even on the full-sky CMB lensing and other secondaries would give a non-zero connected four-point function that would change the variance of the $\mCh_l$ from that calculated here for Gaussian fields.

\section{Multiple maps}
\label{multimap}
In realistic experiments there are often many maps at different frequencies, from different detectors, and/or from different observation periods. Often the noise on these maps can be taken to be independent to an excellent approximation. Here we consider the very simple case where each map has isotropic noise. If there are two maps $a_{lm}^{(1)}$ and $a_{lm}^{(2)}$, each containing sky signal plus noise, the difference map $a_{lm}^{(1)}-a_{lm}^{(2)}$ will be independent of the signal. With $n$ maps, there are therefore $n-1$ linear combinations that do no depend on the signal, and hence can be integrated out of the likelihood function. The remaining uncorrelated linear combination is the inverse-noise weighted combined map
\begin{equation}
a_{lm}^{(t)} \equiv \frac{\sum_{i=1}^n (N_l^{(i)})^{-1}a_{lm}^{(i)}}{\sum_{i=1}^n (N_l^{(i)})^{-1}}.
\end{equation}
A similar argument applies in real space with anisotropic noise. The combined map $\{a_{lm}^{(t)}\}$ is a sufficient statistic for the likelihood function, and the likelihood analysis could therefore be based on $C_l$ estimators from the combined map $a_{lm}^{(t)}$. Alternatively we could consider estimating a set of $\hat{C}_l^{(ij)}$ from all possible combinations of maps
\begin{equation}
\hat{C}_l^{(ij)} = \frac{1}{2l+1} \sum_m a^{(i)}_{lm}{}^* a^{(j)}_{lm},
\end{equation}In the simple case considered above, the optimal linear combination of the $\hat{C}_l^{ij}$ is $\propto \sum_{ij} (N_l^{(i)})^{-1}(N_l^{(j)})^{-1} \hat{C}_l^{(ij)}$, and using this would be equivalent to using the estimator $\hat{C}_l^{(t)}$ from the combined map $a_{lm}^{(t)}$. The likelihood approximations approximations in the main text could be applied directly to realistic pseudo-$C_l$ generalizations of this estimator.

An alternative is to use only the off-diagonal correlations, where $i\ne j$~\cite{Hinshaw:2003ex}. In the simplest case we can define the optimal weighted combination
\begin{equation}
\hat{C}_l^\off \equiv \frac{\sum_{ij} (N_l^{(i)})^{-1}(N_l^{(j)})^{-1} \hat{C}_l^{(ij)}(1-\delta_{ij})}{\sum_{ij} (N_l^{(i)})^{-1} (N_l^{(j)})^{-1}(1-\delta_{ij})}.
\end{equation}
Since $\la \hat{C}_l^\off \ra = C_l$ the estimator is an unbiased estimator of the $C_l$ regardless of the noise. In some instances it might therefore be more robust than including the diagonal correlations, where an error in the noise model can lead to an immediate bias in the estimator. However this estimator is no longer equivalent to a the estimator on the weighted map $a_{lm}^{(t)}$, and has a different distribution. In particular it is not positive definite. If $\hat{C}_l^\off$ are to be used for parameter estimation, in principle it may therefore be necessary to use a different likelihood approximation from those designed for analysing Wishart-like distributions.

To see how different the distribution is we consider the very simplest case of foreground-free full-sky maps where all the maps have identical isotropic noise $N_l^{(i)}=N_l$, and we consider only a single scalar field (no polarization). We can define a $n$-dimensional vector of $a_{lm}^{(i)}$, $\va_{lm}$. The estimator is then
\begin{equation}
\hat{C}_l^\off = \frac{1}{(2l+1)n(n-1)}\sum_{m}\va_{lm}^\dag\left( \ve\ve^\dag - \mI \right) \va_{lm},
\end{equation}
where $\ve$ is a vector of ones, $e_i=1$. The covariance of the $\va_{lm}$ is given by
\begin{equation}
\mM_l \equiv \la \va_{lm}\va_{lm}^\dag \ra = C_l \ve \ve^\dag + N_l\mI.
\end{equation}
The distribution of the $\Coff$ is then given by
\begin{eqnarray}
P(\Coff|C_l,N_l) &=& \int \ud \va_{lm} P(\va_{lm}|\mM_l) \delta\left( \hat{C}_l^\off - \alpha_{ln}\sum_{m}\va_{lm}^\dag\left( \ve\ve^\dag - \mI \right) \va_{lm}\right) \nonumber\\
&=&\frac{1}{2\pi} \int_{-\infty}^{\infty} \ud k \frac{e^{-i k \Coff}}{\left|\mI - 2i k \alpha_{ln} \mM_l(\ve\ve^\dag-\mI)\right|^{l+1/2}},
\end{eqnarray}
where the last line follows from writing the $\delta$-function as a Fourier transform and $\alpha_{ln}^{-1} \equiv (2l+1)n(n-1)$. Substituting for $\mM_l$ and using $|\mI + a \ve\ve^\dag| = 1+ n a$, the characteristic function (Fourier transform of the distribution function) is therefore given by
\begin{equation}
\tilde{P}(k|C_l,N_l) = \frac{1}{\left[ (1+ 2ik\alpha_{ln}N_l)^{n-1}(1-2i k (2l+1)^{-1}(C_l +N_l/n))\right]^{l+1/2}}.
\end{equation}
The quantity $C_l+N_l/n\equiv C_l+N_l^{(t)}$ is just the expectation value of $C_l^{(t)}$ from the optimal map. The distribution of $\Coff$ is therefore the same as that of the variable $\hat{C}_l^{(t)} - \sum_{j=1}^{n-1} \hat{N}_l^{(t)(j)}/(n-1)$, where $\hat{N}_l^{(t)(j)}$ is the estimator from one of $n-1$ independent realizations of the noise. In the limit of many maps, $n \rightarrow \infty$ keeping the total noise $N_l^{(t)}$ fixed, we have
\begin{equation}
\lim_{n\rightarrow\infty} P(\Coff|C_l,N_l) =\frac{1}{2\pi}\int_{-\infty}^{\infty} \ud k \frac{e^{-ik(N_l^{(t)}+\Coff)}}{[ 1-2i k (2l+1)^{-1}(C_l +N_l^{(t)})]^{l+1/2}}.
\end{equation}
This evaluates to the exact full-sky likelihood for $C_l^{(t)}$, so asymptotically with many maps $\Coff + N_l^{(t)}$ has the same distribution as $C_l^{(t)}$, and hence the likelihood can be approximated using the same approximations.

The distribution of $\Coff$ can be calculated analytically for the special case $n=2$ (as for the marginal distribution of $C^{TE}_l$~\cite{Percival:2006ss}), but usually the off-diagonal estimator would be used only when there are several maps. In general the moments and cumulants of the distribution of $\Coff$ can be calculated from the characteristic function, since
\begin{equation}
\left\la (\Coff)^p\right\ra = \left[i^{-p} \frac{\ud^p \tilde{P}(k)}{\ud k^p}\right]_{k=0}\\
\qquad
\kappa_p = \left[i^{-p} \frac{\ud^p \log\tilde{P}(k)}{\ud k^p}\right]_{k=0}.\\
\end{equation}
In particular we have
\begin{eqnarray}
\kappa_1&=&\la \Coff\ra = C_l \\
\kappa_2&=&\la (\Coff-C_l)^2 \ra = \frac{2}{2l+1}\left(  (C_l + N_l^{(t)})^2 + \frac{(N_l^{(t)})^2}{(n-1)}\right)\\
\kappa_3&=&\la (\Coff-C_l)^3 \ra = \frac{8}{(2l+1)^2}\left( (C_l+N_l^{(t)})^3  - \frac{(N_l^{(t)})^3}{(n-1)^2}\right)\\
\kappa_p &=& \frac{2^{p-1}(p-1)!}{(2l+1)^{p-1}}\left( (C_l + N_l^{(t)})^p + (-1)^p
\frac{(N_l^{(t)})^p}{(n-1)^{p-1}}\right).
\end{eqnarray}
The terms involving $(C_l+N_l^{(t)})$ are the equivalent results for $C^{(t)}_l$.
The distribution of $\Coff$ is therefore slightly less skewed than for the optimal estimator, but (as expected) with a slightly broader distribution. The third and higher moments will be close to those for  $C^{(t)}_l$ if $n\gg 1+ N^{(t)}_l/(N^{(t)}_l+C_l)$. We therefore anticipate that if there are enough maps that this criterion is satisfied, $n\gg 2$, the likelihood approximations presented in this paper should also work well using the estimator $\Coff + N_l^{(t)}$.


Note that even though $\Coff$ is unbiased regardless of the noise, the posterior mean of $C_l$ will depend on the noise, and there could therefore be a posterior bias on parameters even if there is no bias directly on the estimators. This bias due to noise error is however suppressed by a factor of $\sim 1/l$ compared the direct bias that would arise from using $C_l^{(t)}$ with an incorrect noise model.

\section{Cut-sky estimators, covariance and exact likelihood}
\label{cutsky}
\subsection{Calculating the CMB cut-sky estimators}

For limited sky coverage the temperature field is observed over only part of the sky. For full-sky observations part of the sky is likely to be dominated by galactic foregrounds, and CMB observations are effectively only available over the region of the sky outside a galactic (and point source) cut. In addition noise properties are generally not uniform across the sky; indeed a cut-sky can be thought of a full-sky observation with infinite noise in the cut region. For these reasons it is useful to define a weighted temperature field $\tilde{T}$ given by
\be
\tilde{T}(\Omega)\equiv W^{T}(\Omega)T(\Omega),
\ee
where $W^{T}$ is a weighting function defined over the whole sky that lies in the range 0 to 1. The simplest weighting function is zero in the cut region and one in the region with useful data; however more general window functions can be useful to obtain more optimal estimators. The pseudo-harmonics $\tilde{a}_{lm}^T$ are then defined by the spherical harmonic transform of $\tilde{T}(\Omega)$. They are related to the underlying un-weighted full-sky coefficients by
\be
\tilde{a}_{l m}^{T}=\sum_{l' m'} W_{l l'}^{m m'} a_{l' m'}^{T},
\ee
where the harmonic window function is defined as
$$W_{l l'}^{m m'}=\int\ud\Omega  W^{T}(\Omega)  Y_{l'm'}(\Omega) Y_{l m}^{*}(\Omega) .
$$
This can also be expressed as~\cite{Hivon:2001jp}
\be
W_{l l'}^{mm'}=\sum_{l'' m''}w_{l'' m''}^{T}\left(\frac{(2l+1)(2l'+1)(2l''+1)}{4\pi}\right)^{1/2}(-1)^{m}\left(\begin{array}{ccc} l & l' & l'' \\ 0 & 0 & 0 \end{array}\right)\left(\begin{array}{ccc} l & l' & l'' \\ -m & m' & m''\end{array}\right),
\ee
with the spherical harmonic transform coefficient of the window function given by
$$
w_{l m}^{T}=\int W^{T}(\Omega)Y_{l m}^{*}(\Omega)d\Omega.
$$

Similarly, for the polarization field the cut-sky pseudo-harmonic modes can be expanded as (see for example~\cite{Lewis:2001hp})
\begin{eqnarray}
\tilde{a}_{l m}^{E}=\sum_{l' m'} (_{+}W_{ll'}^{m m'} a_{l' m'}^{E} + i _{-}W_{l l'}^{mm'} a_{l' m'}^{B}),
\\ \tilde{a}_{l m}^{B}=\sum_{l' m'} (_{+}W_{ll'}^{mm'} a_{l' m'}^{B} - i _{-}W_{ll'}^{mm'} a_{l' m'}^{E}).
\end{eqnarray}
Here
\begin{eqnarray}
_{+}W_{l l'}^{mm'}\equiv\half( _{2}W_{l l'}^{mm'} + _{-2}W_{l l'}^{mm'}),
\\ _{-}W_{l l'}^{mm'}\equiv\half( _{2}W_{l l'}^{mm'} - _{-2}W_{l l'}^{mm'}),
\end {eqnarray}
with the spin weighted harmonic window function for spin $s=\pm 2$ given by
\be
_{s}W_{l l'}^{mm'}=\int d\Omega W_{p}(\Omega) _{s}Y_{l' m'}(\Omega) _{s}Y_{l m}^{*}(\Omega),
\ee
where $_{s}Y_{l m}(\Omega)$ are the spin-weighted harmonic functions. For azimuthal cuts the coupling matrices are diagonal in $m$, so $W_{l l'}^{mm'}=\delta_{mm'}W_{ll'}^m$, and they can be calculated quickly using a set of recursion relations~\cite{Lewis:2001hp}.

The pseudo-$C_l$ power spectra are  defined by
\begin{align} \label{eq}\begin{split}
&\tilde{C}_{l}^{TT}\equiv\frac{1}{2l+1}\sum_{m}\tilde{a}_{l m}^{T}(\tilde{a}_{l m}^{T})^{*} \qquad \tilde{C}_{l}^{TE}\equiv\frac{1}{2l+1}\sum_{m}\tilde{a}_{l m}^{T}(\tilde{a}_{l m}^{E})^{*} \\
 &\tilde{C}_{l}^{EE}\equiv\frac{1}{2l+1}\sum_{m}\tilde{a}_{l m}^{E}(\tilde{a}_{l m}^{E})^{*} \qquad \tilde{C}_{l}^{BB}\equiv\frac{1}{2l+1}\sum_{m}\tilde{a}_{l m}^{B}(\tilde{a}_{l m}^{B})^{*}.
\end{split}\end{align}
Their expectation values are related to the full-sky power spectra via the relation
\be
\left( \begin{array}{c} \la\tilde{C}_{l}^{TT}\ra \\ \la\tilde{C}_{l}^{TE}\ra \\ \la\tilde{C}_{l}^{EE}\ra \\ \la\tilde{C}_{l}^{BB}\ra \end{array} \right) = \sum_{l'} \left( \begin{array}{cccc} \mathcal{M}_{l l'}^{TT} & 0 & 0 & 0 \\ 0 & \mathcal{M}_{l l'}^{TE} & 0 & 0 \\ 0 & 0 & \mathcal{M}_{l l'}^{EE} & \mathcal{M}_{l l'}^{EB} \\ 0 & 0 & \mathcal{M}_{l l'}^{BE} & \mathcal{M}_{l l'}^{BB} \end{array} \right) \left( \begin{array}{c} C_{l'}^{TT} \\ C_{l'}^{TE} \\ C_{l'}^{EE} \\ C_{l'}^{BB} \end{array} \right),
\label{expectation}
\ee
where the coupling matrices are~\cite{Kogut:2003et}
\begin{eqnarray}
\mathcal{M}_{l l'}^{TT} &=& \frac{1}{2l+1}\sum_{mm'}|W_{l l'}^{mm'}|^{2} =
(2l'+1)\Xi_{TT}(l,l',\mathcal{W}^{TT})\\
\mathcal{M}_{l l'}^{TE} &=& \frac{1}{(2l +1)} \sum_{mm'}|W_{l l'}^{(mm')}(_{+}W_{l l'}^{(mm')})| =
(2l'+1)\Xi_{TE}(l,l',\mathcal{W}^{PT})\\
\mathcal{M}_{l l'}^{EE} &=& \mathcal{M}_{l l'}^{BB} = \frac{1}{(2l +1)}\sum_{mm'}|(_{+}W_{l l'}^{(mm')})|^{2} =
(2l'+1)\Xi_{EE}(l,l',\mathcal{W}^{PP})\\
\mathcal{M}_{l l'}^{EB} &=& \mathcal{M}_{l l'}^{BE} = \frac{1}{(2l +1)}\sum_{mm'}|(_{-}W_{l l'}^{(mm')})|^{2} =
(2l'+1)\Xi_{EB}(l,l',\mathcal{W}^{PP}).
\end{eqnarray}
The window function enters via its power spectrum $\mathcal{W}_{l}^{XY}$ given by
\be
 \mathcal{W}_{l}^{XY} = \frac{1}{2l +1}\sum_{m} \omega_{l m}^{X} \omega_{l m}^{Y}{}^*,
\ee
and $X$ and $Y$ being either $T$ or $P$. For isotropic noise tests we only consider $\omega_{l m}^{X} = \omega_{l m}^{Y}$.
The symmetric $\Xi$-matrices are defined by
\begin{equation}
  \Xi_{TT}(l_1, l_2, \tilde W)  \equiv
\sum_{l_3 }  \frac{(2 l_3 + 1)}{4 \pi}\tilde W _{l_3}
{\left ( \begin{array}{ccc}
        l_1 & l_2 & l_3  \\
        0  & 0 & 0
       \end{array} \right )^2},
       \nonumber
\end{equation}

\begin{equation}
  \Xi_{TE}(l_1, l_2, \tilde W)  \equiv
\sum_{l_3 }  \frac{(2 l_3 + 1)}{8 \pi}\tilde W _{l_3} (1 + (-1)^L)
{\left ( \begin{array}{ccc}
        l_1 & l_2 & l_3  \\
        0  & 0 & 0
       \end{array} \right )} {\left ( \begin{array}{ccc}
        l_1 & l_2 & l_3  \\
        -2  & 2 & 0
       \end{array} \right )},
       \nonumber
\end{equation}
\begin{equation}
  \Xi_{EE}(l_1, l_2, \tilde W)  \equiv
\sum_{l_3 }  \frac{(2 l_3 + 1)}{16 \pi}\tilde W _{l_3} (1 + (-1)^L)^2
       {\left ( \begin{array}{ccc}
        l_1 & l_2 & l_3  \\
        -2  & 2 & 0
       \end{array} \right )^2},
       \nonumber
\end{equation}
\begin{equation}
  \Xi_{EB}(l_1, l_2, \tilde W)  \equiv
\sum_{l_3 }  \frac{(2 l_3 + 1)}{16 \pi}\tilde W _{l_3} (1 - (-1)^L)^2
       {\left ( \begin{array}{ccc}
        l_1 & l_2 & l_3  \\
        -2  & 2 & 0
       \end{array} \right )^2},
\label{Xi}
\end{equation}
for $L=l_1+l_2+l_3$.
All other coupling matrices are zero.

Provided that the sky cut is small (the usable region is larger than half the sky), the coupling matrix in Eq~\eqref{expectation} is invertible and pseudo-$C_l$ estimators for the power spectrum are given by (see for example~\cite{Chon:2003gx,Efstathiou:2003dj})
\be
\left( \begin{array}{c}  \Ch^{TT} \\ \Ch^{TE} \\ \Ch^{EE} \\ \Ch^{BB} \end{array} \right) = \left( \begin{array}{cccc} \mathcal{M}^{TT} & 0 & 0 & 0 \\ 0 & \mathcal{M}^{TE} & 0 & 0 \\ 0 & 0 & \mathcal{M}^{EE} & \mathcal{M}^{EB} \\ 0 & 0 & \mathcal{M}^{BE} & \mathcal{M}^{BB} \end{array} \right)^{-1} \left(\begin{array}{c} \tilde{C}^{TT} \\ \tilde{C}^{TE} \\ \tilde{C}^{EE} \\ \tilde{C}^{BB} \end{array} \right).
\ee
The estimators are unbiased, $\la \Ch_l\ra = C_l$.
When the observed area is small the matrix is not invertible. In this case the $C_l$ can be binned into bands to construct band-power estimates of the power spectrum~\cite{Hivon:2001jp} in an analogous manner. Here we shall focus on nearly full-sky observations such as expected from the Planck satellite where estimates can be obtained for each $C_l$ individually.

Unlike in the full-sky case, the exact cut-sky likelihood function cannot be written purely in terms of a set of pseudo-$C_l$ estimators, so the compression of the observed data to the estimators is not lossless. However it can be a good approximation, and the estimators are convenient because the correlations between the $\Ch_l$ induced by the sky cut are accounted for easily.

\subsection{Covariance matrix}

The covariance matrix of the $\tilde{C}_{l}^{TT}$ is given by
\begin{align}\begin{split}
\la\Delta \tilde{C}_{l}^{TT} \Delta \tilde{C}_{l'}^{TT}\ra=\frac{2}{(2l+1)(2l'+1)}\sum_{mm'}\sum_{l_{1} m_{1}}\sum_{l_{2} m_{2}} C_{l_{1}}^{TT}C_{l_{2}}^{TT}W_{l l_{1}}^{mm_{1}}(W_{l' l_{1}}^{m'm_{1}})^{*}W_{l' l_{2}}^{m'm_{2}}(W_{l l_{2}}^{mm_{2}})^{*}.
\end{split}\end{align}

As suggested by Ref.~\cite{Efstathiou:2003dj}, this expression of the $\tilde{C}_{l}^{TT}$ covariance matrix may be simplified for the case of a narrow Galactic cut.  In this case, $C_{l_{1}}^{TT}$ and $C_{l_{2}}^{TT}$ can be replaced with $C_{l}^{TT}$ and $C_{l'}^{TT}$, respectively and then by applying the completeness relation for spherical harmonics ~\cite{AngularMom}, the temperature $\tilde{C}_{l}$'s covariance matrix would be given by
\be
\la\Delta \tilde{C}_{l}^{TT} \Delta \tilde{C}_{l'}^{TT}\ra = 2C_{l}^{TT}C_{l'}^{TT}\Xi_{TT}(l, l',\mathcal{W}^{TT}),
\ee

The covariance matrix of the $\Ch_{l}$-estimators is then given by
\begin{align}\begin{split}
\la\Delta \Ch_{l}^{TT} \Delta \Ch_{l'}^{TT}\ra=\sum_{l_{1}l_{2}}\mathcal{M}_{l l_{1}}^{-1}\mathcal{M}_{l' l_{2}}^{-1}\la \tilde{C}_{l_{1}}^{TT}\tilde{C}_{l_{2}}^{TT}\ra.
\end{split}\end{align}
Unfortunately, the other covariances do not simplify as easily since the completeness relation works only for the spherical harmonics with similar spin.  For our azimuthal tests we use $W^{T}(\Omega)$ that takes values 1 or 0 and approximate the pseudo-covariances by the following
\begin{eqnarray}
\la\Delta \tilde{C}_{l}^{TT} \Delta \tilde{C}_{l'}^{TT}\ra &\approx &2\frac{C_{l}^{TT}C_{l'}^{TT}}{(2 l' +1)}\mathcal{M}_{l l'}^{TT}, \label{TTTT}\\
\la\Delta \tilde{C}_{l}^{TE} \Delta \tilde{C}_{l'}^{TE}\ra &\approx &\frac{\sqrt{C_{l}^{T}C_{l'}^{T}C_{l}^{E}C_{l'}^{E}}}{(2 l' + 1)}\mathcal{M}_{l l'}^{TE} + \frac{C_{l}^{TE}C_{l'}^{TE}}{(2 l' +1)}\mathcal{M}_{l l'}^{TT}, \label{TETE} \\
\la\Delta \tilde{C}_{l}^{EE} \Delta \tilde{C}_{l'}^{EE}\ra &\approx &2\frac{C_{l}^{EE}C_{l'}^{EE}}{(2 l' +1)}\mathcal{M}_{l l'}^{EE} + 2 \frac{C_{l}^{BB}C_{l'}^{BB}}{(2 l' +1)}\mathcal{M}_{l l'}^{EB}, \label{EEEE} \\
\la\Delta \tilde{C}_{l}^{BB} \Delta \tilde{C}_{l'}^{BB}\ra &\approx &2\frac{C_{l}^{BB}C_{l'}^{BB}}{(2 l' +1)}\mathcal{M}_{l l'}^{BB} + 2 \frac{C_{l}^{EE}C_{l'}^{EE}}{(2 l' +1)}\mathcal{M}_{l l'}^{EB}, \label{BBBB} \\
\la\Delta \tilde{C}_{l}^{TT} \Delta \tilde{C}_{l'}^{EE}\ra &\approx &2\frac{C_{l}^{TE}C_{l'}^{TE}}{(2 l' +1)}\mathcal{M}_{l l'}^{TT}, \label{TTEE} \\
\la\Delta \tilde{C}_{l}^{TT} \Delta \tilde{C}_{l'}^{TE}\ra &\approx &  \frac{\sqrt{C_{l}^{TT}C_{l'}^{TT}} (C_{l}^{TE}+C_{l'}^{TE}) \mathcal{M}_{l l'}^{TT}}{(2 l' + 1)},\label{TTTE} \\
\la\Delta \tilde{C}_{l}^{EE} \Delta \tilde{C}_{l'}^{TE}\ra &\approx &  \frac{\sqrt{C_{l}^{EE}C_{l'}^{EE}}(C_{l}^{TE}+C_{l'}^{TE}) \mathcal{M}_{l l'}^{TE}}{(2 l' + 1)},\label{EETE} \\
\la\Delta \tilde{C}_{l}^{EE} \Delta \tilde{C}_{l'}^{BB}\ra &\approx & \frac{\left(\sqrt{C_{l}^{EE}C_{l'}^{EE}}+\sqrt{C_{l}^{BB}C_{l'}^{BB}}\right)^{2}}{2(2 l' + 1)} \mathcal{M}_{l l'}^{EB}. \label{EEBB}
\end{eqnarray}
Note that in the presence of isotropic noise the $C_l$ here include the noise contribution.

At high $l$ one can approximate $\mathcal{M}_{l l'}^{TE}=\mathcal{M}_{l l'}^{EE} = \mathcal{M}_{l l'}^{BB} = \mathcal{M}_{l l'}^{TT}$, since the spin $\pm 2$ harmonics become close to the spin zero ones. Note our approximations in Eqs.~\eqref{TTTE},~\eqref{EETE} differ from those in Ref.~\cite{Brown:2004jn}: since the $C_l^{TE}$ can be negative we require consistency with the exact result on the full-sky rather than forcing these terms to be positive.  Also, note the difference in Eqs.~\eqref{EEEE},~\eqref{BBBB} from those in Ref.~\cite{Brown:2004jn}. More general results applicable with anisotropic noise and general weight function are given in Appendix~\ref{Planck}. More accurate results accounting for the complications of $E$-$B$ mixing are given in Ref.~\cite{Challinor:2004pr}; see also Ref.~\cite{Smith:2006vq}. Note that inaccuracies in the covariance matrix generally only affect the error bars; to this extent accuracy is less crucial than getting the estimators or likelihood function accurate, since there an inaccuracy could introduce biases.

The covariance of the $\hat{C}_{l}$-estimators can be calculated from the $\tilde{C}_l$ covariance using the relevant coupling matrices.

\subsection{Exact likelihood for temperature and polarization}

Although an exact likelihood calculation is prohibitively slow in general, for azimuthal sky cuts the relevant matrices are block-diagonal in $m$ and the calculation is numerically tractable. For the special case of azimuthal cuts we can therefore test cut-sky likelihood approximations against the exact result.

For each $m$ we can define a vector of pseudo-harmonic coefficients
\be
\tilde{\vX} \equiv \left( \begin{array}{c} \tilde{a}_{l m}^{T} \\ \tilde{a}_{l}^{E} + i \tilde{a}_{l m}^{B} \\ \tilde{a}_{l m}^{E} - i\tilde{a}_{l m}^{B} \end{array} \right) = \left( \begin{array}{ccc} W_{l l'}^{(m)} & 0 & 0 \\ 0 & {}_{2}W_{l l'}^{(m)} & 0 \\ 0 & 0 & {}_{-2}W_{l l'}^{(m)} \end{array} \right) \left( \begin{array}{c} a_{l' m}^{T} \\ a_{l' m}^{E}+i a_{l' m}^{B} \\ a_{l' m}^{E}-i a_{l' m}^{B} \end{array} \right),
\ee
which can simply be written as
\be
\tilde{\vX} = \text{diag} \biggl( W_{l l'}^{(m)}, {}_{2}W_{l l'}^{(m)}, {}_{-2}W_{l l'}^{(m)} \biggr) \vX.
\ee
For Gaussian fields $\tilde{\vX}$ is just a  linear combination of Gaussian harmonics, and hence also Gaussian. However due to the sky cut the coupling matrix is not directly invertible, as the $W$-matrices will have eigenvalues very close to zero (corresponding to modes localized in the un-observed region). However we can use a singular value decomposition (SVD) to isolate the observable independent modes following Ref.~\cite{Mortlock00,Lewis:2001hp}. We diagonalize the transformation matrix as $\text{diag} \biggl( W_{l l'}^{(m)}, {}_{2}W_{l l'}^{(m)},{} _{-2}W_{l l'}^{(m)} \biggr) = \mU\mD\mU^{\dagger}$ and define new linear combinations:
\be
\vX' = \hat{\mD}^{-1/2}\hat{\mU}^{\dagger} \tilde{\vX} = \hat{\mD}^{1/2}\hat{\mU}^{\dagger}\vX.
\ee
Here $\hat{\mD}$ denotes the smaller square matrix obtained from $\mD$ by deleting nearly-zero rows and columns. $\hat{U}$ is the corresponding rectangular matrix obtained from $\mU$ by deleting the corresponding columns.

The signal correlation is:
\begin{align}  \begin{split}
\mS = \la\vX' \vX'^{\dagger}\ra &= \hat{D}^{1/2} \hat{U}^{\dagger} \la\vX \vX^{\dagger}\ra \hat{U} \hat{D}^{1/2} \\
&= \hat{D}^{1/2} \hat{U}^{\dagger} \left( \begin{array}{ccc} C_{l}^{TT} & C_{l}^{TE} & C_{l}^{TE} \\ C_{l}^{TE} & C_{l}^{EE}+C_{l}^{BB} & C_{l}^{EE}-C_{l}^{BB} \\ C_{l}^{TE} & C_{l}^{EE}-C_{l}^{BB} & C_{l}^{EE}+C_{l}^{BB} \end{array} \right) \hat{U} \hat{D}^{1/2}.
\end{split} \end{align}

If the noise is isotropic and uncorrelated, this frame structure provides a diagonal noise correlation~\cite{Lewis:2001hp}:

\be
\la\tilde{\vX}_{N} \tilde{\vX}_{N}^{\dagger}\ra = \sigma_{N}^{2} \text{diag}(W_{l l'}^{(m)},2 _{+}W_{l l'}^{(m)},2 _{-}W_{l l'}^{(m)}) \qquad \Rightarrow \qquad \mN = \la\vX_{N}' \vX_{N}'^{\dagger}\ra = \sigma_{N}^{2} \text{diag}(1,2,2),
\ee
where we have considered $\sigma_{N}^{T}{}^{2} = \sigma_{N}^{2}$ and $ \sigma_{N}^{E}{}^{2} = \sigma_{N}^{B}{}^{2} = 2\sigma_{N}^{2}$ for simulation purposes.

Given that the signal and noise are Gaussian, the likelihood function is then given by
\be
 \cll(\{C_{l}^{T},C_{l}^{E},C_{l}^{TE},C_{l}^{B}\}|\vX') \propto
 \frac{\exp[-\half \vX'^{\dagger} (\mS+\mN)^{-1} \vX']}{|\mS+\mN|^{1/2}}.
\label{newlike}
\ee
The only approximation is in the choice of cutoff value for the SVD; for non-zero noise the result is insensitive to this choice as long as it is small.

\section{Anisotropic noise: estimators and test simulation}
\label{Planck}

\subsection{Hybrid Pseudo-$C_l$ estimators with cross-weights}

We consider pixelized maps with anisotropic but uncorrelated pixel noise variance $\sigma_s^2$ (in this section the $C_l$ do not include noise). We generalize the hybrid Pseudo-$C_l$ method of Ref.~\cite{Efstathiou:2006eb} slightly to include Pseudo-$C_l$ estimators from mixed weights, e.g. using a set of Pseudo-$C_l$s
\begin{equation}
\tilde C^{XY,ij}_{l} \equiv \frac{1}{2l+1} \sum_m \tilde{a}^{X,i}_{lm} \tilde{a}^{Y,j}_{lm}{}^*,
\end{equation}
where $\tilde{a}^{X,i}_{lm}$ is defined using weight function $w^i$. For each $X$ and $Y$ there are therefore $n(n+1)/2$ distinct estimators if $X=Y$, or $n^2$ if $X\ne Y$, where $n$ is the number of weight functions. For high signal to noise the best weight function should be close to uniform to minimize cosmic variance, for low signal to noise it should be proportional to the inverse-noise to minimize the noise~\cite{Efstathiou:2003dj}. Combining results from two weight functions, one with uniform and one with inverse-noise weighting, is therefore perhaps the most natural choice, especially if the polarization noise is proportional to the temperature noise in each pixel as we assume for our test simulations. Including the cross-estimator between maps with different weight functions is particularly useful for estimating $C_l^{TE}$: since the polarization noise is much larger than the temperature, over a wide range of scales the cross-estimator between uniform and inverse-noise weighted maps is much better than using uniform/uniform or inverse-noise/inverse-noise. Even for the temperature case there is a range of scales in between noise and signal domination where the cross-estimator can be useful. Including more than two weighting functions seems to gain very little, so we use just two.

The unbiased $\hat{\mC}_l$ estimators are constructed using the coupling matrix
\begin{equation}
\hat{C}_l^{XY,ij} = [\mathcal{M}^{XY,ij}]^{-1}_{ll'} \tilde C_{l'}^{XY,ij},
\end{equation}
where
\begin{equation}
\mathcal{M}^{XY,ij}_{ll'} = (2l'+1)\Xi_{XY}(l,l',\tilde W^{ij}), \qquad \tilde W^{ij} \equiv \frac{1}{2l+1}\sum_m w^i_{lm} w^j_{lm}{}^*,
\end{equation}
and the coupling matrices are defined in Eqs.~\eqref{Xi}.

The noise contribution to the Pseudo-$C_l$ is given, for uncorrelated pixel noise $(\sigma^T_s)^2$, $(\sigma_s^Q)^2$, $(\sigma_s^U)^2$ and pixel area $\Omega_s$, by
\begin{eqnarray}
\tilde{N}_l^{TT,ij} &=& \frac{1}{4\pi} \sum_s (\sigma_s^T)^2 w^i(s) w^j(s) \Omega_s^2
\\
\tilde{N}_l^{EE,ij} &=& \tilde{N}_l^{BB,ij} = \frac{1}{8\pi} \sum_s \left[(\sigma_s^Q)^2+(\sigma_s^U)^2\right] w^i(s) w^j(s) \Omega_s^2,
\end{eqnarray}
with other combinations being zero. We then have $\la \hat{C}_l^{XY,ij}\ra = C_l^{XY} + \sum_{l'}[\mM^{XY,ij}]^{-1}_{l l'}  \tilde{N}_{l'}^{XY,ij}$.

From multiple Pseudo-$C_l$ estimators with different weight functions one can either attempt to apply the likelihood approximations directly to the complete set of estimators, or one can compress into a single hybrid estimator. At low $l$ it is likely to be beneficial to also include more optimal estimators than Pseudo-$C_l$, especially for the polarization~\cite{Efstathiou:2006eb}.

A hybrid pseudo-$C_l$ estimator can be constructed following Ref.~\cite{Efstathiou:2006eb}: this is defined by constructing the best-fit $C_l$ to the multiple estimators by minimizing the Gaussian-approximation to the likelihood using the approximate full covariance. We do this separately for each temperature-polarization spectrum, so that the hybrid estimator is just a linear combination of the individual estimators rather than mixing estimators of different type. Since the polarization noise is higher than for the temperature, we consider cross-spectra of the form $C_l^{TE,ij}$ where $i\ge j$, and the weight functions are ordered so that lower $i$ are more optimal in the case of lower noise. We then have the same number of cross-weight spectra for each of the power spectra. Since the hybrid estimators are just linear combinations of the separate estimators, their covariance can easily be calculated from the coupling matrices and full covariance matrix approximations given below. When including $C_l^{BB}$ we impose a uniform weight function at $l<120$ to minimize E/B mixing effects and ensure that the covariance matrix approximations below remain accurate. This is suboptimal but unbiased; we do not investigate the more difficult problem of optimally constraining the tensor amplitude here.

\subsection{Covariance matrix approximations}

Approximations for some components of the covariance matrices for the Pseudo-$C_l$s were given in Ref.~\cite{Efstathiou:2006eb} for a general pixel-weighting function $w(s)$ (pixels area $\Omega_s$) and anisotropic but uncorrelated instrumental pixel noise $(\sigma^T_s)^2$ and $(\sigma^Q_s)^2 = (\sigma^U_s)^2$. The approximations essentially make as many assumptions as necessary for the result to simplify to the forms given; the approximations should be reasonably accurate for small cuts at high $l$ (where ${}_sY_{lm} \sim Y_{lm}$) and noise-dominated $B$-polarization spectra. Here we summarize these results with slight generalization, and extend to include all the terms needed for the full polarized and correlated estimator covariance. We only consider the case of using Pseudo-$C_l$ estimators from single maps of $T$, $Q$ and $U$ with various weighting; the noise properties of cross-spectra between multiple maps with independent noise are a simple generalization.

Assuming the polarization and temperature noise is uncorrelated, the covariance of the Pseudo-$C_l$ estimators can be estimated using the approximations (for $l\gg 1$ and significant noise so that E-B mixing effects are small and large $\fsky$):

\label{gencovmat}
\begin{multline}
 \langle \Delta \tilde C^{TT,ij}_l \Delta \tilde C^{TT,pq}_{l^\prime}
\rangle   \approx
 C^{TT}_l C^{TT}_{l^\prime} \left[\Xi_{TT}(l, l^\prime, \tilde W^{ (ip)(jq)}) + \Xi_{TT}(l, l^\prime, \tilde W^{(iq)(jp)}) \right] \\
+ (C^T_l C^T_{l^\prime})^{1/2}
\left[ \Xi_{TT}(l, l^\prime, \tilde W^{2T(ip)(jq)})
+ \Xi_{TT}(l, l^\prime, \tilde W^{2T(iq)(jp)})
+ \Xi_{TT}(l, l^\prime, \tilde W^{2T(jq)(ip)})
+ \Xi_{TT}(l, l^\prime, \tilde W^{2T(jp)(iq)})
\right]  \\
+ \Xi_{TT}(l, l^\prime, \tilde W^{TT(ip)(jq)}) + \Xi_{TT}(l, l^\prime, \tilde W^{TT(iq)(jp)}),
\end{multline}
\begin{multline}
\langle \Delta \tilde C^{TE,ij}_l \Delta \tilde C^{TE,pq}_{l^\prime}
\rangle  \approx
 (C^{TT}_l C^{TT}_{l^\prime} C^{EE}_l C^{EE}_{l^\prime})^{1/2}
\Xi_{TE}(l, l^\prime, \tilde W^{(ip)(jq)}) +
 C^{TE}_l C^{TE}_{l^\prime} \Xi_{TT}(l, l^\prime, \tilde W^{(iq)(jp)})   \\+
   \Xi_{TE}(l, l^\prime, \tilde W^{TQ(ip)(jq)}) +
(C^{TT}_l C^{TT}_{l^\prime})^{1/2} \Xi_{TE}(l, l^\prime, \tilde W^{2Q(ip)(jq)}) + (C^{EE}_l
C^{EE}_{l^\prime})^{1/2} \Xi_{TE}(l, l^\prime, \tilde W^{2T(jq)(ip)}),
\end{multline}
\begin{multline}
\langle \Delta \tilde C^{EE,ij}_l \Delta \tilde C^{EE,pq}_{l^\prime} \rangle  \approx
C^{EE}_l C^{EE}_{l^\prime}
 \left[\Xi_{EE}(l, l^\prime, \tilde W^{(ip)(jq)}) + \Xi_{EE}(l, l^\prime, \tilde W^{(iq)(jp)})\right] \\
+(C^{EE}_l C^{EE}_{l^\prime})^{1/2} \left[
\Xi_{EE}(l, l^\prime, \tilde W^{2Q(ip)(jq)}) +
\Xi_{EE}(l, l^\prime, \tilde W^{2Q(iq)(jp)}) +
\Xi_{EE}(l, l^\prime, \tilde W^{2Q(jp)(iq)}) +
\Xi_{EE}(l, l^\prime, \tilde W^{2Q(jq)(ip)}) \right] \\
 +  \Xi_{EE}(l, l^\prime, \tilde W^{QQ(ip)(jq)}) +  \Xi_{EE}(l, l^\prime, \tilde W^{QQ(iq)(jp)}),
\end{multline}
\begin{multline}
\langle \Delta \tilde C^{BB,ij}_l \Delta \tilde C^{BB,pq}_{l^\prime}\rangle  \approx   C^{BB}_l C^{BB}_{l^\prime}
\left[ \Xi_{EE}(l, l^\prime, \tilde W^{(ip)(jq)}) + \Xi_{EE}(l, l^\prime, \tilde W^{(iq)(jp)})\right]\\
   +(C^{BB}_lC^{BB}_{l^\prime})^{1/2}  \left[
\Xi_{EE}(l, l^\prime, \tilde W^{2Q(ip)(jq)}) +
\Xi_{EE}(l, l^\prime, \tilde W^{2Q(iq)(jp)}) +
\Xi_{EE}(l, l^\prime, \tilde W^{2Q(jp)(iq)}) +
\Xi_{EE}(l, l^\prime, \tilde W^{2Q(jq)(ip)}) \right] \\
 + \Xi_{EE}(l, l^\prime, \tilde W^{QQ(ip)(jq)}) +  \Xi_{EE}(l, l^\prime, \tilde W^{QQ(iq)(jp)}),
\end{multline}
\begin{multline}
\langle \Delta \tilde C^{EE,ij}_l \Delta \tilde C^{BB,pq}_{l^\prime}
\rangle  \approx
\left[(C^{EE}_l C^{EE}_{l^\prime})^{1/2} + (C^{BB}_l C^{BB}_{l^\prime})^{1/2} \right]^2
\half \left[ \Xi_{EB}(l, l^\prime, \tilde W^{(ip)(jq)}) +  \Xi_{EB}(l, l^\prime, \tilde W^{(iq)(jp)})\right]
\\ \qquad +(C^{EE}_lC^{EE}_{l^\prime})^{1/2}  \left[
\Xi_{EB}(l, l^\prime, \tilde W^{2Q(ip)(jq)}) +
\Xi_{EB}(l, l^\prime, \tilde W^{2Q(iq)(jp)}) +
\Xi_{EB}(l, l^\prime, \tilde W^{2Q(jp)(iq)}) +
\Xi_{EB}(l, l^\prime, \tilde W^{2Q(jq)(ip)}) \right] \\
 +\Xi_{EB}(l, l^\prime, \tilde W^{QQ(ip)(jq)}) +  \Xi_{EB}(l, l^\prime, \tilde W^{QQ(iq)(jp)}),
\end{multline}
\begin{multline}
 \langle \Delta \tilde C^{TT,ij}_l \Delta \tilde C^{TE,pq}_{l^\prime}
\rangle   \approx
\half (C^{TT}_l C^{TT}_{l'})^{1/2} (C^{TE}_l + C^{TE}_{l'})
\left[ \Xi_{TT}(l,l',\tilde W^{(ip)(jq)}) + \Xi_{TT}(l,l',\tilde W^{(iq)(jp)})\right]\\ +
 \half(C^{TE}_l + C^{TE}_{l^\prime}) \left[ \Xi_{TT}(l, l^\prime, \tilde W^{2T(ip)(jq)}) + \Xi_{TT}(l, l^\prime, \tilde W^{2T(jp)(iq)})\right],
\end{multline}
\begin{multline}
 \langle \Delta \tilde C^{EE,ij}_l \Delta \tilde C^{TE,pq}_{l^\prime}
\rangle   \approx
 \half (C^{EE}_l C^{EE}_{l'})^{1/2} (C^{TE}_l + C^{TE}_{l'}) \left[\Xi_{EE}(l,l',\tilde W^{(ip)(jq)}) + \Xi_{EE}(l,l',\tilde W^{(iq)(jp)})\right]\\
 +  \half (C^{TE}_l + C^{TE}_{l^\prime}) \left[ \Xi_{EE}(l, l^\prime, \tilde W^{2Q(ip)(jq)}) + \Xi_{EE}(l, l^\prime, \tilde W^{2Q(jp)(iq)})\right],
\end{multline}
\begin{equation}
 \langle \Delta \tilde C^{TT,ij}_l \Delta \tilde C^{EE,pq}_{l^\prime}
\rangle   \approx C^{TE}_l C^{TE}_{l'} \left[  \Xi_{TT}(l,l',\tilde W^{(ip)(jq)}) +  \Xi_{TT}(l,l',\tilde W^{(ip)(jq)})\right],
\end{equation}
where the various window functions appearing are determined by the power spectra
\begin{equation}
\tilde W^{(ij)(pq)}_l = \frac{1}{2l+1} \sum_m w^{ij}_{lm} w^{pq}_{lm}{}^*
\end{equation}
\begin{equation}
 \tilde W^{TT(ij)(pq)}_l =\frac{1}{2l+1} \sum_m (w^{T,ij}_{l m}  w^{T,pq}_{l m}{}^*),
 \qquad
 \tilde W^{TQ(ij)(pq)}_l \equiv \tilde W^{TU(ij)(pq)}_l =  \frac{1}{2l+1} \sum_m (w^{T,ij}_{l m}  w^{Q,pq}_{l m}{}^*),  \qquad
\end{equation}
\begin{equation}
 \tilde W^{2T(ij)(pq)}_l = \frac{1}{2l+1}\sum_m ( w^{ij}_{l m}  w^{T,pq}_{l m}{}^*), \qquad
 \tilde W^{2Q(ij)(pq)}_l \equiv \tilde W^{2U(ij)(pq)}_l = \frac{1}{2l+1} \sum_m (w^{(ij)}_{l m}  w^{Q,pq}_{l m}{}^*),
\end{equation}
\begin{equation}
 \tilde W^{QQ(ij)(pq)}_l \equiv \tilde W^{QU(ij)(pq)}_l \equiv
\tilde W^{UU(ij)(pq)}_l = \frac{1}{2l+1} \sum_m w^{Q,ij}_{lm}w^{Q,pq}_{lm}{}^* ,
\end{equation}
and the harmonic coefficients are given as sums over pixels with area $\Omega_s$ as
\begin{equation}
 w^{ij}_{l m} = \sum_s  w^i(s) w^j(s)  \Omega_s Y_{l m}(s)^* , \quad
 w^{T,ij}_{l m} = \sum_s (\sigma^T_s)^2 w^i(s)w^j(s) \Omega_s^2  Y_{l m}(s)
\end{equation}
 \begin{equation}
  w^{Q,ij}_{l m}
\equiv w^{U,ij}_{l m} = \sum_s (\sigma^Q_s)^2 w^i(s) w^j(s) \Omega_s^2  Y_{l m}(s).
\end{equation}

At the level of approximation considered here $\Xi_{EE} \sim \Xi_{TT} \sim \Xi_{TE}$, so there is some ambiguity in which particular form to use in the approximations. Note that the contribution of $E$ to the $\tilde C_l^B$ covariance is neglected, which is a poor approximation when the noise is not dominant; more accurate approximations are given in Ref.~\cite{Challinor:2004pr}. If the $B$-polarization contribution to the variance becomes important relative to the noise, the non-Gaussianity of the lensed $B$-polarization field also becomes an issue (see e.g. Ref.~\cite{Smith:2005ue}). For Planck noise levels the $B$-lensing signal is well below the noise and $E/B$ mixing effects are also well below the noise at $l\agt 100$.

The covariance matrix for the $\hat{C}_l$ estimators is determined straightforwardly by applying the inverse coupling matrix to the above results. The covariance of the hybrid estimator is then just a contraction of the full multi-estimator covariance with the hybrid mixing matrix.

\subsection{Test simulations}

The diagonal of the covariance matrix approximations given above agree very well with simulations at $l\agt 30$ if the weight map does not have too much small scale power. The covariance approximations are more sensitive to small scale power in the noise and weights than the coupling matrices; for this reason we use a smoother mask and noise map than is needed to obtain an accurate coupling matrix. This avoid numerical issues in our tests so that we can focus on any errors due to the likelihood approximations. We use a \healpix\footnote{\url{http://www.eso.org/science/healpix/}}~\cite{Gorski:2004by} pixelization at $\nside=2048$, upgrading the simulated Planck noise map~\cite{Ashdown:2007ta} and convolving it with $7\arcmin$ Gaussian kernel so that it is smooth on this scale. For the mask we take the WMAP kp2 map, upgrade to  $\nside=2048$ ($12\times 2048^2$ pixels), smooth with $7\arcmin$ kernel, set negative pixels to zero, and smooth again with a $7\arcmin$ kernel. This gives point source cuts that still go to essentially zero, while having edges smoothly tapering to one. To calculate the Pseudo-$C_l$ estimators we take $w^1$ as uniform weighting (multiplied by the cut), and a regularized inverse-noise weighting given by $w^2(s)\propto 1/(\sigma_s^2 + \min(\sigma_s^2))$, smoothed with a $7\arcmin$ kernel and then multiplied by the cut.
We use the same weight functions for temperature and polarization, and take $(\sigma^Q_s)^2=(\sigma^U_s)^2=4(\sigma^T_s)^2$ for simplicity. Gaussian simulations are done to $\lmax=2200$ with zero monopole and dipole. The simulation code is available on the web\footnote{\url{http://cosmologist.info/cosmomc/CMBLike.html}}.

\subsection{Lensed simulation}
\label{lensing}

\begin{figure}
 \epsfig{file=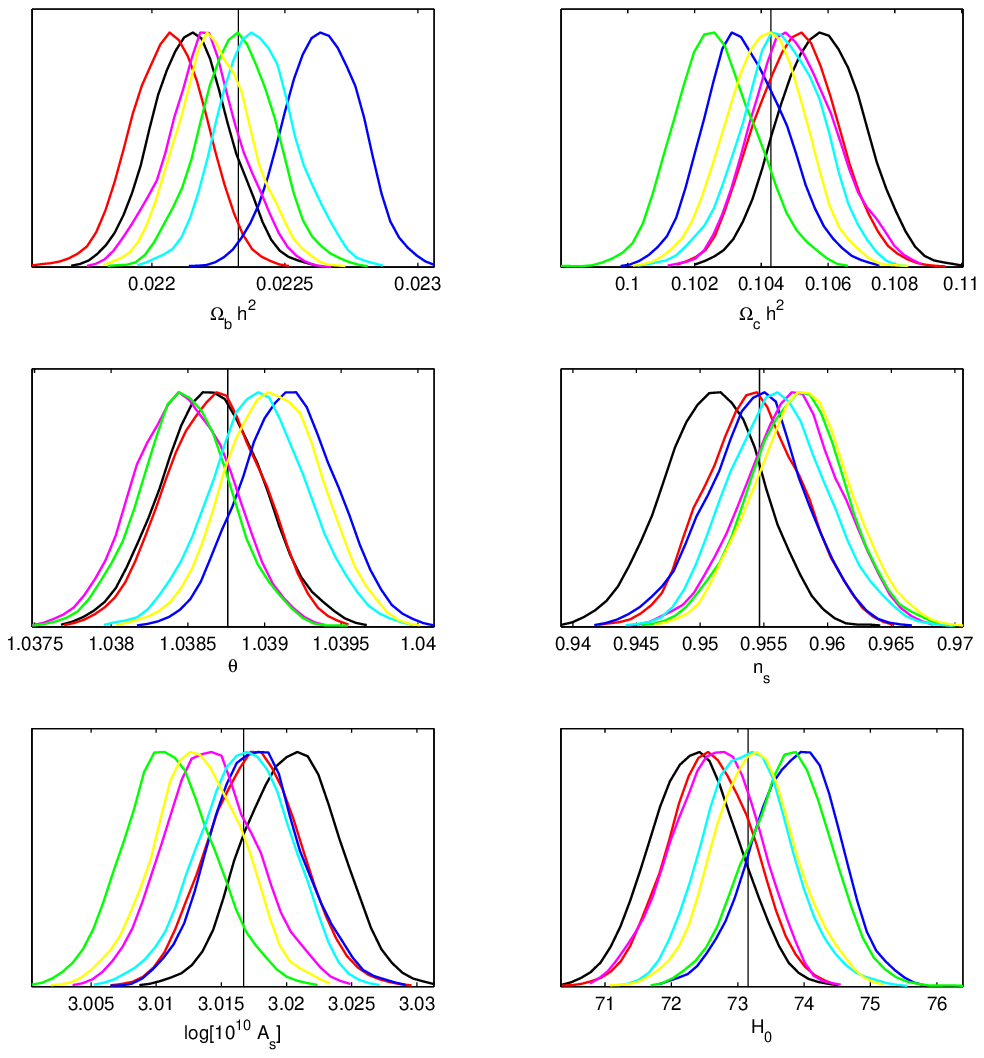, width=12cm}
 \caption{Simulated parameter constraints from eight lensed CMB realizations using the new likelihood approximation with $E$ and $T$ (and cross) hybrid pseudo-$C_l$ estimators at $30\le l \le 2000$. The covariance was calculated using only the lensed power spectra. Input parameter values are marked with vertical lines, and the reionization optical depth was fixed. Including $B$ estimators has virtually no effect at Planck noise levels.}
\label{fig:lensing} \end{figure}

The largest non-linear effect on intermediate scales is expected to be that of CMB lensing~\cite{Lewis:2006fu}. Detailed modelling of the non-Gaussian distribution induced by this effect is beyond the scope of this paper, however for Planck noise levels the non-Gaussianity can be neglected to good approximation when performing parameter analyses from the lensed CMB power spectra~\cite{Hu:2001fa}. The effect of lensing on the power spectrum is many percent, and must be included to obtain correct parameters with Planck. We update the LensPix code~\cite{Lewis:2005tp} to quickly simulate high-resolution lensed maps accurately. Our simulation method is as follows: 1. we simulate a \healpix\ map of a realization of the lensing deflection angle from a Gaussian realization of the lensing potential; 2. Divide the sphere into a number of slices separated by lines at constant polar angle $\theta$, and assign each slice to a different processor (with some overlap given by the largest $\theta$-deflection); 3. each processor simulates a Gaussian unlensed CMB map over its assigned slice on an equicylindrical grid; 4. interpolate from the equicylindrical grid to the deflected positions corresponding to the centre of \healpix\ pixels offset by the deflection angles. Equations used for simulating gradient maps, deflecting points along geodesics, and appropriately rotating Stokes parameters are given in Ref.~\cite{Lewis:2005tp}. Our updated code is publicly available\footnote{\url{http://cosmologist.info/lenspix}}.

For our simulation we use $\nside=2048$, and generate equicylindrical unlensed grids with points at 6144 different $\theta$ values ($\text{interp\_factor} = 1.5$, effectively the same resolution as \healpix\ at $\nside=2048$). The number of $\phi$-pixels is chosen for each slice to be of the form $2^n3^m$ (for integer $n,m$) so that FFTs can be performed quickly, with lowest spacing roughly the same as the spacing in $\theta$. To interpolate we use an extended cubic interpolation algorithm TOMS760~\cite{Akima96}; this is significantly slower than a basic bicubic interpolation scheme, but more accurate and stable --- it ensures our results converge as the number of equicylindrical pixels is increased. Averaged over simulations our simulated lensed CMB power spectra then agree at the $0.1\%$-level with theoretical expectations for the same $\lmax$~\cite{Challinor:2005jy,Lewis:2006fu}.  Other simulation method are discussed in Refs.~\cite{Das:2007eu,Carbone:2007yy,Smith:2007rg,Hirata:2004rp}, though non-linear evolution effects are minor at Planck noise levels. Since the unlensed CMB is not band limited but contains residual power at $l\ge 2000$ our method does not rely on band-limited interpolations and works directly with maps that contain power up to the highest simulated $\lmax$. On a modern few-node cluster lensed maps with polarization can be simulated in a few minutes.

Figure~\ref{fig:lensing} shows parameter estimation constraints generated using a set of simulated lensed maps with Planck-like noise, and modelling the covariance as in the unlensed case simply by using the lensed power spectra instead of the unlensed ones. A more optimal analysis would use the non-Gaussian information in the lensed field to indirectly constrain the lensing potential and hence cosmological parameters (see e.g. references in~\cite{Lewis:2006fu}), though it is unclear how much can be gained in the presence of real-world complications.


\begin{thebibliography}{44}
\expandafter\ifx\csname natexlab\endcsname\relax\def\natexlab#1{#1}\fi
\expandafter\ifx\csname bibnamefont\endcsname\relax
  \def\bibnamefont#1{#1}\fi
\expandafter\ifx\csname bibfnamefont\endcsname\relax
  \def\bibfnamefont#1{#1}\fi
\expandafter\ifx\csname citenamefont\endcsname\relax
  \def\citenamefont#1{#1}\fi
\expandafter\ifx\csname url\endcsname\relax
  \def\url#1{\texttt{#1}}\fi
\expandafter\ifx\csname urlprefix\endcsname\relax\def\urlprefix{URL }\fi
\providecommand{\bibinfo}[2]{#2}
\providecommand{\eprint}[2][]{\url{#2}}

\bibitem[{\citenamefont{Wandelt et~al.}(2004)\citenamefont{Wandelt, Larson, and
  Lakshminarayanan}}]{Wandelt:2003uk}
\bibinfo{author}{\bibfnamefont{B.~D.} \bibnamefont{Wandelt}},
  \bibinfo{author}{\bibfnamefont{D.~L.} \bibnamefont{Larson}},
  \bibnamefont{and}
  \bibinfo{author}{\bibfnamefont{A.}~\bibnamefont{Lakshminarayanan}},
  \bibinfo{journal}{Phys. Rev.} \textbf{\bibinfo{volume}{D70}},
  \bibinfo{pages}{083511} (\bibinfo{year}{2004}), \eprint{astro-ph/0310080}.

\bibitem[{\citenamefont{Larson et~al.}(2007)}]{Larson:2006ds}
\bibinfo{author}{\bibfnamefont{D.~L.} \bibnamefont{Larson}}
  \bibnamefont{et~al.}, \bibinfo{journal}{Astrophys. J.}
  \textbf{\bibinfo{volume}{656}}, \bibinfo{pages}{653} (\bibinfo{year}{2007}),
  \eprint{astro-ph/0608007}.

\bibitem[{\citenamefont{Chu et~al.}(2005)}]{Chu:2004zp}
\bibinfo{author}{\bibfnamefont{M.}~\bibnamefont{Chu}} \bibnamefont{et~al.},
  \bibinfo{journal}{Phys. Rev.} \textbf{\bibinfo{volume}{D71}},
  \bibinfo{pages}{103002} (\bibinfo{year}{2005}), \eprint{astro-ph/0411737}.

\bibitem[{\citenamefont{Bond et~al.}(2000)\citenamefont{Bond, Jaffe, and
  Knox}}]{Bond:1998qg}
\bibinfo{author}{\bibfnamefont{J.~R.} \bibnamefont{Bond}},
  \bibinfo{author}{\bibfnamefont{A.~H.} \bibnamefont{Jaffe}}, \bibnamefont{and}
  \bibinfo{author}{\bibfnamefont{L.~E.} \bibnamefont{Knox}},
  \bibinfo{journal}{Astrophys. J.} \textbf{\bibinfo{volume}{533}},
  \bibinfo{pages}{19} (\bibinfo{year}{2000}), \eprint{astro-ph/9808264}.

\bibitem[{\citenamefont{Verde et~al.}(2003)}]{Verde:2003ey}
\bibinfo{author}{\bibfnamefont{L.}~\bibnamefont{Verde}} \bibnamefont{et~al.},
  \bibinfo{journal}{Astrophys. J. Suppl.} \textbf{\bibinfo{volume}{148}},
  \bibinfo{pages}{195} (\bibinfo{year}{2003}), \eprint{astro-ph/0302218}.

\bibitem[{\citenamefont{Smith et~al.}(2006)\citenamefont{Smith, Challinor, and
  Rocha}}]{Smith:2005ue}
\bibinfo{author}{\bibfnamefont{S.}~\bibnamefont{Smith}},
  \bibinfo{author}{\bibfnamefont{A.}~\bibnamefont{Challinor}},
  \bibnamefont{and} \bibinfo{author}{\bibfnamefont{G.}~\bibnamefont{Rocha}},
  \bibinfo{journal}{Phys. Rev.} \textbf{\bibinfo{volume}{D73}},
  \bibinfo{pages}{023517} (\bibinfo{year}{2006}), \eprint{astro-ph/0511703}.

\bibitem[{\citenamefont{Percival and Brown}(2006)}]{Percival:2006ss}
\bibinfo{author}{\bibfnamefont{W.~J.} \bibnamefont{Percival}} \bibnamefont{and}
  \bibinfo{author}{\bibfnamefont{M.~L.} \bibnamefont{Brown}},
  \bibinfo{journal}{Mon. Not. Roy. Astron. Soc.}
  \textbf{\bibinfo{volume}{372}}, \bibinfo{pages}{1104} (\bibinfo{year}{2006}),
  \eprint{astro-ph/0604547}.

\bibitem[{\citenamefont{G\'{o}rski}(1994)}]{Gorski94}
\bibinfo{author}{\bibfnamefont{K.~M.} \bibnamefont{G\'{o}rski}},
  \bibinfo{journal}{Astrophys. J. Lett.} \textbf{\bibinfo{volume}{430}},
  \bibinfo{pages}{85} (\bibinfo{year}{1994}), \eprint{astro-ph/9403066}.

\bibitem[{\citenamefont{Slosar et~al.}(2004)\citenamefont{Slosar, Seljak, and
  Makarov}}]{Slosar:2004fr}
\bibinfo{author}{\bibfnamefont{A.}~\bibnamefont{Slosar}},
  \bibinfo{author}{\bibfnamefont{U.}~\bibnamefont{Seljak}}, \bibnamefont{and}
  \bibinfo{author}{\bibfnamefont{A.}~\bibnamefont{Makarov}},
  \bibinfo{journal}{Phys. Rev.} \textbf{\bibinfo{volume}{D69}},
  \bibinfo{pages}{123003} (\bibinfo{year}{2004}), \eprint{astro-ph/0403073}.

\bibitem[{\citenamefont{Page et~al.}(2007)}]{Page:2006hz}
\bibinfo{author}{\bibfnamefont{L.}~\bibnamefont{Page}} \bibnamefont{et~al.}
  (\bibinfo{collaboration}{WMAP}), \bibinfo{journal}{Astrophys. J. Suppl.}
  \textbf{\bibinfo{volume}{170}}, \bibinfo{pages}{335} (\bibinfo{year}{2007}),
  \eprint{astro-ph/0603450}.

\bibitem[{\citenamefont{Kamionkowski et~al.}(1997)\citenamefont{Kamionkowski,
  Kosowsky, and Stebbins}}]{Kamionkowski:1996ks}
\bibinfo{author}{\bibfnamefont{M.}~\bibnamefont{Kamionkowski}},
  \bibinfo{author}{\bibfnamefont{A.}~\bibnamefont{Kosowsky}}, \bibnamefont{and}
  \bibinfo{author}{\bibfnamefont{A.}~\bibnamefont{Stebbins}},
  \bibinfo{journal}{Phys. Rev.} \textbf{\bibinfo{volume}{D55}},
  \bibinfo{pages}{7368} (\bibinfo{year}{1997}), \eprint{astro-ph/9611125}.

\bibitem[{\citenamefont{Gupta and Nagar}(1999)}]{Gupta99}
\bibinfo{author}{\bibfnamefont{A.}~\bibnamefont{Gupta}} \bibnamefont{and}
  \bibinfo{author}{\bibfnamefont{D.}~\bibnamefont{Nagar}},
  \emph{\bibinfo{title}{Matrix Variate Distributions}}
  (\bibinfo{publisher}{Chapman \& Hall}, \bibinfo{year}{1999}), ISBN
  \bibinfo{isbn}{1584880465}.

\bibitem[{\citenamefont{Planck}(2006)}]{unknown:2006uk}
\bibinfo{author}{\bibnamefont{Planck}} (\bibinfo{collaboration}{Planck})
  (\bibinfo{year}{2006}), \eprint{astro-ph/0604069}.

\bibitem[{\citenamefont{Lewis and Bridle}(2002)}]{Lewis:2002ah}
\bibinfo{author}{\bibfnamefont{A.}~\bibnamefont{Lewis}} \bibnamefont{and}
  \bibinfo{author}{\bibfnamefont{S.}~\bibnamefont{Bridle}},
  \bibinfo{journal}{Phys. Rev.} \textbf{\bibinfo{volume}{D66}},
  \bibinfo{pages}{103511} (\bibinfo{year}{2002}), \eprint{astro-ph/0205436}.

\bibitem[{\citenamefont{Martin and Ringeval}(2004)}]{Martin:2003sg}
\bibinfo{author}{\bibfnamefont{J.}~\bibnamefont{Martin}} \bibnamefont{and}
  \bibinfo{author}{\bibfnamefont{C.}~\bibnamefont{Ringeval}},
  \bibinfo{journal}{Phys. Rev.} \textbf{\bibinfo{volume}{D69}},
  \bibinfo{pages}{083515} (\bibinfo{year}{2004}), \eprint{astro-ph/0310382}.

\bibitem[{\citenamefont{Mortlock et~al.}(2002)\citenamefont{Mortlock,
  Challinor, and Hobson}}]{Mortlock00}
\bibinfo{author}{\bibfnamefont{D.~J.} \bibnamefont{Mortlock}},
  \bibinfo{author}{\bibfnamefont{A.~D.} \bibnamefont{Challinor}},
  \bibnamefont{and} \bibinfo{author}{\bibfnamefont{M.~P.}
  \bibnamefont{Hobson}}, \bibinfo{journal}{MNRAS}
  \textbf{\bibinfo{volume}{330}}, \bibinfo{pages}{405} (\bibinfo{year}{2002}),
  \eprint{astro-ph/0008083}.

\bibitem[{\citenamefont{Lewis et~al.}(2002)\citenamefont{Lewis, Challinor, and
  Turok}}]{Lewis:2001hp}
\bibinfo{author}{\bibfnamefont{A.}~\bibnamefont{Lewis}},
  \bibinfo{author}{\bibfnamefont{A.}~\bibnamefont{Challinor}},
  \bibnamefont{and} \bibinfo{author}{\bibfnamefont{N.}~\bibnamefont{Turok}},
  \bibinfo{journal}{Phys. Rev.} \textbf{\bibinfo{volume}{D65}},
  \bibinfo{pages}{023505} (\bibinfo{year}{2002}), \eprint{astro-ph/0106536}.

\bibitem[{\citenamefont{Efstathiou}(2004)}]{Efstathiou:2003dj}
\bibinfo{author}{\bibfnamefont{G.}~\bibnamefont{Efstathiou}},
  \bibinfo{journal}{Mon. Not. Roy. Astron. Soc.}
  \textbf{\bibinfo{volume}{349}}, \bibinfo{pages}{603} (\bibinfo{year}{2004}),
  \eprint{astro-ph/0307515}.

\bibitem[{\citenamefont{Hinshaw et~al.}(2007)}]{Hinshaw:2006ia}
\bibinfo{author}{\bibfnamefont{G.}~\bibnamefont{Hinshaw}} \bibnamefont{et~al.}
  (\bibinfo{collaboration}{WMAP}), \bibinfo{journal}{Astrophys. J. Suppl.}
  \textbf{\bibinfo{volume}{170}}, \bibinfo{pages}{288} (\bibinfo{year}{2007}),
  \eprint{astro-ph/0603451}.

\bibitem[{\citenamefont{Tegmark}(1997)}]{Tegmark:1996qt}
\bibinfo{author}{\bibfnamefont{M.}~\bibnamefont{Tegmark}},
  \bibinfo{journal}{Phys. Rev.} \textbf{\bibinfo{volume}{D55}},
  \bibinfo{pages}{5895} (\bibinfo{year}{1997}), \eprint{astro-ph/9611174}.

\bibitem[{\citenamefont{Wandelt et~al.}(2001)\citenamefont{Wandelt, Hivon, and
  Gorski}}]{Wandelt:2000av}
\bibinfo{author}{\bibfnamefont{B.~D.} \bibnamefont{Wandelt}},
  \bibinfo{author}{\bibfnamefont{E.}~\bibnamefont{Hivon}}, \bibnamefont{and}
  \bibinfo{author}{\bibfnamefont{K.~M.} \bibnamefont{Gorski}},
  \bibinfo{journal}{Phys. Rev.} \textbf{\bibinfo{volume}{D64}},
  \bibinfo{pages}{083003} (\bibinfo{year}{2001}), \eprint{astro-ph/0008111}.

\bibitem[{\citenamefont{Hivon et~al.}(2002)}]{Hivon:2001jp}
\bibinfo{author}{\bibfnamefont{E.}~\bibnamefont{Hivon}} \bibnamefont{et~al.},
  \bibinfo{journal}{\apj} \textbf{\bibinfo{volume}{567}}, \bibinfo{pages}{2}
  (\bibinfo{year}{2002}), \eprint{astro-ph/0105302}.

\bibitem[{\citenamefont{Hansen et~al.}(2002)\citenamefont{Hansen, Gorski, and
  Hivon}}]{Hansen:2002zq}
\bibinfo{author}{\bibfnamefont{F.~K.} \bibnamefont{Hansen}},
  \bibinfo{author}{\bibfnamefont{K.~M.} \bibnamefont{Gorski}},
  \bibnamefont{and} \bibinfo{author}{\bibfnamefont{E.}~\bibnamefont{Hivon}},
  \bibinfo{journal}{Mon. Not. Roy. Astron. Soc.}
  \textbf{\bibinfo{volume}{336}}, \bibinfo{pages}{1304} (\bibinfo{year}{2002}),
  \eprint{astro-ph/0207464}.

\bibitem[{\citenamefont{Brown et~al.}(2005)\citenamefont{Brown, Castro, and
  Taylor}}]{Brown:2004jn}
\bibinfo{author}{\bibfnamefont{M.~L.} \bibnamefont{Brown}},
  \bibinfo{author}{\bibfnamefont{P.~G.} \bibnamefont{Castro}},
  \bibnamefont{and} \bibinfo{author}{\bibfnamefont{A.~N.}
  \bibnamefont{Taylor}}, \bibinfo{journal}{Mon. Not. Roy. Astron. Soc.}
  \textbf{\bibinfo{volume}{360}}, \bibinfo{pages}{1262} (\bibinfo{year}{2005}),
  \eprint{astro-ph/0410394}.

\bibitem[{\citenamefont{Efstathiou}(2006)}]{Efstathiou:2006eb}
\bibinfo{author}{\bibfnamefont{G.}~\bibnamefont{Efstathiou}},
  \bibinfo{journal}{Mon. Not. Roy. Astron. Soc.}
  \textbf{\bibinfo{volume}{370}}, \bibinfo{pages}{343} (\bibinfo{year}{2006}),
  \eprint{astro-ph/0601107}.

\bibitem[{\citenamefont{Smith and Zaldarriaga}(2007)}]{Smith:2006vq}
\bibinfo{author}{\bibfnamefont{K.~M.} \bibnamefont{Smith}} \bibnamefont{and}
  \bibinfo{author}{\bibfnamefont{M.}~\bibnamefont{Zaldarriaga}},
  \bibinfo{journal}{Phys. Rev.} \textbf{\bibinfo{volume}{D76}},
  \bibinfo{pages}{043001} (\bibinfo{year}{2007}), \eprint{astro-ph/0610059}.

\bibitem[{\citenamefont{Szapudi et~al.}(2001)\citenamefont{Szapudi, Prunet,
  Pogosyan, Szalay, and Bond}}]{Szapudi:2000xj}
\bibinfo{author}{\bibfnamefont{I.}~\bibnamefont{Szapudi}},
  \bibinfo{author}{\bibfnamefont{S.}~\bibnamefont{Prunet}},
  \bibinfo{author}{\bibfnamefont{D.}~\bibnamefont{Pogosyan}},
  \bibinfo{author}{\bibfnamefont{A.~S.} \bibnamefont{Szalay}},
  \bibnamefont{and} \bibinfo{author}{\bibfnamefont{J.~R.} \bibnamefont{Bond}},
  \bibinfo{journal}{Astrophys. J. Lett.} \textbf{\bibinfo{volume}{548}},
  \bibinfo{pages}{115} (\bibinfo{year}{2001}), \eprint{astro-ph/0010256}.

\bibitem[{\citenamefont{Chon et~al.}(2004)\citenamefont{Chon, Challinor,
  Prunet, Hivon, and Szapudi}}]{Chon:2003gx}
\bibinfo{author}{\bibfnamefont{G.}~\bibnamefont{Chon}},
  \bibinfo{author}{\bibfnamefont{A.}~\bibnamefont{Challinor}},
  \bibinfo{author}{\bibfnamefont{S.}~\bibnamefont{Prunet}},
  \bibinfo{author}{\bibfnamefont{E.}~\bibnamefont{Hivon}}, \bibnamefont{and}
  \bibinfo{author}{\bibfnamefont{I.}~\bibnamefont{Szapudi}},
  \bibinfo{journal}{Mon. Not. Roy. Astron. Soc.}
  \textbf{\bibinfo{volume}{350}}, \bibinfo{pages}{914} (\bibinfo{year}{2004}),
  \eprint{astro-ph/0303414}.

\bibitem[{\citenamefont{Hinshaw et~al.}(2003)}]{Hinshaw:2003ex}
\bibinfo{author}{\bibfnamefont{G.}~\bibnamefont{Hinshaw}} \bibnamefont{et~al.},
  \bibinfo{journal}{Astrophys. J. Suppl.} \textbf{\bibinfo{volume}{148}},
  \bibinfo{pages}{135} (\bibinfo{year}{2003}), \eprint{astro-ph/0302217}.

\bibitem[{\citenamefont{Challinor and Chon}(2005)}]{Challinor:2004pr}
\bibinfo{author}{\bibfnamefont{A.}~\bibnamefont{Challinor}} \bibnamefont{and}
  \bibinfo{author}{\bibfnamefont{G.}~\bibnamefont{Chon}},
  \bibinfo{journal}{Mon. Not. Roy. Astron. Soc.}
  \textbf{\bibinfo{volume}{360}}, \bibinfo{pages}{509} (\bibinfo{year}{2005}),
  \eprint{astro-ph/0410097}.

\bibitem[{\citenamefont{{Ashdown} et~al.}(2007)\citenamefont{{Ashdown},
  {Baccigalupi}, {Balbi}, {Bartlett}, {Borrill}, {Cantalupo}, {de Gasperis},
  {G{\'o}rski}, {Heikkil{\"a}}, {Hivon} et~al.}}]{Ashdown:2007ta}
\bibinfo{author}{\bibfnamefont{M.~A.~J.} \bibnamefont{{Ashdown}}},
  \bibinfo{author}{\bibfnamefont{C.}~\bibnamefont{{Baccigalupi}}},
  \bibinfo{author}{\bibfnamefont{A.}~\bibnamefont{{Balbi}}},
  \bibinfo{author}{\bibfnamefont{J.~G.} \bibnamefont{{Bartlett}}},
  \bibinfo{author}{\bibfnamefont{J.}~\bibnamefont{{Borrill}}},
  \bibinfo{author}{\bibfnamefont{C.}~\bibnamefont{{Cantalupo}}},
  \bibinfo{author}{\bibfnamefont{G.}~\bibnamefont{{de Gasperis}}},
  \bibinfo{author}{\bibfnamefont{K.~M.} \bibnamefont{{G{\'o}rski}}},
  \bibinfo{author}{\bibfnamefont{V.}~\bibnamefont{{Heikkil{\"a}}}},
  \bibinfo{author}{\bibfnamefont{E.}~\bibnamefont{{Hivon}}},
  \bibnamefont{et~al.}, \bibinfo{journal}{\aap} \textbf{\bibinfo{volume}{471}},
  \bibinfo{pages}{361} (\bibinfo{year}{2007}), \eprint{astro-ph/0702483}.

\bibitem[{\citenamefont{Bennett et~al.}(2003)}]{Bennett:2003bz}
\bibinfo{author}{\bibfnamefont{C.~L.} \bibnamefont{Bennett}}
  \bibnamefont{et~al.} (\bibinfo{collaboration}{WMAP}),
  \bibinfo{journal}{Astrophys. J. Suppl.} \textbf{\bibinfo{volume}{148}},
  \bibinfo{pages}{1} (\bibinfo{year}{2003}), \eprint{astro-ph/0302207}.

\bibitem[{\citenamefont{Gorski et~al.}(2005)}]{Gorski:2004by}
\bibinfo{author}{\bibfnamefont{K.~M.} \bibnamefont{Gorski}}
  \bibnamefont{et~al.}, \bibinfo{journal}{Astrophys. J.}
  \textbf{\bibinfo{volume}{622}}, \bibinfo{pages}{759} (\bibinfo{year}{2005}),
  \eprint{astro-ph/0409513}.

\bibitem[{\citenamefont{Kogut et~al.}(2003)}]{Kogut:2003et}
\bibinfo{author}{\bibfnamefont{A.}~\bibnamefont{Kogut}} \bibnamefont{et~al.},
  \bibinfo{journal}{Astrophys. J. Suppl.} \textbf{\bibinfo{volume}{148}},
  \bibinfo{pages}{161} (\bibinfo{year}{2003}), \eprint{astro-ph/0302213}.

\bibitem[{\citenamefont{Varshalovich et~al.}(1988)\citenamefont{Varshalovich,
  Moskalev, and Khersonskii}}]{AngularMom}
\bibinfo{author}{\bibfnamefont{D.~A.} \bibnamefont{Varshalovich}},
  \bibinfo{author}{\bibfnamefont{A.~N.} \bibnamefont{Moskalev}},
  \bibnamefont{and} \bibinfo{author}{\bibfnamefont{V.~K.}
  \bibnamefont{Khersonskii}}, \emph{\bibinfo{title}{Quantum Theory of Angular
  Momentum}} (\bibinfo{publisher}{Word Scientific, Singapore},
  \bibinfo{year}{1988}), ISBN \bibinfo{isbn}{9971509962}.

\bibitem[{\citenamefont{Lewis and Challinor}(2006)}]{Lewis:2006fu}
\bibinfo{author}{\bibfnamefont{A.}~\bibnamefont{Lewis}} \bibnamefont{and}
  \bibinfo{author}{\bibfnamefont{A.}~\bibnamefont{Challinor}},
  \bibinfo{journal}{Phys. Rept.} \textbf{\bibinfo{volume}{429}},
  \bibinfo{pages}{1} (\bibinfo{year}{2006}), \eprint{astro-ph/0601594}.

\bibitem[{\citenamefont{Hu}(2001)}]{Hu:2001fa}
\bibinfo{author}{\bibfnamefont{W.}~\bibnamefont{Hu}}, \bibinfo{journal}{Phys.
  Rev.} \textbf{\bibinfo{volume}{D64}}, \bibinfo{pages}{083005}
  (\bibinfo{year}{2001}), \eprint{astro-ph/0105117}.

\bibitem[{\citenamefont{Lewis}(2005)}]{Lewis:2005tp}
\bibinfo{author}{\bibfnamefont{A.}~\bibnamefont{Lewis}},
  \bibinfo{journal}{Phys. Rev.} \textbf{\bibinfo{volume}{D71}},
  \bibinfo{pages}{083008} (\bibinfo{year}{2005}), \eprint{astro-ph/0502469}.

\bibitem[{\citenamefont{Akima}(1996)}]{Akima96}
\bibinfo{author}{\bibfnamefont{H.}~\bibnamefont{Akima}}, \bibinfo{journal}{ACM
  Trans. Math. Softw.} \textbf{\bibinfo{volume}{22}}, \bibinfo{pages}{357}
  (\bibinfo{year}{1996}), ISSN \bibinfo{issn}{0098-3500}.

\bibitem[{\citenamefont{Challinor and Lewis}(2005)}]{Challinor:2005jy}
\bibinfo{author}{\bibfnamefont{A.}~\bibnamefont{Challinor}} \bibnamefont{and}
  \bibinfo{author}{\bibfnamefont{A.}~\bibnamefont{Lewis}},
  \bibinfo{journal}{Phys. Rev.} \textbf{\bibinfo{volume}{D71}},
  \bibinfo{pages}{103010} (\bibinfo{year}{2005}), \eprint{astro-ph/0502425}.

\bibitem[{\citenamefont{Das and Bode}(2007)}]{Das:2007eu}
\bibinfo{author}{\bibfnamefont{S.}~\bibnamefont{Das}} \bibnamefont{and}
  \bibinfo{author}{\bibfnamefont{P.}~\bibnamefont{Bode}}
  (\bibinfo{year}{2007}), \eprint{arXiv:0711.3793 [astro-ph]}.

\bibitem[{\citenamefont{Carbone et~al.}(2008)\citenamefont{Carbone, Springel,
  Baccigalupi, Bartelmann, and Matarrese}}]{Carbone:2007yy}
\bibinfo{author}{\bibfnamefont{C.}~\bibnamefont{Carbone}},
  \bibinfo{author}{\bibfnamefont{V.}~\bibnamefont{Springel}},
  \bibinfo{author}{\bibfnamefont{C.}~\bibnamefont{Baccigalupi}},
  \bibinfo{author}{\bibfnamefont{M.}~\bibnamefont{Bartelmann}},
  \bibnamefont{and}
  \bibinfo{author}{\bibfnamefont{S.}~\bibnamefont{Matarrese}},
  \bibinfo{journal}{Mon. Not. Roy. Astron. Soc.}
  \textbf{\bibinfo{volume}{388}}, \bibinfo{pages}{1618} (\bibinfo{year}{2008}),
  \eprint{0711.2655}.

\bibitem[{\citenamefont{Smith et~al.}(2007)\citenamefont{Smith, Zahn, and
  Dore}}]{Smith:2007rg}
\bibinfo{author}{\bibfnamefont{K.~M.} \bibnamefont{Smith}},
  \bibinfo{author}{\bibfnamefont{O.}~\bibnamefont{Zahn}}, \bibnamefont{and}
  \bibinfo{author}{\bibfnamefont{O.}~\bibnamefont{Dore}},
  \bibinfo{journal}{Phys. Rev.} \textbf{\bibinfo{volume}{D76}},
  \bibinfo{pages}{043510} (\bibinfo{year}{2007}), \eprint{arXiv:0705.3980
  [astro-ph]}.

\bibitem[{\citenamefont{Hirata et~al.}(2004)\citenamefont{Hirata, Padmanabhan,
  Seljak, Schlegel, and Brinkmann}}]{Hirata:2004rp}
\bibinfo{author}{\bibfnamefont{C.~M.} \bibnamefont{Hirata}},
  \bibinfo{author}{\bibfnamefont{N.}~\bibnamefont{Padmanabhan}},
  \bibinfo{author}{\bibfnamefont{U.}~\bibnamefont{Seljak}},
  \bibinfo{author}{\bibfnamefont{D.}~\bibnamefont{Schlegel}}, \bibnamefont{and}
  \bibinfo{author}{\bibfnamefont{J.}~\bibnamefont{Brinkmann}},
  \bibinfo{journal}{Phys. Rev.} \textbf{\bibinfo{volume}{D70}},
  \bibinfo{pages}{103501} (\bibinfo{year}{2004}), \eprint{astro-ph/0406004}.

\end{thebibliography}
\providecommand{\aj}{Astron. J. }\providecommand{\apj}{Astrophys. J.
  }\providecommand{\apjl}{Astrophys. J.
  }\providecommand{\mnras}{MNRAS}\providecommand{\aap}{Astron.
  Astrophys.}\providecommand{\aj}{Astron. J. }\providecommand{\apj}{Astrophys.
  J. }\providecommand{\apjl}{Astrophys. J.
  }\providecommand{\mnras}{MNRAS}\providecommand{\aap}{Astron. Astrophys.}


\end{document}